\documentclass{emulateapj}

\shorttitle{Coherent structures in solar wind}
\shortauthors{Perrone et al.}

\usepackage{color}
\begin{document}

\title{Compressive coherent structures at ion scales in the slow solar wind}

\author{D. Perrone$^{1,2}$, O. Alexandrova$^1$, A. Mangeney$^1$, M. Maksimovic$^1$, C. Lacombe$^1$, V. Rokoto$^1$, J. C. Kasper$^3$, D. Jovanovic$^4$}
\affil{$^1$ LESIA, Observatoire de Paris, PSL Research University, CNRS, Sorbonne Universités, UPMC Univ. Paris 06, Univ. Paris Diderot, Sorbonne Paris Cité. \\
$^2$ European Space Agency, Science and Robotic Exploration Directorate, ESAC, Madrid, Spain. \\
$^3$ University of Michigan, Ann Arbor, MI, United States. \\
$^4$ Institute of Physics, University Belgrade, Belgrade, Serbia. 
}

\begin{abstract}
We present a study of magnetic field fluctuations, in a slow solar wind stream, close to ion scales, where an increase of the level of magnetic compressibility is observed. Here, the nature of these compressive fluctuations is found to be characterized by coherent structures. Although previous studies have shown that current sheets can be considered as the principal cause of intermittency at ion scales, here we show for the first time that, in the case of the slow solar wind, a large variety of coherent structures contributes to intermittency at proton scales, and current sheets are not the most common. Specifically, we find compressive ($\delta b_{\|} \gg \delta b_{\perp}$), linearly polarized structures in the form of magnetic holes, solitons and shock waves. Examples of Alfv\'enic structures ($\delta b_{\perp} > \delta b_{\|}$) are identified as current sheets and vortex-like structures. Some of these vortices have $ \delta b_{\perp} \gg \delta b_{\|}$, as in the case of Alfv\'en vortices, but the majority of them are characterized by $\delta b_{\perp} \gtrsim \delta b_{\|}$. Thanks to multi-point measurements by Cluster spacecraft, for about 100 structures, we could determine the normal, the propagation velocity and the spatial scale along this normal. Independently of the nature of the structures, the normal is always perpendicular to the local magnetic field, meaning that $k_{\perp} \gg k_{\parallel}$. The spatial scales of the studied structures are found to be between 2 and 8 times the proton gyroradius. Most of them are simply convected by the wind, but 25\% propagate in the plasma frame. Possible interpretations of the observed structures and the connection with plasma heating are discussed.
\end{abstract}

\section{Introduction}

The interplanetary medium, a plasma that is almost collisionless, magnetized and, despite being highly ionized, neutral, can be considered the best natural laboratory to study the dynamical behavior of turbulent plasmas. `In situ' spacecraft measurements generally reveal that the solar wind plasma is usually in a state of fully-developed turbulence, where electromagnetic fields and plasma properties have a very large number of  excited degrees of freedom \cite[]{bru05, mar06}. 

The large scales are essentially incompressible and the fluctuations of magnetic field and plasma velocity are often highly correlated, so that at times they can be thought of as nearly perfect Alfv\'en waves \cite[]{gos09,mat14}. The power spectra, in the inertial range of the turbulent cascade, manifest a behavior reminiscent of the Kolmogorov power law for fluid turbulence \cite[]{kol41,mat82,tu95}. Then, the turbulent cascade extends to smaller spatial scales down to a wavelength range where kinetic effects govern the plasma dynamics. At these scales, around the proton characteristic lengths, different physical processes come into play, leading to a change in the spectral shape \cite[]{lea98,lea00,bal05,smi06,ale07,ale08,bou12}. Although the origin of such variations is not yet well understood, a recent analysis by \cite{bru14} has shown that the spectral slope above the frequency break is strongly related to the wind speed and to the power of the fluctuations within the inertial range; i.e. steeper spectra are found when speed and power are higher. Moreover, another important aspect, recovered at these scales, is an enhancement of magnetic compressive fluctuations \cite[]{ale07,ale08,ham08,sal12,kiy13}, but the nature of the compressive component remains uncertain. At shorter scales (smaller than the proton characteristic scales and up to a fraction of the electron scales), the energy continues to be transferred and another spectrum is observed, whose interpretation is still controversial \cite[]{ale09,ale12,sah10,sah13}.

Although the physical mechanisms of solar wind turbulence have been matter of investigation for many decades, some of the primary problems concerning the nature of turbulent fluctuations along the turbulent cascade and the dissipation in a collisionless plasma still remain a puzzle. A fundamental question is whether space plasma turbulence can be considered as a mixture of quasi-linear waves, as Alfv\'en waves at MHD scales \cite[]{mat14} and whistlers or kinetic Alfv\'en waves at kinetic scales \cite[]{bal05,pod11,he11,sal12,lac14}, or if the turbulence is strong with formation of coherent structures responsible for intermittency and departure from self-similarity \cite[]{bis93,fri95}, or even if the coexistence of both waves and structures is a more realistic vision  \cite[]{new01,rob13}. 

Intermittency phenomenon is the manifestation of non uniform and not homogeneous dissipation of the energy of a turbulent system \cite[]{fri95}. Different analyses have been focused on the characterization of intermittency in the solar wind \cite[]{bur91,bur93,mar94,car95,car96,hor97,vel99,vel99b,bru03,sor01,sor05}. Many of these studies are centered around the departure of the probability distribution functions (PDFs) from the Gaussian statistics or the anomalous scaling of structure functions and the relative deviations from self-similarity. These approaches are purely statistical and geometrical  and do not provide any information on the physics of intermittency. An efficient way to study intermittency is given by wavelet analysis method, through a decomposition of a signal into time-frequency space \cite[]{far92}. While the Fourier transform is inherently nonlocal (due to the nature of trigonometrical functions) and the information is completely delocalized among the spectral coefficients, wavelet transforms are able to catch localized events  in time and frequency. It is important to note, however, that the choice of the mother function of the wavelet transform may favor the selection of different events: for example, the Haar wavelet (which is similar to a step function) will find mostly sharp discontinuities in the signal, while the Morlet mother function will favor wave packet-like fluctuations.  

In hydrodynamic turbulence, the intermittency has been observed in the appearance of localized coherent structures in the vorticity field, spontaneously produced by nonlinear dynamics \cite[]{she90,fri95}. The characteristic length of these filaments of vorticity is of the order of the energy injection scale, while the cross-section is of the order of the dissipation scale. On the other hand, the appearance of coherent structures has also been observed in turbulent plasma systems. In the solar wind, in fact, the magnetic field has been found to be spatially characterized by abrupt changes, related to changes in plasma characteristics (e.g. particle heating). Their investigation has led to the discovering of a large variety of structures, with different properties. Recently, many efforts have been put forward to understand the nature of coherent structures by using `in situ' measurements \cite[]{bru01,bru07,ale04,ale06,sun05,ale08b,sal09,osm11,osm12,gre12,per12,zhd12,tes13,gre14} and numerical data analysis \cite[]{gre08,wan12a,wan12b,ser12,ser14,gre12b,wu13,per13,per14,hay15}. 

During the last years, the search for discontinuities in solar wind, by using the Haar wavelet \cite[]{vel99} or the partial variance of increments (PVI) technique \cite[]{gre08}, resulted in finding one-dimensional current sheets in different regimes of the turbulent cascade \cite[]{vel99b,bru01,osm11,osm12,per12,gre12,gre14}. An extensive study of these structures has shown that the current sheets are almost incompressible and pressure balanced, with the component of maximum variation which changes sign and is perpendicular to the mean magnetic field. Different examples of current sheets have been also recovered in other turbulent regions of the interplanetary medium, such as in the Earth's magnetosheath \cite[]{ret07,sun07,cha15}, showing that thin current sheets are important sites of energy dissipation and particle heating.

Recently, \cite{rob13}, using the $k$-filtering technique \cite[]{pin91} and wave polarization analysis, have given some observational indications that the fast solar wind turbulence may be populated by kinetic Alfv\'en waves, small scale current sheets and by Alfv\'en vortices. The Alfv\'en vortex is one of the non-linear solutions of the ideal incompressible MHD equations and it is characterized by magnetic and velocity fluctuations mostly perpendicular to the unperturbed magnetic field \cite[]{pet92}. Signatures of Alfv\'en vortices have been observed in the solar wind by \cite{ver03} at scales of order of minutes. At scales close to the proton spectral break, Alfv\'en vortices have been observed for the first time in Earth's and then in Saturn's magnetosheaths \cite[]{ale06,ale08b}. Recently, large amplitude Alfv\'en vortex-like structures at ion scales have been detected in a fast solar wind stream by \cite{lio15}. A multi-satellite analysis of one well-defined Alfv\'en vortex in the slow wind can be found in \cite{rob15}.

Current sheets and Alfv\'en vortices are coherent structures with $ \delta b_{\perp} \gg \delta b_{\|}$. However, there exist also very compressive structures with $ \delta b_{\|} \gg \delta b_{\perp}$, namely magnetic solitons or magnetic holes. The term magnetic hole was introduced by \cite{tur77} to indicate a localized depression in the magnitude of the interplanetary magnetic field. More recently, \cite{bau99} suggested an alternative description of magnetic hole, using the concept of magnetic soliton. Events of this class are usually believed to result from the mirror instability; that requires high plasma beta and a perpendicular temperature anisotropy \cite[]{win94,erd96,gen09}. According to previous studies, these events are characterized by a localized change in magnetic field accompanied by simultaneous changes in plasma density and kinetic pressure. The typical scales of these events cover a range from 5 seconds to several tens of seconds, which corresponds to a thickness from about 10 to several tens of proton inertial lengths ($\lambda_p$).

\cite{ree06} presented observational evidence for a class of structures that has the characteristics of solitons, using magnetic field measurements from the Ulysses magnetometer. These events appear as pulses in the magnetic field magnitude, associated with a rotation in the field direction itself. The duration of the impulsive events is of the order of 30 seconds, with the rotation lasting longer than the field enhancement. An approximate scale size of about 30 $\lambda_p$ has been determined, in the direction parallel to the minimum variance direction, assuming that these structures are convected by the wind.

Multi-spacecraft observations of solitary structures have been performed by \cite{sta03}, thanks to {\em Cluster} measurements in the magnetopause boundary layer, showing the existence of a slow-mode magnetosonic soliton with a short duration of 10 seconds, observed by two spacecraft. This structure propagates close to a direction perpendicular to the mean magnetic field, with a speed of about 250 km/s with respect to the satellites (the velocity in the background medium cannot be completely determined because the structure is observed only by two spacecraft) and it has a perpendicular size of 1000-2000 km. More recently, \cite{ste07} presented a first statistical study of a magnetic hole plasma signature in a steady solar wind, using {\em Wind} observations. These structures have been shown to be pressure-balanced events with similar properties from fluid to kinetic scales (from about $10^3$ to 10 proton Larmor radii, $\rho_p$).

The aim of the present work is to study the nature of the turbulent fluctuations around the proton scales in the slow solar wind turbulent cascade, using high-time resolution magnetic field data from {\em Cluster} spacecraft. For this purpose we will use different tools of time series analysis. First, we will apply Morlet wavelet transform in order to check if our signal is homogeneous or intermittent in time and frequencies. We also use wavelets to determine spectral properties of the signal and as a pass-band filter, to focus on the fluctuations around proton scales only. These fluctuations are more compressive than those within the inertial range. This property motivates us to focus our study on events with a finite compressive component. Second, in order to select intermittent events without any a priori idea of their shape (to be the most general possible), we use magnetic fluctuations around proton scales rather than the wavelet coefficients at these scales. As a result, we find that $\sim40\%$ of the analyzed time interval ($\sim 600$ events of several seconds during $\sim2$~h) is covered by coherent structures of different nature. Among them we observe linearly polarized compressive events like  magnetic hole-, soliton-like or shock structures; linearly polarized Alfv\'enic events (i.e. with dominant transverse fluctuations $\delta b_{\perp} > \delta b_{\|}$) like current sheets and elliptically polarized Alfv\'enic events, which look like magnetic vortices. Using the four satellites of {\em Cluster}, we are able to characterize 109 (out of 600) structures in terms of the orientation of their normals and propagation in the plasma frame.

The paper is organized as follows: in Section \ref{sec:data} the selected data interval is described in terms of plasma parameters and turbulent behavior; in Section \ref{sec:ident} the selection method for intermittent coherent events is presented and different coherent structures are given as examples; in Section \ref{sec:statistic} we show the results of statistical studies, using one- and multi-satellite approaches.

\section{Studied Time Interval}
\label{sec:data}

\begin{figure}
\begin{center}
\includegraphics [width=7.5cm]{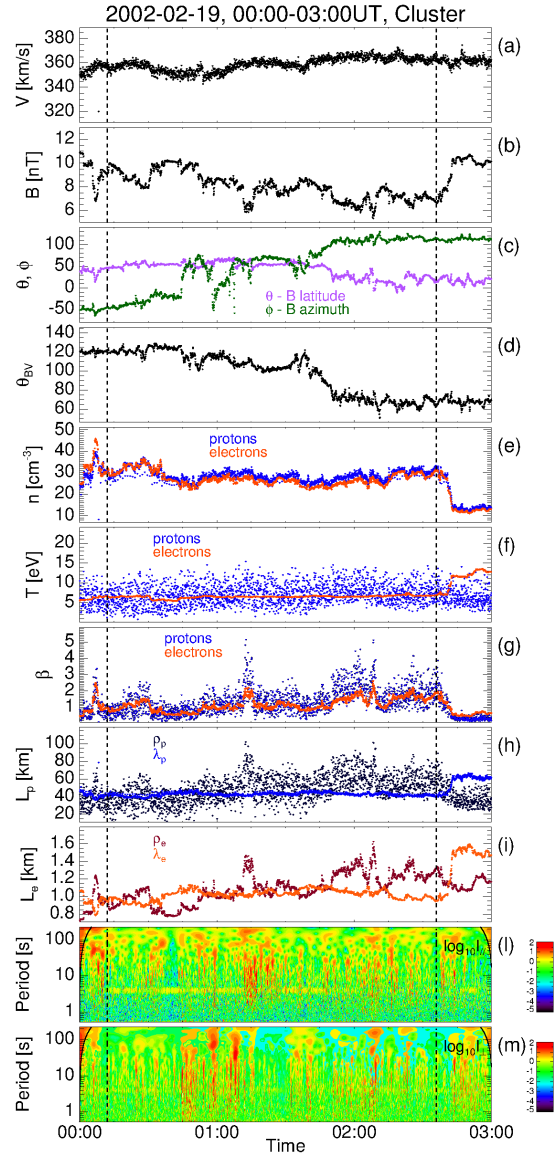}
\caption{Overview of solar wind data for the time interval 00:00-03:00 UT on February $19^{th}$, 2002 from {\em Cluster}. From top to bottom: magnitude of $V$ (a) and $B$ (b), latitude ($\theta$, purple dots) and azimuth ($\phi$, green dots) angles of $B$ (c) and $\theta_{BV}$ (d). Proton (blue dots) and electron (red dots) density (e), temperature (f) and plasma beta (g). Characteristic lengths for protons $L_p$ (h) and electrons $L_e$ (i): $\rho_i$ (dark lines) and $\lambda_i$ (light lines), with $i=p,e$. Logarithmic contour plots of LIM, $I(\tau,t)$ (see text), for parallel (l) and perpendicular (m) magnetic field fluctuations.
Vertical dashed lines denote the time interval 00:12-02:36 UT used in the present analysis. }
\label{fg:stream}
\end{center}
\end{figure}

The present study is based on `in situ' measurements from the {\em Cluster} mission in an interval of time (February 19$^{th}$, 2002) when the spacecraft was at apogee and in pure solar wind plasma. The same interval of three hours (00:00-03:00 UT) has been analyzed by \cite{bal05} and by \cite{kel05}. 

Different plasma experiments on board the {\em Cluster} spacecraft have been considered in order to characterize the observed solar wind stream. In particular, we consider high-resolution magnetic field data given by the fluxgate magnetometers (FGM) on {\em Cluster 1} (C1), with a sampling time of $22$ Hz \cite[]{bal01}. Proton data have been obtained from the Hot Ion Analyser (HIA) sensor of the Cluster Ion Spectrometry (CIS) experiment on C1 with a resolution of 4 seconds \cite[]{rem01}. For the electrons, two different experiments have been taken into account: Plasma Electron and Current Experiment (PEACE) on C2 for electron temperature (no well-resolved PEACE data on C1 are available for the present interval) with a resolution of 4 seconds \cite[]{szi01} and Waves of High Frequency and Sounder for Probing of the Electron Density by Relaxation (WHISPER) on C1 with a resolution of $1.5$ seconds \cite[]{dec01}. 

An overview of the considered time interval is summarized in Figure~\ref{fg:stream}. 
Panels (a) and (b) show the magnitude of velocity and magnetic fields, respectively, where the bulk velocity in the spacecraft frame has been corrected for the $\sim 30$ km/s aberration produced by the orbital speed of the spacecraft and Earth around the Sun. The correction consists of subtracting $\sim 30$ km/s from the $y_{GSE}$ component, because the $y_{GSE}$ axis is anti-parallel to the direction of Earth's motion.
The present data set is representative of a slow solar wind, characterized by a mean velocity of about 360 km/s and a mean magnetic field magnitude of about 9~nT. In panel (c), the temporal evolution of the latitude ($\theta$, purple dots) and the azimuth ($\phi$, green dots) angles of the magnetic field is displayed. A change in the behavior of these quantities is observed around 01:50 UT. $\theta$ is almost constant before this instant ($\sim 50^{\circ}$), then it starts to vary between $0^{\circ}$ and $50^{\circ}$. Conversely, the azimuthal component widely changes before this instant: $\phi$ starts at $-50^{\circ}$, then abrupt magnetic field reversals are observed. After 01:50 UT, $\phi$ reaches $120^{\circ}$ and remains almost steady. At the same time, a variation from $125^{\circ}$ to $55^{\circ}$ in the angle between magnetic and velocity fields, $\theta_{BV}$, is observed (panel (d)).

Figure~\ref{fg:stream}, panels (e) and (f), present the density and the temperature, for protons (blue dots) and electrons (red dots), respectively. The two populations have about the same mean value for the total temperature. The large fluctuations in the proton temperature are essentially due to the digitization in energy, that for protons is $\sqrt{m_i/m_e}$ times worse than for electrons (I. Dandouras, Private Communication 2015). The mean plasma density is about 25-30 cm$^{-3}$ before 02:45~UT and drops to about 10 cm$^{-3}$ after this time.
This density jump coincides with the jumps in temperature and in magnetic field magnitude, indicating a change of the solar wind stream. In this study we focus on the time interval 00:12-02:36~UT (between the two vertical dashed lines) which represents a more or less homogeneous solar wind.

Figure~\ref{fg:stream}, panel (g), shows the plasma beta for protons (blue dots) and electrons (red dots), defined as the ratio between proton/electron kinetic pressure and magnetic pressure. Even if the values for $\beta_p$ are scattered by the inaccuracy in the temperature determination, the mean value is about 1.5. In \cite{bal05}, the averaged value for the proton beta is $\sim 5$ (about three times greater than in our case). This discrepancy comes from the fact that  \citet{bal05} used CIS/CODIF data on C4 (Stuart Bale, Private Communication, 2015) and we use here CIS/HIA data, more appropriate for the solar wind measurements. Although CODIF on C4 was operating on the low-sensitivity side during the solar wind mode, meaning that the solar wind beam has been correctly detected, CODIF is less accurate than HIA in the solar wind, due to the time-of-flight principle of operation. In fact, in solar wind, protons can saturate the CODIF instrument and, therefore, the moments of the ion distribution functions, produced by CODIF can be incorrect \cite[]{rem97}.

Using the particle and field data, we could also determine the total pressure (not shown here). We found it to be almost constant during the studied time interval, with a mean value of $8 \cdot 10^{-2}$ nPa. 

Figure~\ref{fg:stream}, panel (h) shows the proton characteristic lengths, such as the Larmor radius and the inertial length. Panel (i) shows the same characteristic lengths for the electrons. The Larmor radii, $\rho_{p,e}$ (dark dots), are defined as the ratio between the perpendicular thermal speeds ($v_{th}^{(p,e)}=\sqrt{2kT_{\perp}^{(p,e)}/m_{p,e}}$), and the particle cyclotron frequencies, $\omega_c^{(p,e)}$. The inertial lengths, $\lambda_{p,e}$ (light dots) are defined as the ratio between the light speed, $c$, and the particle plasma frequencies, $\omega_p^{(p,e)}$. 

Figure~\ref{fg:stream}, panels (l) and (m), show the evolution of the energy of magnetic fluctuations (parallel and perpendicular to ${\bf B}_0$, respectively) in time and at different scales, normalized at each time point by a mean spectrum over the whole time interval. 

The decomposition in time, $t$, and scales, $\tau$, is done using the wavelet transform: 
\begin{equation}
\label{eq:coeff}
{\mathcal W}_i(\tau,t) = \sum_{j=0}^{N-1} B_i(t_j)\psi^{*}[(t_j-t)/\tau] \ ,
\end{equation}  
where $B_i(t_j)$ is the $i$-th component of the magnetic field and  $\psi^{*}$ is the conjugate of a wavelet function. The mother function used in the present analysis is the Morlet wavelet:
\begin{equation}
\psi(u) = 2^{1/2}\pi^{-1/4} \cos(\omega_0 u) \exp(-u^2/2) \ , 
\end{equation}
which consists of a plane wave modulated by a Gaussian, where $\omega_0$ is the non dimensional frequency and is taken to be 6 to satisfy the admissibility condition \cite[]{far92}.

The compressive fluctuations (panel (l)) are approximated by the variations of magnetic field magnitude, so the corresponding energy is
\begin{equation}
{\mathcal W}_{\|}^2(\tau,t) = {\mathcal W}_{|B|}^2(\tau,t).  
\end{equation}  
Independently of the definition of a mean magnetic field, $B_0$, this approximation is valid when the level of fluctuations is much lower than $B_0$ ($\delta {B}/B_0 \ll 1$). 
In this case, we can write 
\begin{eqnarray}
 |B|^2 =  ({\bf B\cdot B}) =  ({\bf B_0} + \delta {\bf B})^2  \simeq  B_0^2 + 2 \delta {\bf B} \cdot {\bf B_0} \; , 
\end{eqnarray}
then, the variations of the field amplitude is 
\begin{equation}
\delta |B|^2 =  |B|^2 - B_0^2 =  2 \delta {\bf B} \cdot {\bf B_0}  = 2 \delta B_{\|}B_0 \simeq \delta B_{\|}^2. 
\end{equation}

Knowing the total energy of magnetic fluctuations as a function of time and scale
\begin{equation}
{\mathcal W}_{{\bf B}}^2(\tau,t) = \sum_{j} {\mathcal W}_{j}^2(\tau,t), \; j=x,y,z, 
\end{equation}  
we define the energy of Alfv\'enic (or transverse to the mean field) fluctuations, independently of $B_0$, as 
\begin{equation}
\label{eq:W2-perp}
{\mathcal W}_{\perp}^2(\tau,t) = {\mathcal W}_{{\bf B}}^2(\tau,t) -  {\mathcal W}_{\|}^2(\tau,t).  
\end{equation}  

The normalization used in Figure~\ref{fg:stream}, panels (l) and (m), is the following:
\begin{equation}
I_{\|,\perp}(\tau,t) = \frac{| {\mathcal W_{\|,\perp}}(\tau,t) |^2}{\langle| {\mathcal W_{\|,\perp}}(\tau,t) |^2 \rangle_t},
\end{equation}
where the angled brackets indicate the time average. In the literature it is called the {\it Local Intermittency Measure} (LIM) \cite[]{far92}. It is worth pointing out that the horizontal light band, in panel (l), around 4 seconds is due to the spin satellite frequency. The curved black lines, on each side of the plots, represent the cone of influence, where the Morlet coefficients are affected by edge effects \cite[]{tor98}.

Local Intermittency Measure representation helps to see small  (and less energetic) scales  in more detail. In Figure~\ref{fg:stream}, panels (l) and (m), one observes a non-homogeneous distribution of magnetic energy in time with appearance of localized energetic events covering a range of scales: an inherent property of intermittent coherent  structures (Frisch, 1995).  We will study these structures in the next section of the paper. Before that, we consider  statistical  properties of the studied turbulent flow.  
 
\begin{figure}
\begin{center}
\includegraphics [width=8.8cm]{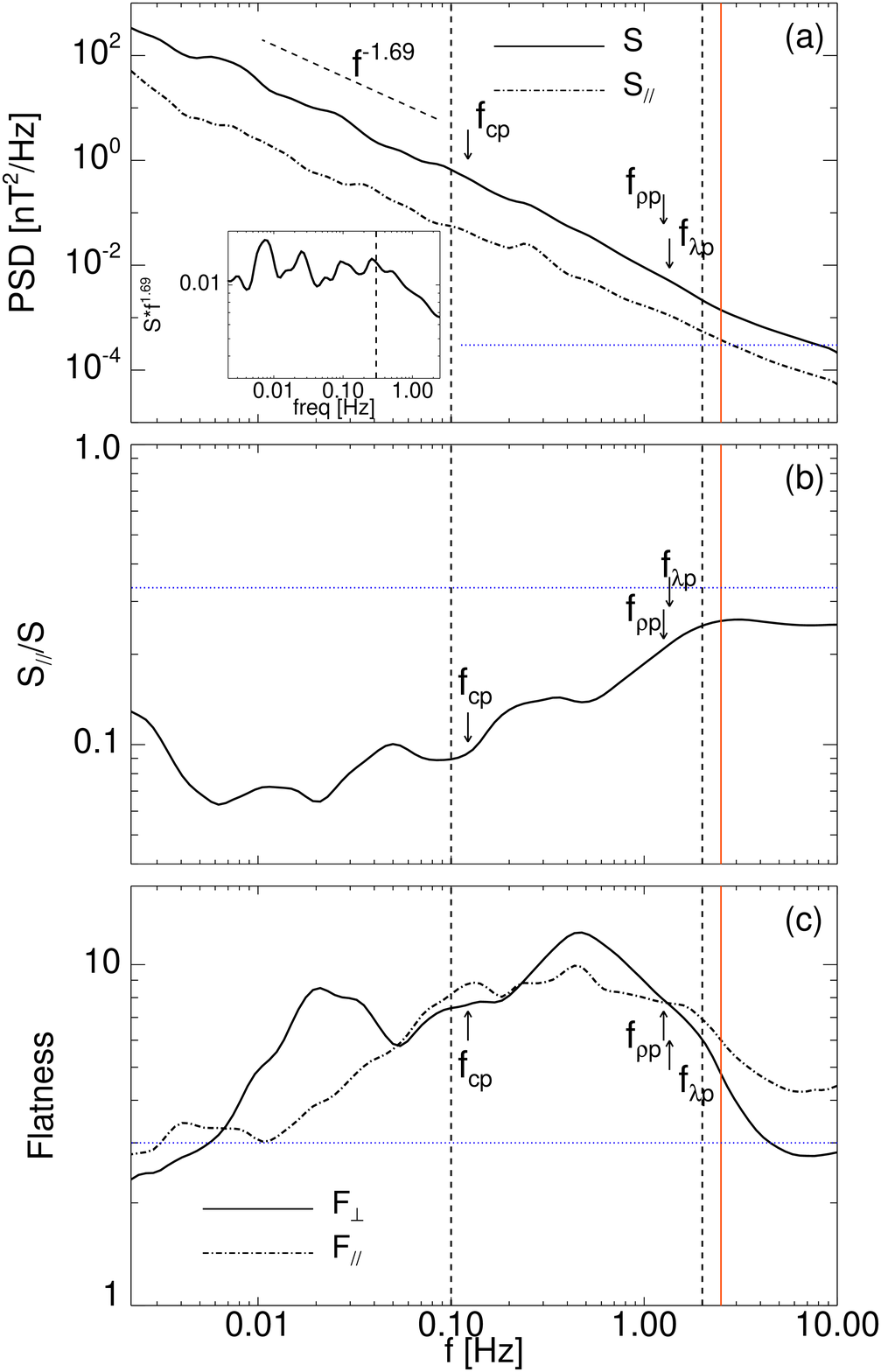}
\caption{Panel (a): PSD of total ($S$, solid line) and compressive ($S_{\parallel}$, dot-dashed line) magnetic fluctuations. The horizontal blue-dotted line indicates the SNR=3. The insert in panel (a) shows the compensated spectrum $S * f^{1.69}$. Panel (b): Level of compressibility of magnetic fluctuations ($S_{\parallel}/S$) as a function of frequency. The horizontal blue-dotted line refers to the isotropic case. Panel (c): Flatness for Alfv\'enic ($F_{\perp}$, solid line) and compressive ($F_{\parallel}$, dot-dashed line) magnetic fluctuations. The value of the flatness for a standard normal distribution (horizontal blue-dotted line) is given as reference. The vertical red-solid line indicates the maximum resolved frequency for the spectra ($f_{max}=2.5$ Hz).}
\label{fg:spectra}
\end{center}
\end{figure}

Figure~\ref{fg:spectra}(a) shows the total power spectral density (PSD) of magnetic fluctuations $S(f)=\sum_{i=x,y,z} S_i(f)$ (solid line). $S_i$ is the PSD  of $B_i$ component, defined as 
\begin{equation}
S_i(\tau) = \frac{2 \delta t}{N} \sum_{j=0}^{N-1} | {\mathcal W}_i(\tau,t_j) |^2 \ , \ \ \ i=x,y,z
\end{equation}
where $\delta t$ represents the time spacing. The frequency dependence is easily obtained using the $f=1/\tau$ relationship.
In the MHD range, $S(f)$ manifests the typical behavior of the Kolmogorov power law, $f^{-1.69}$.  Then, in-between the proton cyclotron frequency $f_{cp}$ and Doppler shifted proton Larmor radius $f_{\rho_p}=v_{sw}/2\pi {\rho_p}$ and proton inertia length $f_{\lambda_p}=v_{sw}/2\pi {\lambda_p}$ (estimated under the assumption of wave-vector parallel to the plasma flow, see the arrows in the plot), at $\sim 0.3$ Hz, the spectrum changes its slope. This is more visible in the insert of panel (a), where a compensated spectrum $S(f)\cdot f^{1.69}$ is displayed.

In order to study kinetic scales, we should, first, determine the frequency range where the measurements are not affected by the instrumental noise. The digitization of the Cluster/FGM instrument in the solar wind mode is $10^{-2}$~nT, therefore in the spectrum it appears at $10^{-4}$~nT$^2$/Hz for a one-component spectrum. So, for the total PSD, $S_{noise}=3\cdot 10^{-4}$~nT$^2$/Hz. We fix the maximum resolved frequency for the spectra at $f_{max} \simeq 2.5$~Hz (vertical red-solid line), that corresponds to a signal to noise ratio (SNR) for a one-component spectrum equal to 3 (and for the total spectrum, equal to 9), see the horizontal blue-dotted line. 
Therefore, in the kinetic range, we can study the frequency range $f = [0.3,2.5]$~Hz, that is nearly one decade. Within this range, the total PSD becomes steeper, with a spectral slope of $\simeq -2$. It is in agreement with recent results of \cite{bru14} for slow solar wind spectra.

The PSD of the compressive fluctuations 
\begin{equation}
S_{\parallel}(\tau)=\frac{2 \delta t}{N} \sum_{j=0}^{N-1} | {\mathcal W}_{\|}(\tau,t_j) |^2
\end{equation}
is shown by dash-dotted line in Figure~\ref{fg:spectra}(a). It follows $S(f)$ in the MHD range of scales. A small bump of compressive energy around 0.25~Hz corresponds to the satellite spin, visible in the Morlet scalogram (panel (l) of Figure~\ref{fg:stream}) as a horizontal band around $\tau=4$~s.


Figure~\ref{fg:spectra}(b) shows the level of compressibility of magnetic turbulent fluctuations as a function of frequency, defined as $S_{\parallel}(f)/S(f)$. The horizontal blue-dotted line indicates the isotropic case $S_{\parallel}= S/3 $. 
The level of compressive fluctuations starts to increase at the end of the MHD range and continues around proton characteristic scales as was already observed by \cite{ale08}, \cite{sal12} and \cite{kiy13}. In our study, unfortunately, we have no information on sub-ion scales compressibility, because the maximal frequency is very close to (but higher than) the highest ion characteristic frequency ($f_{\lambda p}$). However, with our data we can study in detail a frequency range around all ion scales, i.e. $[0.1,2]$~Hz (see vertical black-dashed lines), where the increase of compressibility is observed.

Figure~\ref{fg:spectra}(c) shows the fourth order moment of compressive (dot-dashed line) and Alfv\'enic (solid line) magnetic fluctuations, defined as
\begin{equation}\label{eq:flatper}
F_{\|,\perp}(\tau) = \frac{\langle {\tilde{\mathcal W_{\|,\perp}}}(\tau,t)^4 \rangle}{\langle {\tilde{\mathcal W_{\|,\perp}}}(\tau,t)^2 \rangle^2} \ , 
\end{equation}
where $\tilde{\mathcal W}$ is the real part of the wavelet coefficients. The value for the flatness of a standard normal distribution is 3, indicated on the plot by an horizontal blue-dotted line. The study of solar wind turbulence suggests that the intermittency increases when considering smaller and smaller scales or, equivalently, higher and higher frequencies, starting from MHD scales \cite[]{bru03}. In our case, we are not able to catch non-Gaussian contributions at large scales because of the limited length of the data set ($\sim 2$ hours). However, we observe that, at the end of the MHD range, both curves of flatness depart from the value of flatness of the normal distribution. Note that, here, we calculate flatness of Alfv\'enic fluctuations using the definition (\ref{eq:W2-perp}). If we project $\delta {\bf B}$ in the mean field frame (with the mean field ${\bf B}_0$ defined as a mean over the total interval of study), we can calculate the flatness of two perpendicular components. We find that the flatness $F_{\perp}$, shown in Figure~\ref{fg:spectra}(c), corresponds to the median of the two perpendicular components. 

Between the proton characteristic scales, the flatness of compressive fluctuations, $F_{\|}(f)$, becomes more or less constant while the flatness of transverse fluctuations $F_{\perp}(f)$ reaches its maximum at $\sim 0.4-0.5$~Hz and then starts to decrease. Another local maximum of $F_{\perp}$ is observed around 0.02~Hz, that is about $0.2f_{cp}$, i.e. the frequency where Alfv\'en Ion Cyclotron (AIC) waves can be unstable. (However, AIC waves are out of the scope of the present paper.) The observed fluctuating behavior of  flatness reflects the non-homogeneous distribution of turbulent fluctuations, as observed at the Morlet wavelet scalograms of Figure~\ref{fg:stream}.

In the following part of the paper, we will focus on a range of scales just around ion scales, $f \in [0.1,2]$~Hz (denoted in Figure~\ref{fg:spectra} by vertical dashed lines), which corresponds to a time scale range $\tau \in [0.5,10]$~s.

\section{Identification of Intermittent Events}
\label{sec:ident}

\begin{figure}
\begin{center}
\includegraphics [width=8.6cm]{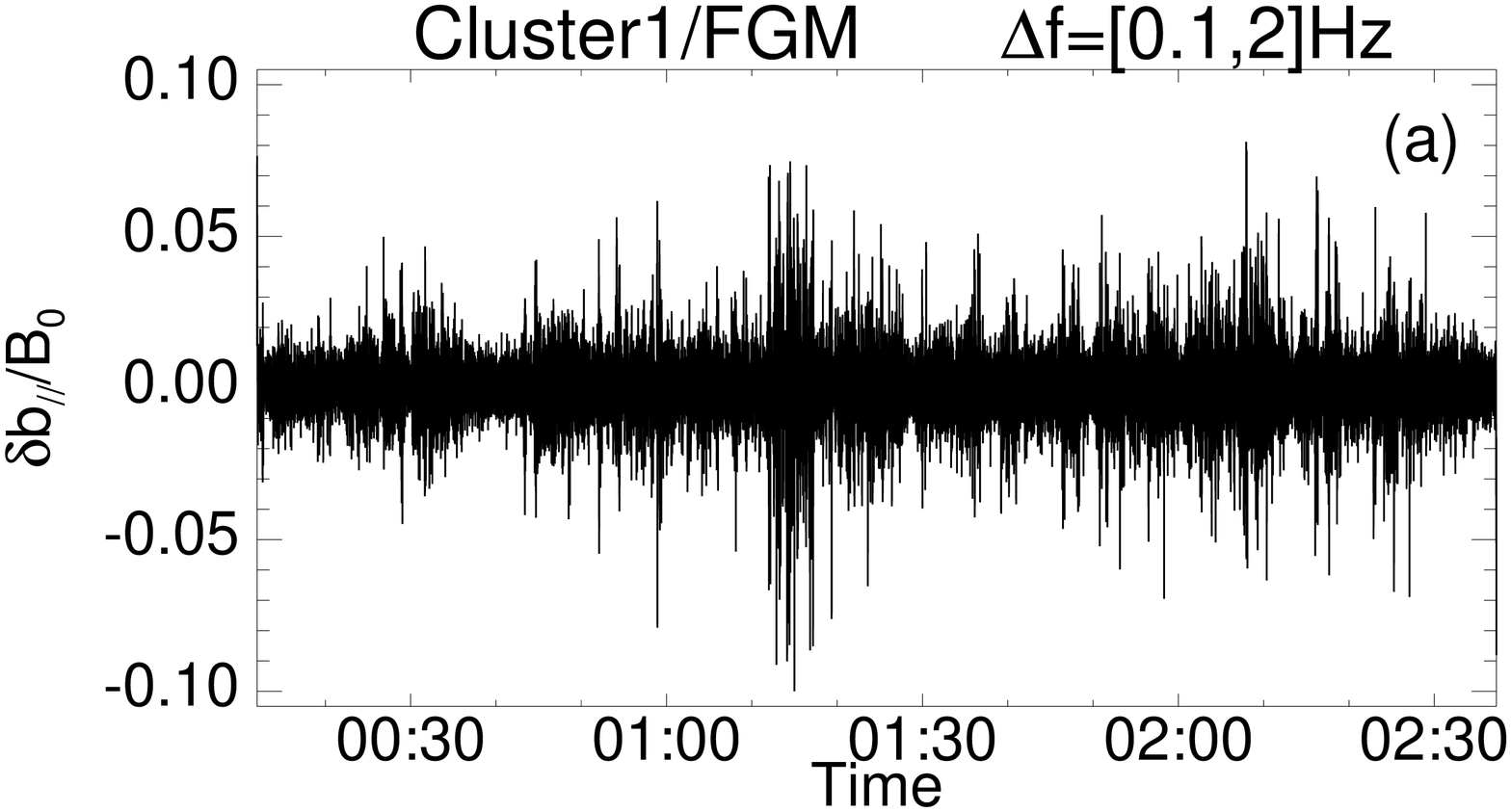}
\includegraphics [width=8.6cm]{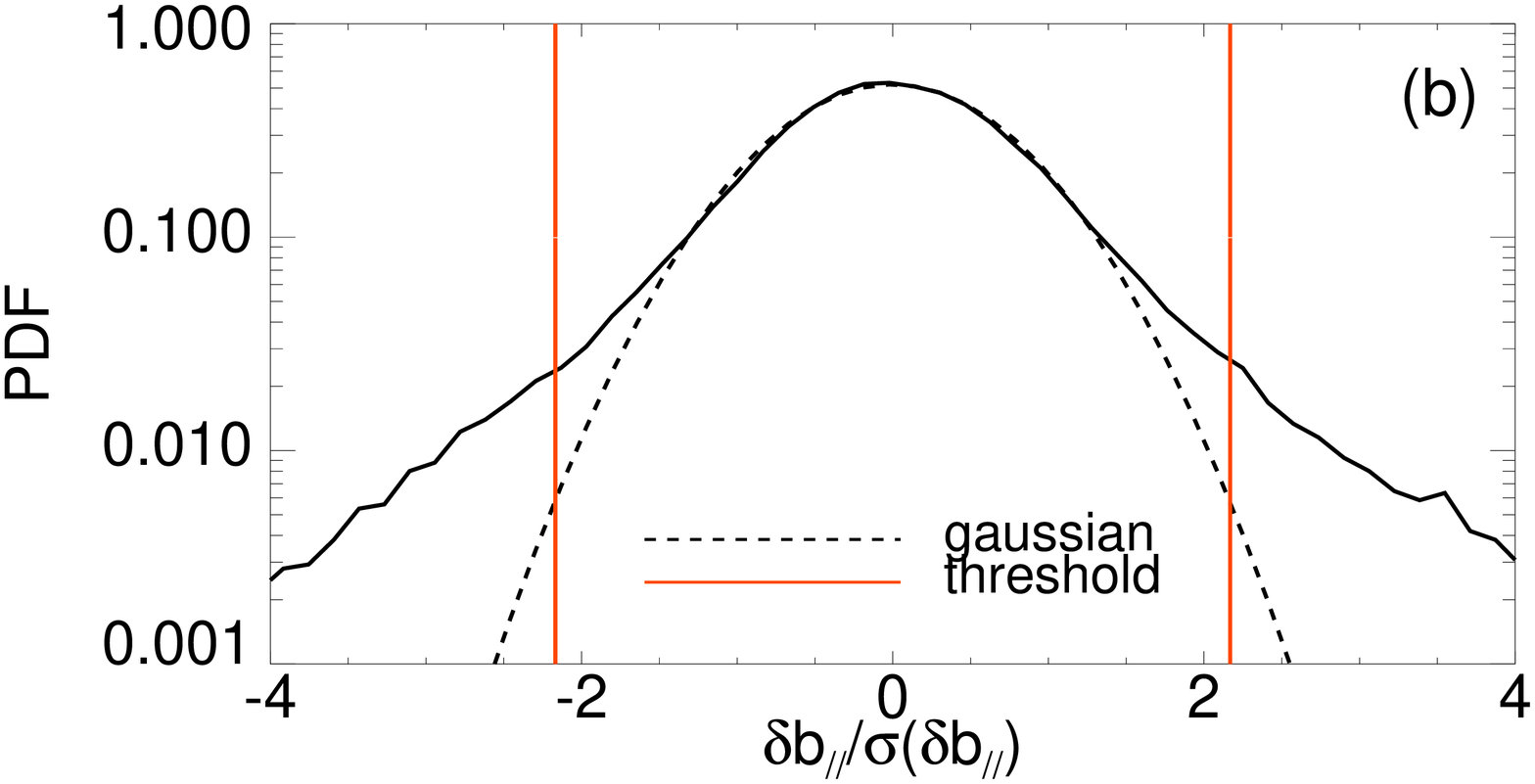}
\includegraphics [width=8.6cm]{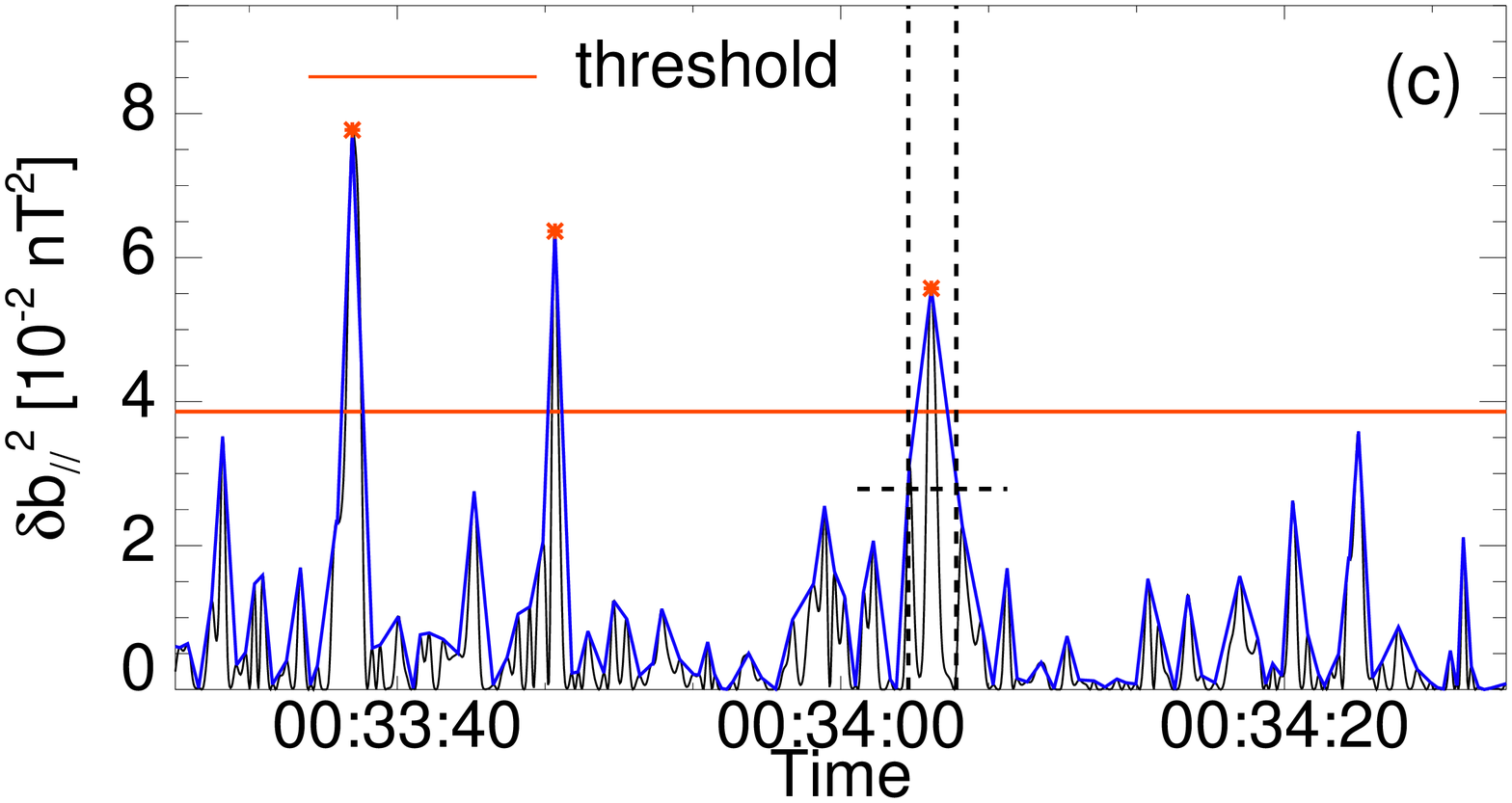}
\caption{Panel (a): Time evolution of compressive fluctuations, $\delta b_{\|}$, in the frequency range $[0.1,2]$ Hz, normalized to the mean magnetic field $B_0$. Panel (b): PDF of $\delta b_{\|}$, normalized to $\sigma(\delta b_{\|})$, (black-solid line) and the corresponding Gaussian fit (black-dashed line). 
The vertical red-solid lines indicate the position of 3 standard deviations of the Gaussian fit, used to determine the threshold in panel (c). Panel (c): Zoom of the compressive energy, $\delta b^2_{\|}$. The blue-solid line is an envelope of the magnetic energy, the red stars indicate the maximum of the energy of the intermittent events (i.e. events with the energy over the threshold, shown by the red solid line) and the dashed-lines show how to define the characteristic temporal scale of an event (width at half height).}
\label{fg:recon}
\end{center}
\end{figure}

\subsection{Method} 
\label{sec:met}

The magnetic field fluctuations in a particular scale range can be defined using a band pass filter based on the wavelet transform \cite[]{tor98,he12,rob13}
\begin{equation}\label{eq:db}
\delta b_i(t) = \frac{\delta j \delta t^{1/2}}{C_{\delta} \psi_0(0)} 
\sum_{j=j_1}^{j_2} \frac{\tilde{\mathcal W_i}(\tau_j,t) }{\tau_j^{1/2}} \ , 
\end{equation}
where $j$ is the scale index and $\delta j$ is the constant step in scales; the factor $\psi_0(0) =\pi^{1/4}$ and the value of the constant $C_{\delta}$, that is derived from the reconstruction of a $\delta$ function using the Morlet wavelet, is 0.776 \cite[]{tor98}. Here we use $\tau(j_1) = 0.5$~s and $\tau(j_2) = 10$~s to study scales (and frequencies) around ion scales.
 
As we have seen in Figure~\ref{fg:spectra}(b), at scales around ion scales, the compressibility increases. Let us consider these compressive fluctuations, which we denote $\delta b_{\|}$. Figure~\ref{fg:recon}(a) displays the time evolution of $\delta b_{\|}$, defined by eq.~(\ref{eq:db}) with $\mathcal W_{\|}$, and normalized to the mean magnetic field over the whole time interval under study, $\delta b_{\|}/B_0$. The PDF of $\delta b_{\|}$, normalized to its own standard deviation $\sigma(\delta b_{\|})$, is shown in panel~(b) (black-solid line) and it is compared to the corresponding Gaussian fit (black-dashed line): the present non-Gaussian tails are characteristic of some intermittency or inhomogeneity of the turbulence \cite[]{fri95}. The vertical red-solid lines indicate the position of 3 standard deviations of the Gaussian fit, that include $99.7$ \% of the Gaussian contribution. All the events that exceed this limit contribute to the non-Gaussian part of the PDF. This value will be used as a threshold to select non-Gaussian compressive intermittent events. 

Figure~\ref{fg:recon}(c) shows a zoom of $\delta b^2_{\|}(t)$, during 1 minute (black-solid line). An envelope of the energy of magnetic fluctuations is indicated by the blue-solid line and defined as the smooth curve outlining the extremes of the oscillating signal. The corresponding  threshold in the energy ($\delta b^2_{\|}= 3.9\cdot 10^{-2}$~nT$^2$) is shown by the horizontal red-solid line. The maxima of the energy of the intermittent events over this threshold are marked by red stars. We define the width of an event $\Delta \tau{'}$ as the time range between the two minima of the envelope, containing a maximum of the energy over the threshold. Then the characteristic temporal scale of an event, $\Delta \tau$, can be defined as the width at half height (intersection of the black-dashed lines in panel (c)).

During the one minute time interval shown in Figure~\ref{fg:recon}(c), we observe 3 intermittent events. For the whole time interval under study we get about 600 events. The characteristic time scale of these events varies in the range $\Delta \tau \in [0.25,7]$~s and the width $\Delta \tau{'}\in [0.75,7.5]$~s. 

\begin{figure}
\begin{center}
\includegraphics [width=8.6cm]{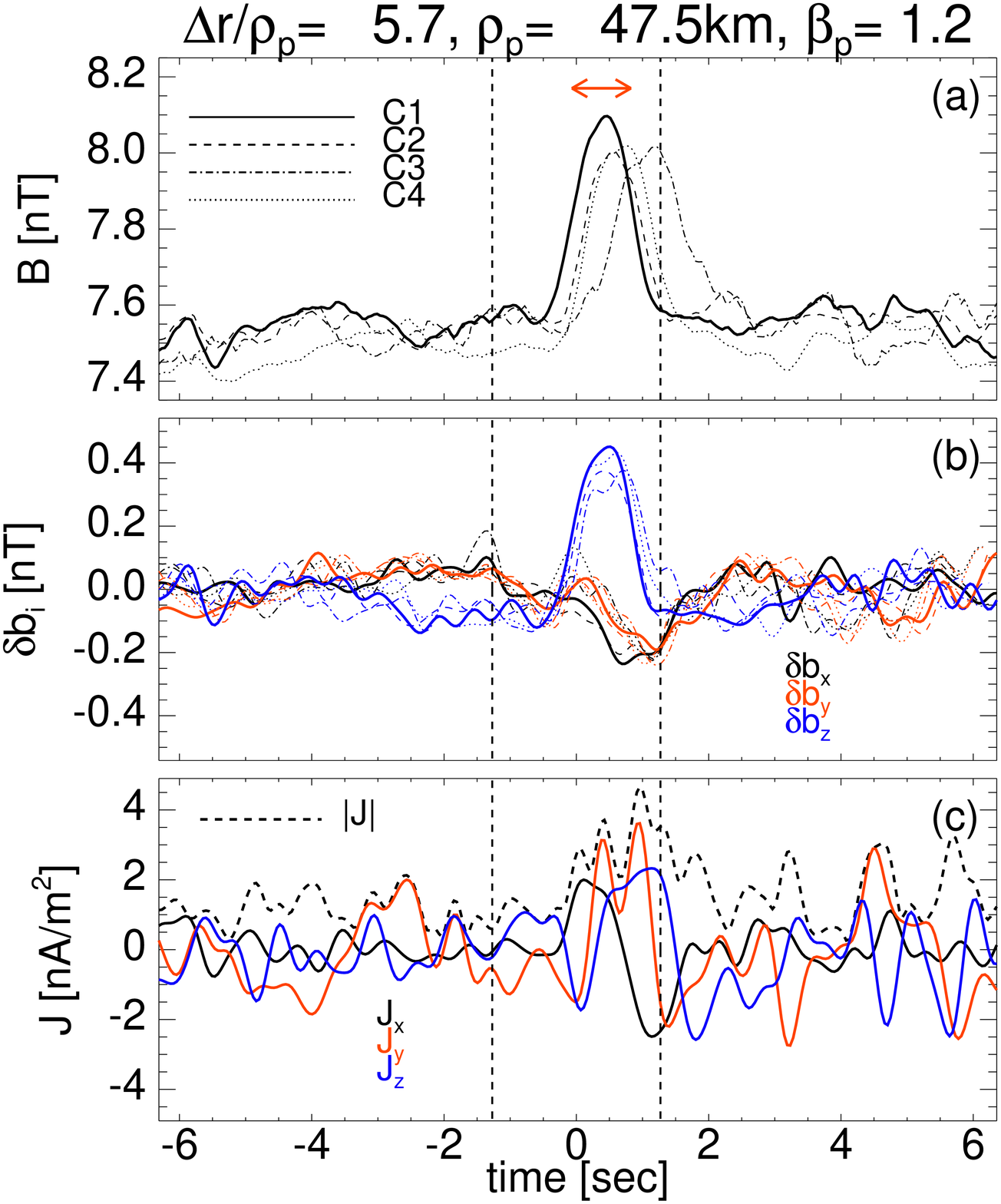}
\includegraphics [width=8.6cm]{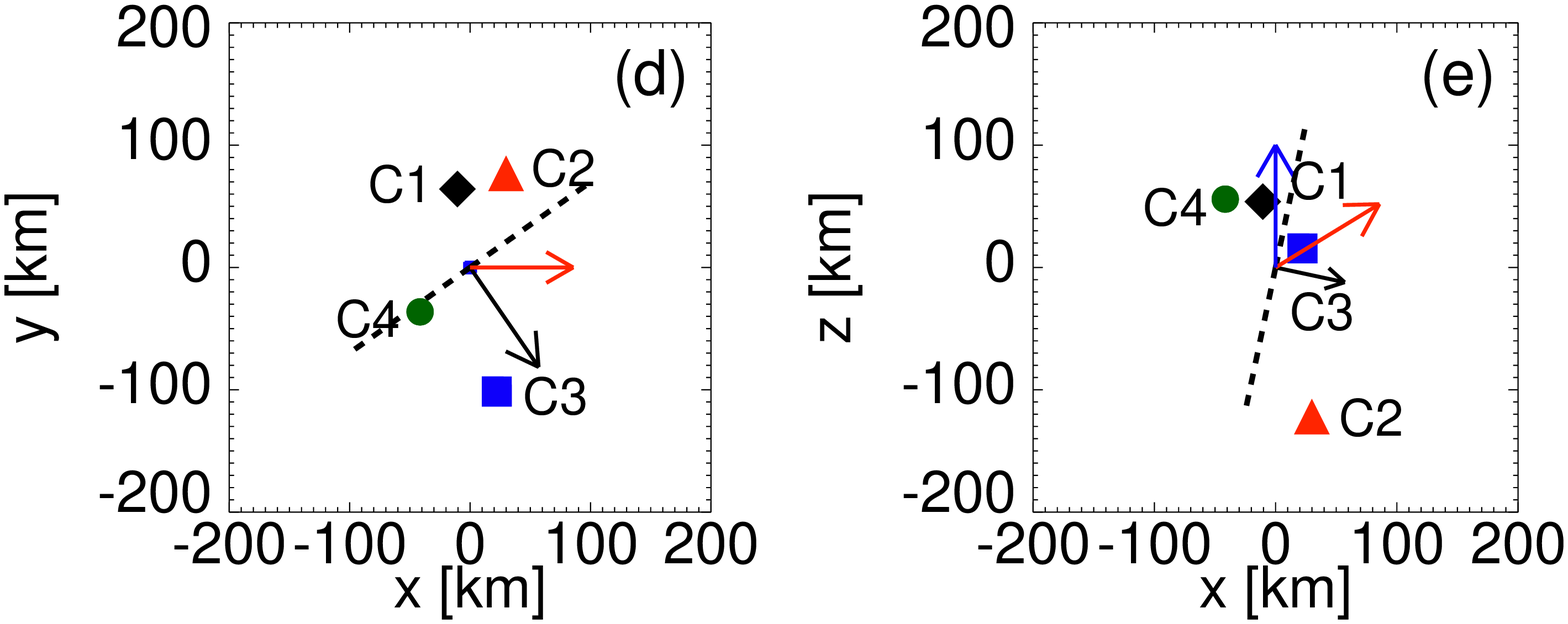}
\caption{Example of linearly polarized compressive soliton-like structure, centered at 02:10:42.5~UT. Panel (a): modulus of the large scale magnetic field observed by the four {\em Cluster} satellites (different style lines). The red double arrow indicates $\Delta \tau$, corresponding to $\Delta r$ defined in Section \ref{sub11}.
Panel (b): components of magnetic fluctuations defined by eq.~(\ref{eq:db}), in $BV$-frame. The time of each satellite is shifted taking into account the time delays with respect to C1.
Panel (c): modulus (black-dashed line) and components (in $BV$-frame) of the current density. The vertical black-dashed lines indicate $\Delta \tau{'}$, corresponding to the total extension of the structure ($\Delta r{'}= 16.3 \rho_p$). 
Panels (d) and (e): Configuration of  {\it Cluster} satellites in $BV$-frame: black diamonds for C1, red triangles for C2, blue squares for C3 and green circles for C4. The arrows indicate the direction of the normal (black), local flow (red) and local magnetic field (blue), while the black-dashed lines represent the plane of the structure.}
\label{fg:sol}
\end{center}
\end{figure}
\begin{figure}
\begin{center}
\includegraphics [width=8.6cm]{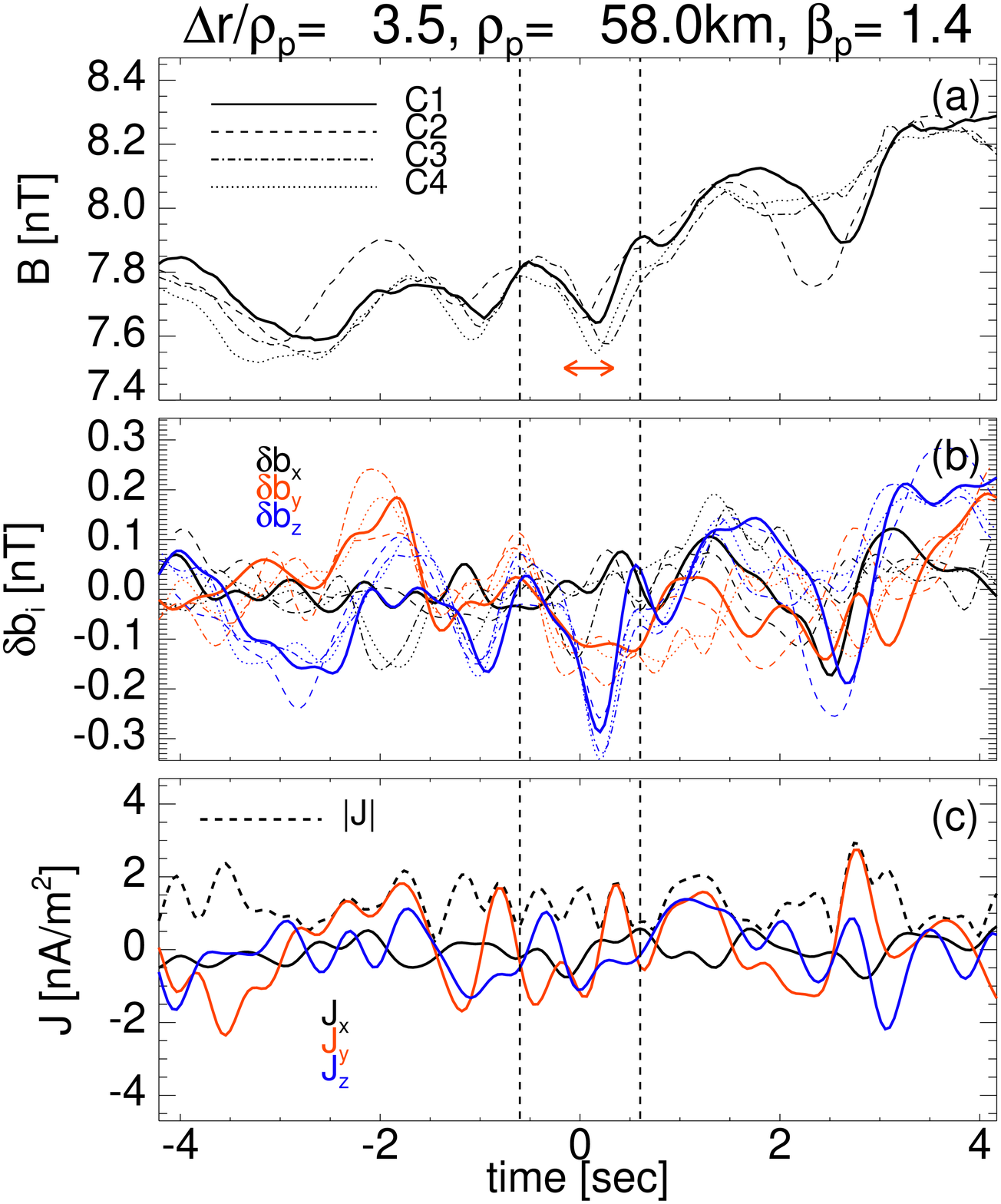}
\includegraphics [width=8.6cm]{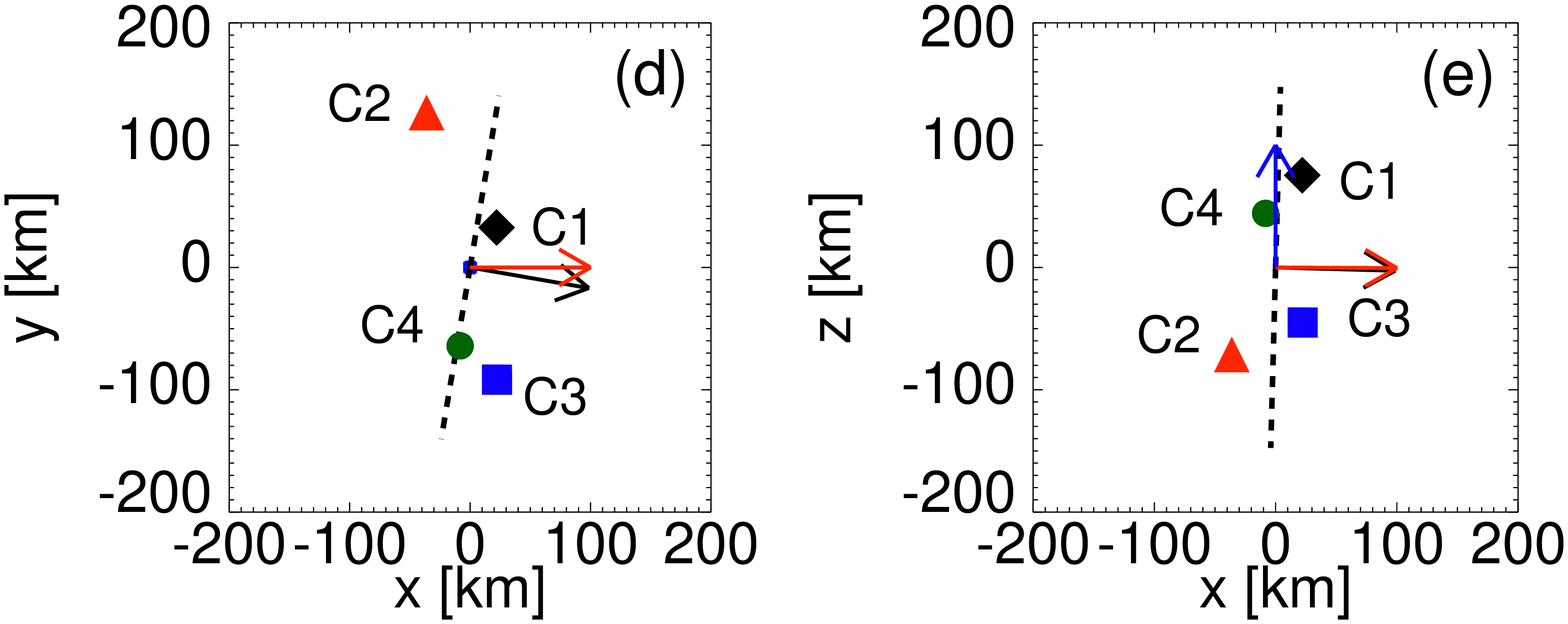}
\caption{Example of linearly polarized and compressive chain of magnetic hole-like structures, centered at 01:47:15.7~UT and with $\Delta r{'}= 8.6 \rho_p$. The panels are the same as in Figure~\ref{fg:sol}.}
\label{fg:hole}
\end{center}
\end{figure}

\begin{figure}
\begin{center}
\includegraphics [width=8.6cm]{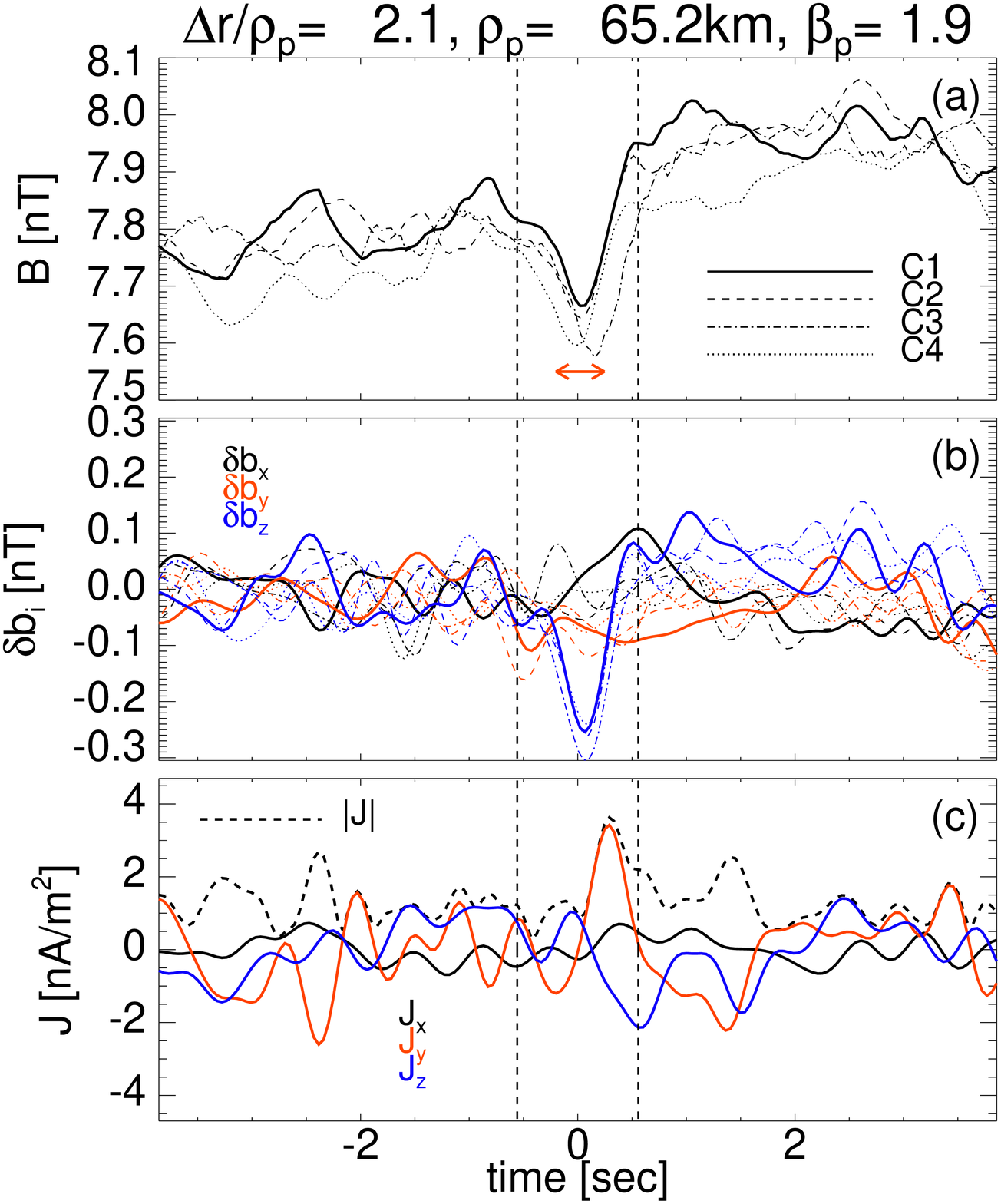}
\includegraphics [width=8.6cm]{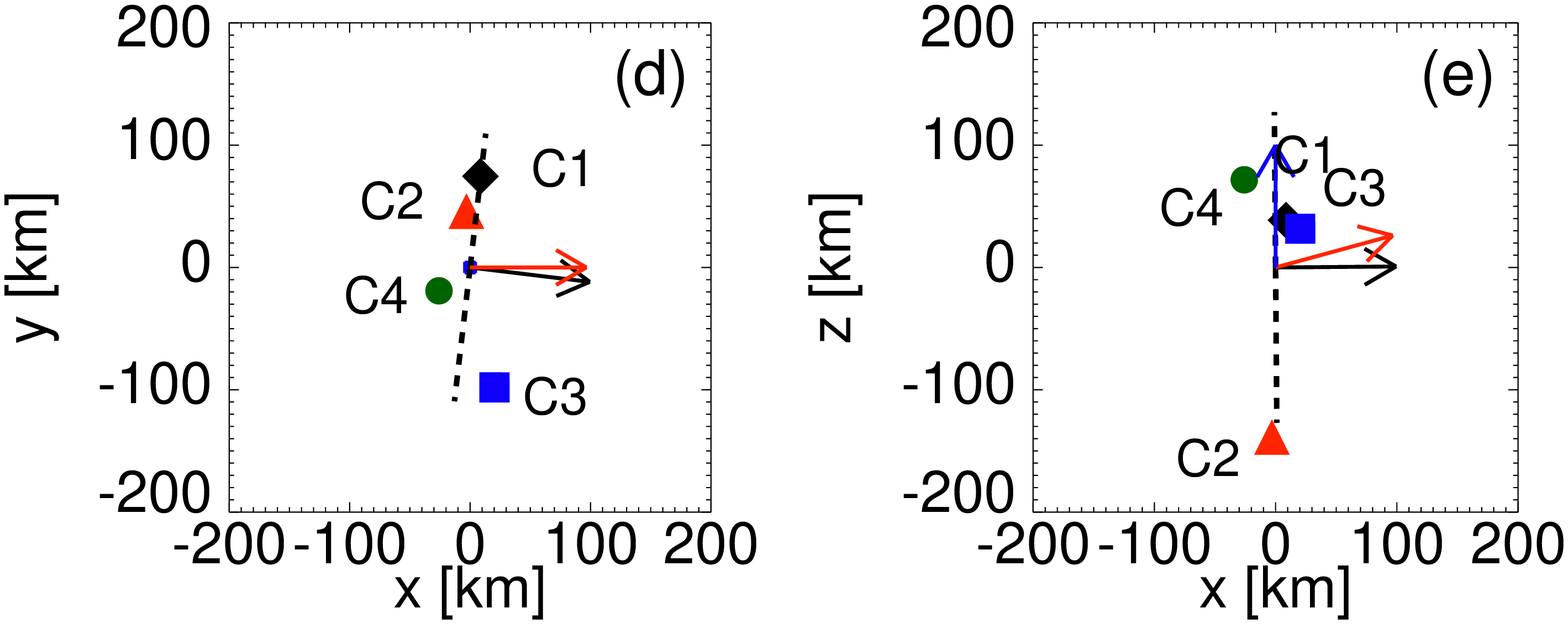}
\caption{Example of linearly polarized and compressive solitary magnetic hole-like structure, centered at 02:14:08.2~UT and with $\Delta r{'}= 5.3 \rho_p$. The panels are the same as in Figure~\ref{fg:sol}.}
\label{fg:hole_bis}
\end{center}
\end{figure}

\begin{figure}
\begin{center}
\includegraphics [width=8.6cm]{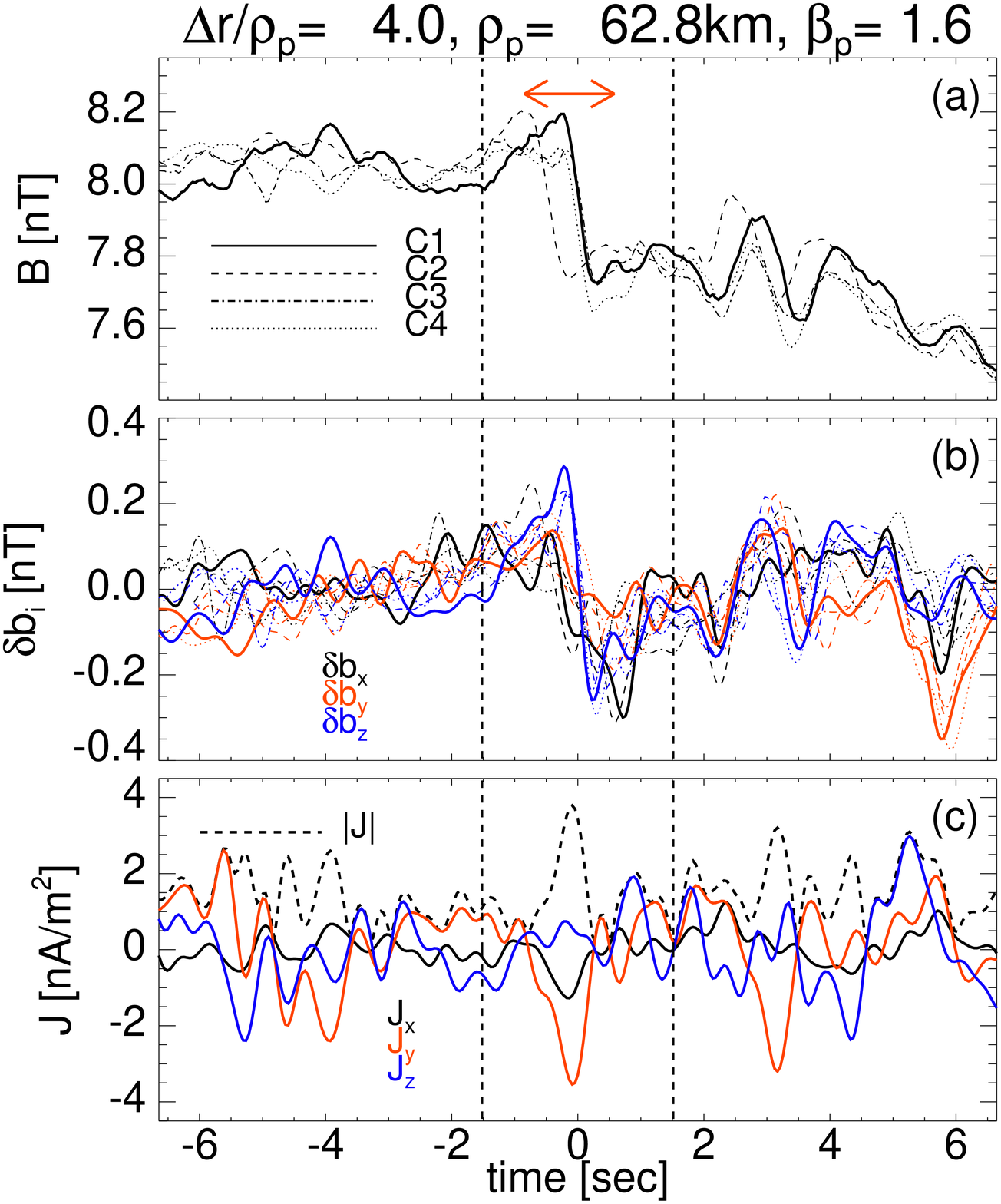}
\includegraphics [width=8.6cm]{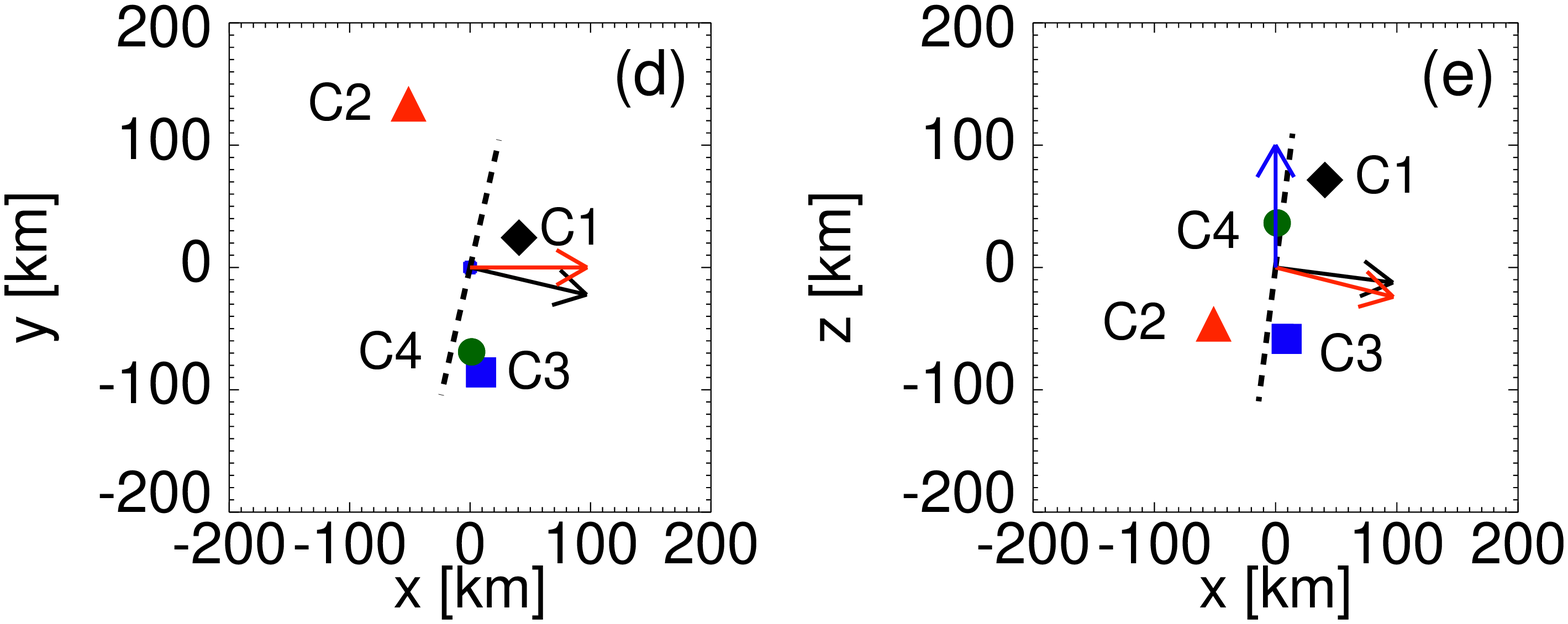}
\caption{Example of linearly polarized compressive shock-like structure, centered at 01:17:44.9~UT and with $\Delta r{'}= 8.4 \rho_p$. The panels are the same as in Figure~\ref{fg:sol}.}
\label{fg:sho}
\end{center}
\end{figure}
\begin{figure}
\begin{center}
\includegraphics [width=8.6cm]{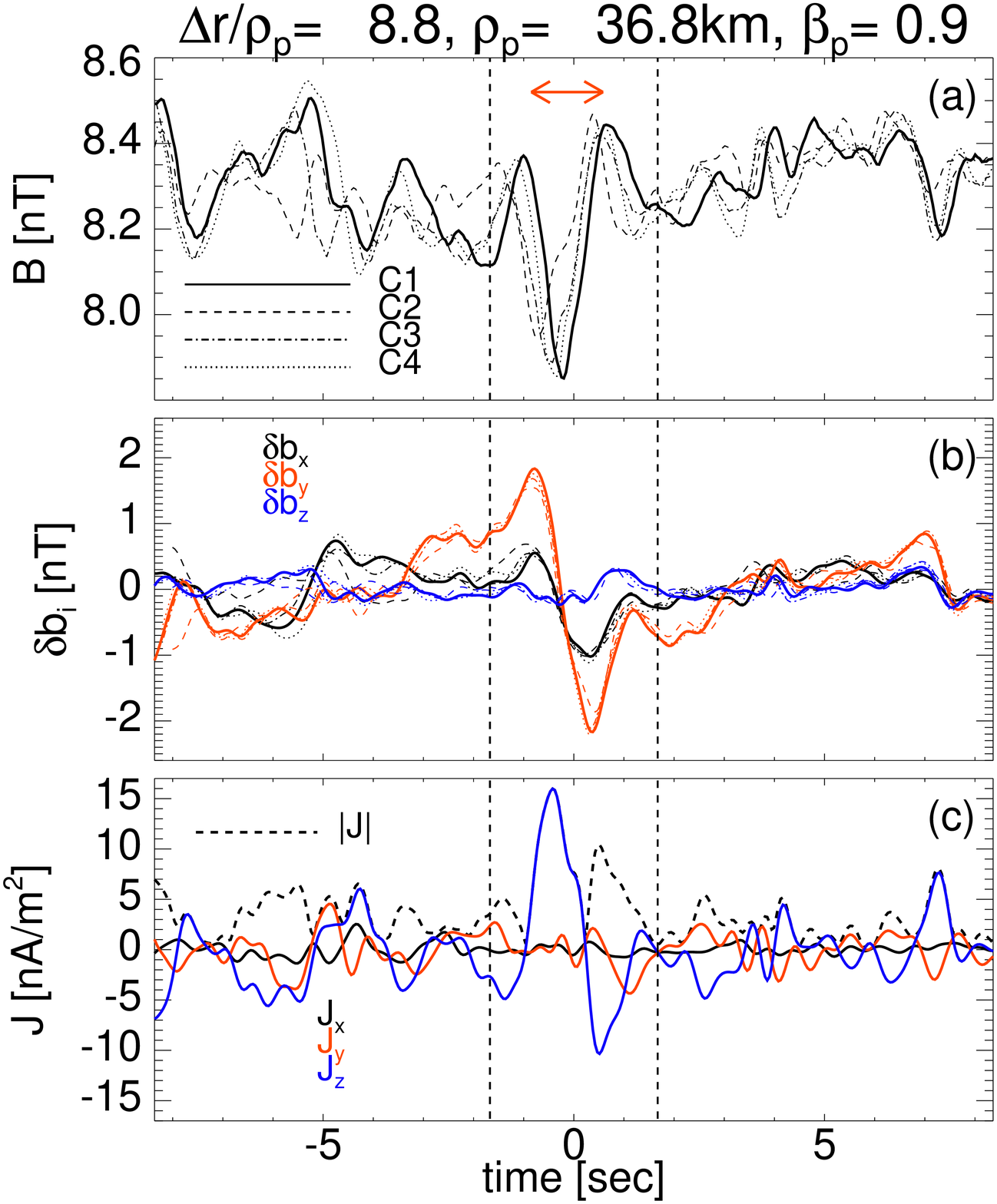}
\includegraphics [width=8.6cm]{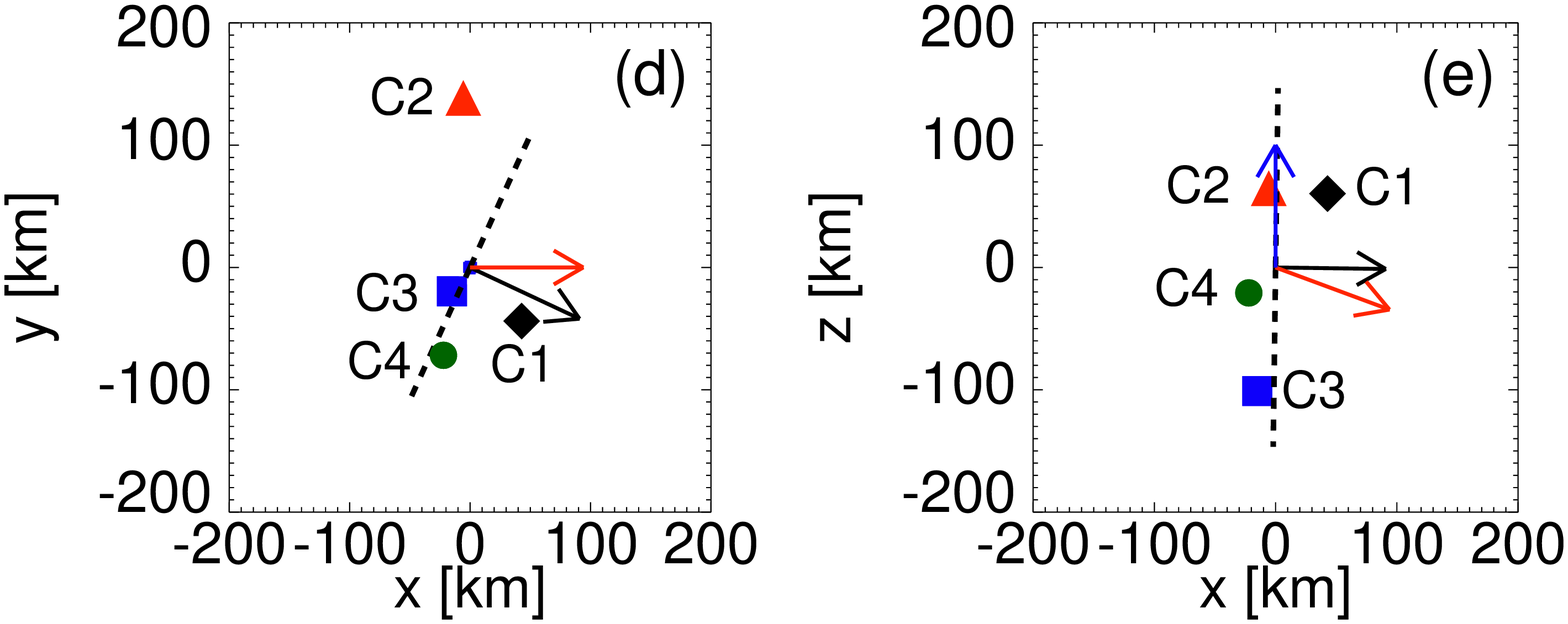}
\caption{Example of a current sheet, centered at 01:07:37.8~UT and with $\Delta r{'}= 30.1 \rho_p$. The panels are the same as in Figure~\ref{fg:sol}.}
\label{fg:cur}
\end{center}
\end{figure}
\begin{figure}
\begin{center}
\includegraphics [width=8.6cm]{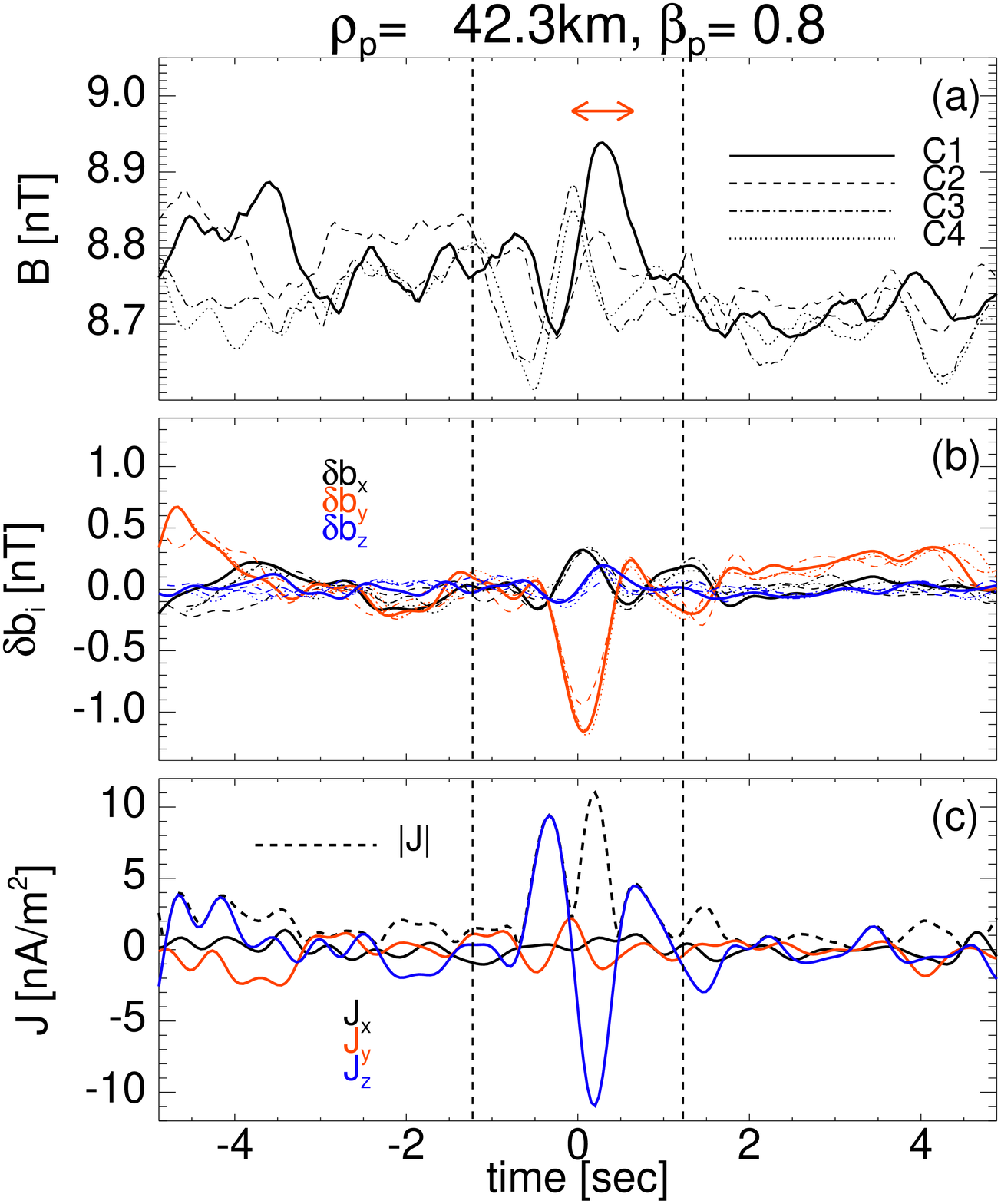}
\includegraphics [width=8.6cm]{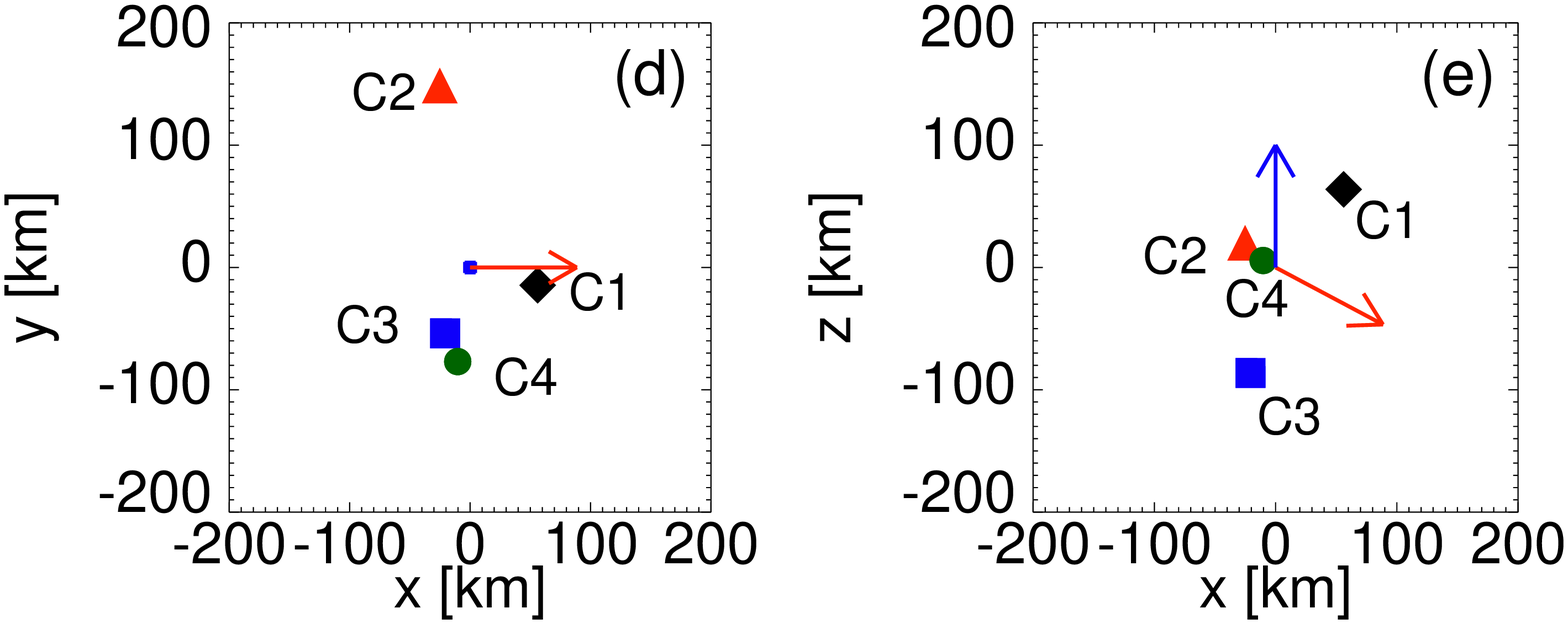}
\caption{Example of solitary Alfv\'en vortex, centered at 00:56:16.1~UT. The panels are the same as in Figure~\ref{fg:sol}, except for panels (d) and (e), where no indication about the direction of the normal and the plane of the structure is given.}
\label{fg:vor_alf}
\end{center}
\end{figure}
\begin{figure}
\begin{center}
\includegraphics [width=8.6cm]{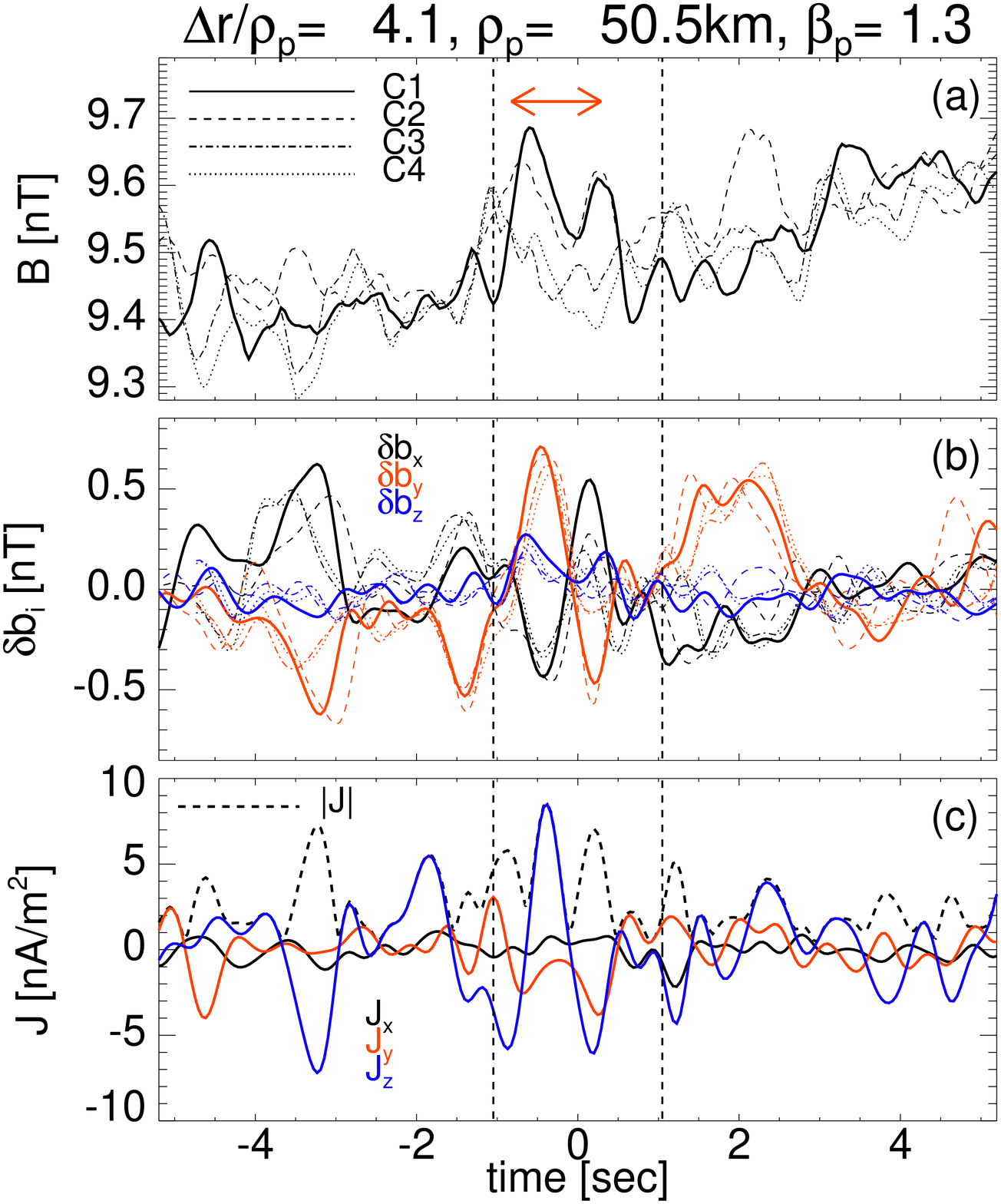}
\includegraphics [width=8.6cm]{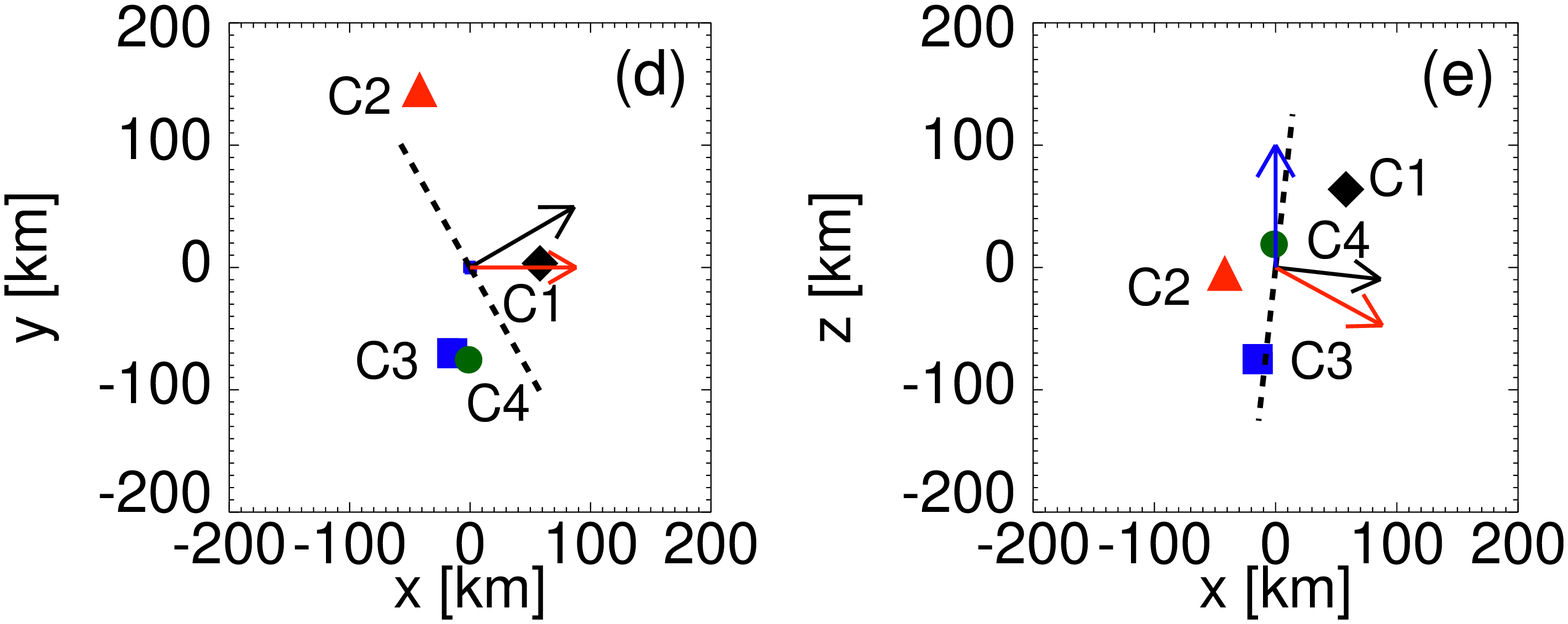}
\caption{Example of Alfv\'en vortex-like structure, centered at 00:49:58.5~UT and with $\Delta r{'}= 7.7 \rho_p$. The panels are the same as in Figure~\ref{fg:sol}.}
\label{fg:vor}
\end{center}
\end{figure}
\begin{figure}
\begin{center}
\includegraphics [width=8.5cm]{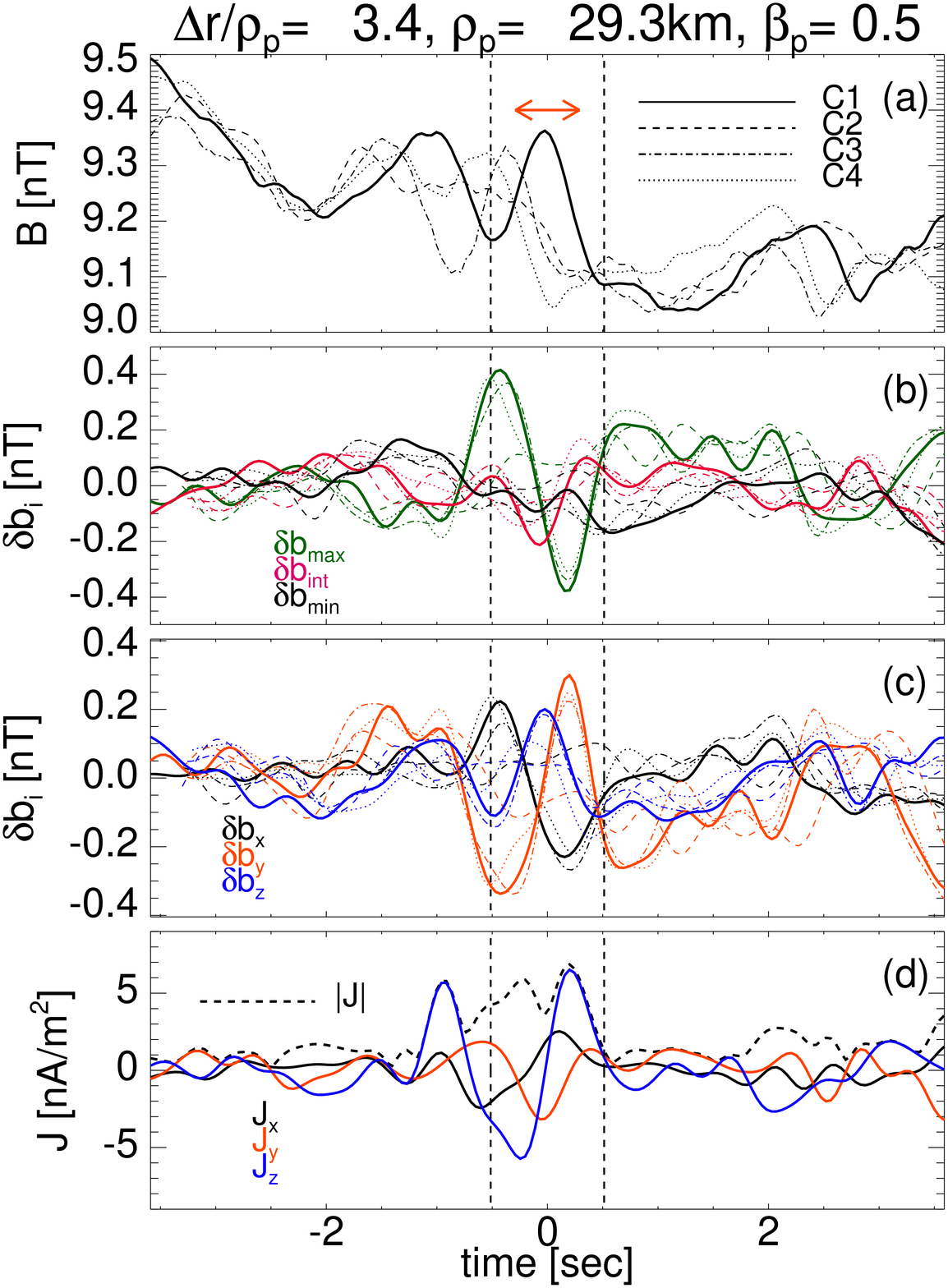}
\includegraphics [width=8.5cm]{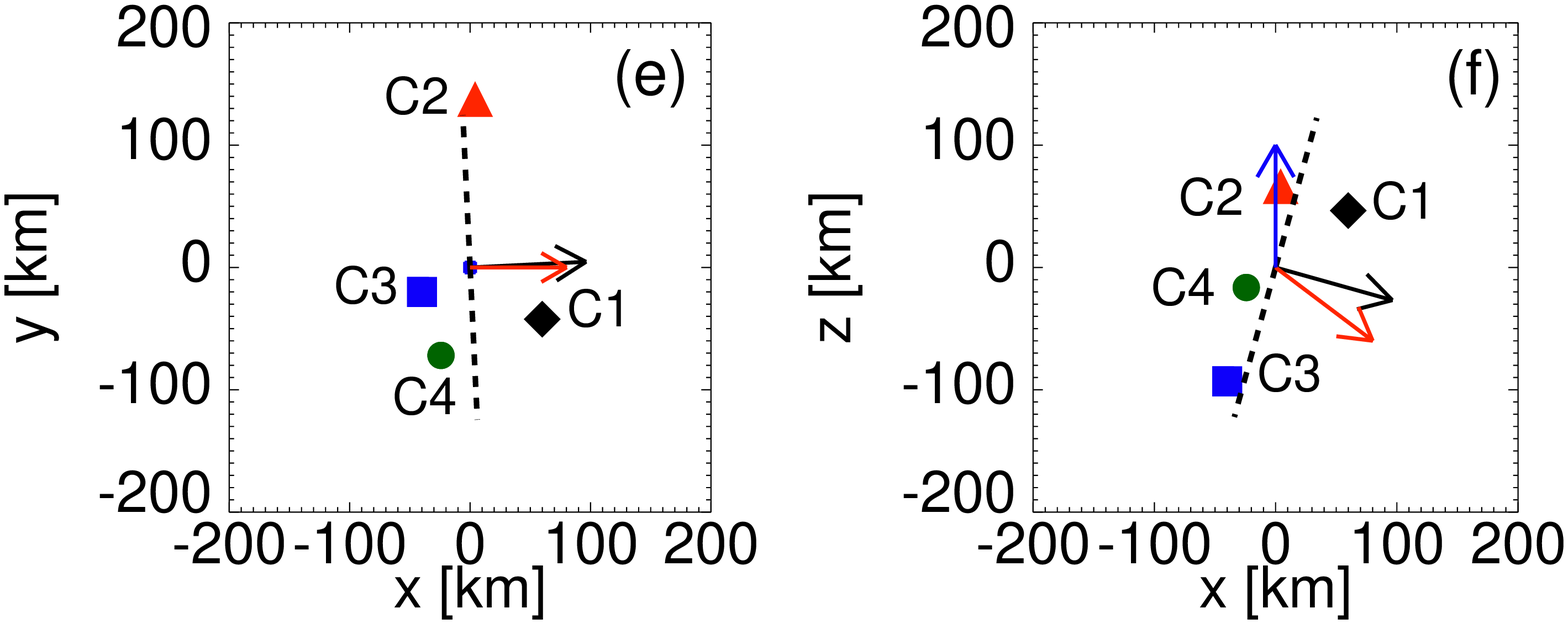}
\caption{Example of compressive vortex-like structure, centered at 00:31:10.4~UT. Panel (a): modulus of the large scale magnetic field observed by the four {\em Cluster} satellites (different style lines). The red double arrow indicates $\Delta \tau$, corresponding to $\Delta r$.
Panel (b): components of magnetic fluctuations defined by eq.~(\ref{eq:db}), in minimum variance frame. The maximum direction is in green, the intermediate in red and the minimum in black. The time of each satellite is shifted taking into account the time delays with respect to C1.
Panel (c): same representation of panel (b), but in $BV$-frame.
Panel (d): modulus (black-dashed line) and components (in $BV$-frame) of the current density. The vertical black-dashed lines indicate $\Delta \tau{'}$ of the structure, corresponding to $\Delta r{'}= 6.1 \rho_p$.
Panels (e) and (f): Configuration of {\it Cluster} in $BV$-frame: black diamonds for C1, red triangles for C2, blue squares for C3 and green circles for C4. The arrows indicate the direction of the normal (black), local flow (red) and local magnetic field (blue), while the black-dashed lines represent the plane of the structure.}
\label{fg:vor_com}
\end{center}
\end{figure}

\subsection{Examples} 
\label{sec:ex}

By performing a minimum variance analysis around the 600 events, we could identify by eye six different families. Figures~\ref{fg:sol}--\ref{fg:vor_com} show some examples of these families, in different time ranges (from about 4 to 7 $\Delta \tau{'}$), depending on the presence of other events outside the considered one. These intermittent events are well localized in time and have regular magnetic field profiles. We can identify them as coherent structures. For a subset of 109 structures (which also contains the examples in Figures~\ref{fg:sol}--\ref{fg:cur} and Figures~\ref{fg:vor}--\ref{fg:vor_com}), we were able to study the orientation of their normals and the propagation in the plasma frame.


\subsubsection{Strongly compressive structures} 
\label{sec:compressible}


\paragraph{{\bf Soliton and magnetic holes}}

The first three examples of coherent structures are shown in Figures~\ref{fg:sol}--\ref{fg:hole_bis}. Panels (a) display the modulus of the raw magnetic field measurements, as observed by the four satellites, where the FGM noise at $f>2.5$~Hz is taken off. In the following part of the paper, we refer to it as the large scale magnetic field. The different line styles correspond to different satellites of {\em Cluster}, as indicated in the legend. The red double arrow indicates $\Delta \tau$, i.e. the characteristic temporal scale of the structures (width at half height). 

The structures appear as an impulse (increase or decrease) in the ambient magnetic field, looking like a magnetic hump or soliton and magnetic holes, respectively. By looking outside the width of these structures ($\Delta \tau{'}$, indicated by the two vertical dashed lines), we observe that, while the soliton is a quite solitary hump (Figure~\ref{fg:sol}), the magnetic hole in Figure~\ref{fg:hole} seems to be only one structure in a chain of structures. A careful inspection of the 600 structures shows that usually magnetic holes appear in the plasma as a chain of compressive structures, while solitons are observed as isolated structures. However, few examples of solitary depression are also observed and an example is given in Figure~\ref{fg:hole_bis}.

In panels (b), magnetic fluctuations $\delta b_i$ (with $i=x,y,z$), defined by eq.~(\ref{eq:db}), are shown in the reference frame which takes into account the directions of the local mean magnetic field ${\bf b_0}$ and flow velocity ${\bf v_0}$ defined within each structure time scale $\Delta \tau{'}$ (time between two vertical dashed lines):  $z$ is aligned with the local ${\bf b_0}$,  ${\bf e_z} = {\bf e_b}$ (blue lines), $x$ is aligned with ${\bf v_0}$ in the plane perpendicular to ${\bf b_0}$, ${\bf e_x} = ({\bf e_b}\times{\bf e_v})\times{\bf e_b}$ (black lines) and $y$ closes the right-hand reference frame, ${\bf e_y} = {\bf e_b}\times{\bf e_x} $(red lines). Below, we refer to this frame as $BV$--frame. The time of each satellite is shifted taking into account the time delays with respect to C1. 

All the structures shown in Figures~\ref{fg:sol}--\ref{fg:hole_bis} are strongly compressive: the maximal variation is $\delta b_z$, as observed in panels (b). In order to quantify the compressibility of the structures, we evaluate a compressibility parameter, $\xi_{\parallel}$, defined as the ratio between parallel and perpendicular contributions. In particular, 
\begin{equation}
\label{eq:com_par}
\xi_{\parallel} = \sqrt{ \frac{\max(\delta b_z^2)}{\max(\delta b_x^2+\delta b_y^2)} } \ ,
\end{equation}
where the maximum of the magnetic components is evaluated within $\Delta \tau{'}$. For the selected compressive structures, we found $\xi_{\parallel}=1.6$ for the soliton, 2.1 for the chain of magnetic holes and 1.8 for the solitary magnetic hole.

Minimum variance analysis applied to these structures confirms the previous results: the direction of the maximal variation ${\bf e_{max}}$ is close to the direction of ${\bf b_0}$. In particular, the angle between these two directions, $\theta_{max}$, is $\simeq 15^{\circ}$ for the soliton, $\simeq 16^{\circ}$ for the chain and $\simeq 12^{\circ}$ for the solitary magnetic hole. Furthermore, the minimum variance ${\bf e_{min}}$ is strictly perpendicular to ${\bf b_0}$ for both examples of magnetic holes and nearly perpendicular for the soliton-like structure ($\theta_{min} \simeq 83^{\circ}$ and $\theta_{int} \simeq 77^{\circ}$).

Panel~(c) in Figures~\ref{fg:sol}--\ref{fg:hole_bis} displays the evolution of the current density ${\bf J}$, calculated using the curlometer technique \cite[]{dun88,dun02}, based on four-point measurements of {\em Cluster}. The three components of ${\bf J}$ are given in the $BV$--frame; the modulus, $|{\bf J}|$, is shown by dashed line. The curlometer technique works well in our case, as far as the four satellites are inside the same event during most of the time of the width of the structure $\Delta \tau{'}$. The factor  $Q= div {\bf B}/{curl}{\bf B}$ is usually used as a quality factor of the calculation of gradients. In our case it is very good ($Q\ll 1$ within $\Delta \tau{'}$). Note however that $div {\bf B}$ and $curl {\bf B}$  include different gradients, so it is difficult to assess wether $Q$ can indeed play the role of a quality factor [Gerard Chanteur, Private Communication, 2015]. The current density ${\bf J}$ is almost perpendicular to ${\bf b_0}$ in the case of the holes, while it is more oblique in the case of the soliton. 

Finally, panels (d) and (e) show the configuration of the four {\it Cluster} satellites in the $BV$--frame, by using different symbols and colors, as indicated in the caption of the Figure~\ref{fg:sol}. The arrows display the directions of the normal of the structures, ${\bf n}$ (black), determined by using the timing method (see Section \ref{sub11}), of ${\bf v_0}$ (red) and of ${\bf b_0}$ (blue). Moreover, the black-dashed lines indicate the plane of the structures. For the magnetic holes, we have ${\bf n}$ strictly perpendicular to ${\bf b_0}$, while for the soliton we have $\Theta_{nB} = 83^{\circ}\pm 15^{\circ}$. Moreover, if $\bf n$ of the chain of magnetic holes and for the solitary magnetic hole is nearly aligned with the solar wind flow speed in $(x,y)$--plane ($\Theta_{nV_{\perp}}=10^{\circ}\pm 10^{\circ}$ and $7^{\circ}\pm 8^{\circ}$, respectively), $\bf n$ of the magnetic soliton is oblique to it ($\Theta_{nV_{\perp}}=55^{\circ}\pm 20^{\circ}$). The propagation velocity of the structures along $\bf n$ in the plasma rest frame (see Section \ref{sub12}) is $\mathcal{V}_0=(55 \pm 92)$~km/s and $-(35 \pm 71)$~km/s for the chain of magnetic holes and for the solitary magnetic hole, respectively, and $\mathcal{V}_0=(151 \pm 97)$~km/s for the soliton ($V_A=32$~km/s and $V_F=54$~km/s). While the magnetic holes are simply convected, in the limit of errors, by the wind, the soliton-like structure has a finite velocity in the plasma rest frame.

The normal to the structures was determined assuming that the structure is locally planar, i.e. that holes and soliton may have an infinite front in the plane perpendicular to ${\bf n}$. The results show that ${\bf n}$ is perpendicular to ${\bf b_0}$ and the latter is in the plane (see panels (e)). However, this front seems to be perturbed or finite, especially in case of the magnetic holes. Indeed, for example from Figure~\ref{fg:hole}, one can see that the different satellites observe different amplitudes: satellite C2 (red triangles) sees the event first and the smallest amplitude, then, C4 (green circles) and C1 (black diamonds) see the signal, nearly at the same time, but with different amplitudes, and C3 (blue squares) is the last to observe the signal, seeing almost the same amplitude observed by C4. Such variation in amplitude cannot be explained by an infinite plane; in that case, all satellites would see the same amplitude in each point of the plane. Therefore, the structure is not perfectly planar. If a magnetic hole is a cylinder (or a cigar) with an axis along ${\bf b_0}$, variations of the amplitude from one satellite to another is related to the fact that different satellites cross the structure at different distances from its axis. Along the axis the signal is expected to be the same, as it is indeed observed on C3 and C4, separated along ${\bf b_0}$ by $\sim 100$~km and very close in the perpendicular plane.
 
In the case of the magnetic soliton, the amplitudes of the magnetic fluctuations (see Figure~\ref{fg:sol}(b)) are nearly the same on the four satellites, indicating that the topology of the structure is not far from the planar front. This front is going through  C2 and C4 in (x,y)--plane (see panel (d)). Note that these satellites observe the same signal at the same time. 
 
To conclude on the geometry of the discussed compressive structures, a comparison of the signals on the four satellites with different geometrical  models of holes and solitons should be done (a subject of our future work).
 
In terms of plasma parameters: the soliton is observed for $\beta_p \simeq 1.2$, while the magnetic holes appear at higher value of local plasma beta (1.4 for the chain and 1.9 for the solitary magnetic hole). The time localization is different as well: the magnetic soliton is nearly two times larger with respect to the magnetic holes. In terms of normalized spatial scales (see Section \ref{sub11}), the magnetic hole in the chain is $\sim 3.5 \rho_p$, while the solitary magnetic hole is $\sim 2 \rho_p$; and the soliton-like structure is $\sim 6 \rho_p$. These structures differ also by the values of the local proton temperature anisotropy $A=T_{\perp}/T_{\|}$: within the magnetic holes $A \simeq 2$, while within the soliton the protons are nearly isotropic ($A \simeq 1.1$). 

In the subset of 109 structures, we have observed 10~magnetic holes, considering both magnetic holes in the chains and solitary magnetic holes, and 6~solitons. The magnetic holes present the characteristics of mirror mode structures \cite[]{sou08,gen09}: high values of temperature anisotropy and plasma beta, and they are simply convected by the flow in the limits of the errors. Moreover, if we evaluate the mirror parameter, $C_M$\footnote{The values of the perpendicular plasma beta and of the temperature have to be considered in a region of the ambient plasma near the structures and in our case we evaluate them at the borders of the structures.}, as defined in eq.(4) of \cite{gen09}, we obtain the result that almost all the magnetic holes are observed under mirror unstable plasma conditions ($C_M>1$). It is worth pointing out that $C_M$ is obtained for bi-Maxwellian distribution functions, in the low-frequency, long-wavelength limit of the Vlasov-Maxwell equations and in the case of cold electrons. Moreover, the particle detector used for the analysis (CIS/HIA) bins the ions according to their energy per charge ratio. Therefore, protons and alpha particles are generally mixed and moments are averaged. For the structures shown in Figures~\ref{fg:hole} and \ref{fg:hole_bis}, we have $C_M=1.7$ and $1.9$, respectively. However, some examples of magnetic holes under stable mirror conditions ($C_M<1$) are also found. These observations could have different explanations. First of all, as we do the evaluation of $C_M$ automatically for 600 structures, it is possible that the borders of some structures, especially for the chains of magnetic holes, are not properly determined. Therefore, the evaluation of $C_M$ could be mistaken. On the other hand, it is also possible that the holes, found in stable conditions, are generated elsewhere, where the plasma was unstable, and convected away by the wind. 

Otherwise, the observed magnetic solitons have moderate plasma beta and almost isotropic ion temperatures. In most cases, their propagation velocities are different from zero and are comparable with the velocity of fast mode and/or proton thermal speed. The physics of the soliton-like structures appears completely different from the nature of the mirror mode structures. Moreover, the evaluation of the mirror parameter gives $C_M<1$. In particular, for the example in Figure~\ref{fg:sol}, we have $C_M = 0.1$. However it could be also possible that these structures are the result of the plasma relaxation after the mirror instability.

\paragraph{{\bf Shock}}

Another example of compressive coherent structures is shown in Figure~\ref{fg:sho}. The panels are the same as those in Figure~\ref{fg:hole}.
Here, we observe an abrupt decrease of the magnetic field modulus $|{\bf B}|$ (panel (a)). The four satellites observe nearly the same relative amplitude decrease, except the satellite C3, which observed a smaller amplitude gradient. The principal fluctuation $\delta b_z$ is nearly the same on the four satellites, with small differences.  This decrease looks like a shock wave. So, it is expected to be a planar structure. However, differences in the amplitudes of magnetic components on four satellites indicate that the shock front is not perfectly planar, but it probably has ripples or it undergoes a reformation process, e.g. \cite[]{kra13}.

Here, the plasma beta is $\beta_p \simeq 1.6$ and the ion temperature anisotropy is $A \simeq 2.1$ on both sides of the decrease. Particle measurements on {\it Cluster} have 4 seconds time resolution and so there are only one-two points of measurements within the event. Indeed, sometimes it is possible to use the satellite potential fluctuations (with 5 measurements per second time resolution) as a proxy of the electron density $n_e$ \cite[]{ped95,ped01,bal03}. 
However, this method can give information about $n_e$ in the range $10^{-2}$--$10$ cm$^{-3}$ \cite[]{gus97}. In our case, the mean plasma density is about $25$--$30$ cm$^{-3}$. Therefore, this method cannot be applied for this particular time interval.

The normal of the structure is  quasi-perpendicular to ${\bf b_0}$ ($\Theta_{nB}=83^{\circ}\pm 6^{\circ}$) and it is almost aligned with ${\bf v_0}$  ($\Theta_{nV_{\perp}}=15^{\circ}\pm 11^{\circ}$), as observed in panels (d) and (e). Therefore, the plane of the structure contains ${\bf b_0}$ and it is perpendicular to ${\bf v_0}$ (see panel (e)). Moreover, the current density, ${\bf J}$, shown in panel (c), is almost perpendicular to ${\bf b_0}$. The velocity of propagation in the plasma frame is $\mathcal{V}_0=-(172 \pm 41)$~km/s. This corresponds to Mach numbers $M_F = \mathcal{V}_0/V_{F} = 2.8$ and $M_A = \mathcal{V}_0/V_A = 4.9$. 

A conclusive interpretation of this structure is difficult without high resolution density and temperature measurements. However,  its strongly compressive nature ($\xi_{\parallel}=0.95$) and  high values of Mach numbers are compatible with the fast magnetosonic shock wave. Among 109 events, we have found only 3~examples of such shock waves.

\subsubsection{Alfv\'enic structures} 
\label{sec:alfvenic}

Together with compressive structures (such as holes, solitons and shocks), we have also detected Alfv\'enic structures ($\delta b_{\perp} > \delta b_{\|}$), which have localized, more or less pronounced, compressive fluctuations.


\paragraph{{\bf Current sheet}}

The first example of an Alfv\'enic structure is shown in Figure~\ref{fg:cur} (the format of the figure is the same as for the previous examples). Here, the principal variation of the magnetic field is $\delta b_y$; $\delta b_x$ has also regular variation but with small amplitude, while $\delta b_z$ is  $\simeq 0$ (see panel (b)). The 3 components reduce (almost) to zero in the center of the structure, where the large scale magnetic field has its local minimum (panel (a)). This is a property of a current sheet.
${\bf J}$ is essentially parallel to ${\bf b}_0$ (panel (c)). The normal to the current sheet ${\bf n}$ is perpendicular to ${\bf b_0}$, and it is  oblique to the $V_{sw}$, $\Theta_{nV\perp}=25^{\circ}\pm 14^{\circ}$ (panels (d) and (e)). Its thickness, estimated from the four satellites analysis (see Section \ref{sub11} for more details), is $\sim 9\rho_p$.
The four satellites observe the same amplitudes of the fluctuations (see panel (b)), that is consistent with the planar geometry.
The velocity of this structure in the plasma frame is $\mathcal{V}_0=(24 \pm 88)$~km/s. Therefore, it is convected by the flow, as expected for a current sheet. It is observed for $\beta_p \simeq 1$ and anisotropy $A \simeq 0.6$. 9~examples of current sheets are found in the subset of 109 structures, characterized by $\beta_p \lesssim 1$ and $T_{\parallel} > T_{\perp}$. Different characteristic sizes are found, from $\sim 4 \rho_p$ to $\sim 11 \rho_p$.


\paragraph{{\bf Vortex structures}}

Finally, Figures~\ref{fg:vor_alf}--\ref{fg:vor_com} show three examples of coherent structures, which look like vortices. They are characterized by a local increase of the background magnetic field, observed by the four satellites (panels (a)).  The principal spatial gradients are $\nabla _{\perp} \gg \nabla_{\|}$, as shown by the timing analysis, for the structures in Figures~\ref{fg:vor} and \ref{fg:vor_com}, which gives ${\bf n} \perp {\bf b_0}$ (see panels (d) and (e) of Figure~\ref{fg:vor} and panels (e) and (f) of Figure~\ref{fg:vor_com}). 

In the first two cases, Figures~\ref{fg:vor_alf} and \ref{fg:vor}, the principal variations of ${\bf \delta b}$ are almost in the plane perpendicular to ${\bf b_0}$ ($\xi_{\parallel} = 0.16$ and $0.33$, respectively) and the current density, ${\bf J}$, displayed in panels (c), is along ${\bf b_0}$, as in the case of an Alfv\'en vortex \cite[]{pet92,ale08_n}. The variations of the magnetic magnitude and components from one satellite to another are similar to what is observed for dipolar Alfv\'en vortices in the Earth's magnetosheath \citep{ale06} and compatible with a cylindrical structure, crossed by the four satellites along different paths.

Unfortunately, for the Alfv\'enic vortex in Figure~\ref{fg:vor_alf}, that is more isolated than the other, the assumptions of the timing method are not verified; therefore, the normal and the velocity of this structure cannot be properly determined. Otherwise, we have an indication about the direction and the velocity of propagation for the other example of an Alfv\'enic vortex (Figure~\ref{fg:vor}). For this structure, the velocity of propagation along the normal and in the plasma rest frame is $\mathcal{V}_0=-(95 \pm 35)$~km/s with $\Theta_{nV_{\perp}}=30^{\circ}\pm 10^{\circ}$ and the spatial scale is about $\sim 4 \rho_p$. To compare with the model of an Alfv\'en vortex \cite[]{pet92,ale08_n}, it is useful to also evaluate its diameter. This can be estimated as the spatial scale corresponding to $\Delta \tau{'}$ (see Section \ref{sub11}), that is about $\sim 8 \rho_p$. In terms of plasma parameters, this vortex is observed for $\beta_p \simeq 1.3$ and an isotropic ion distribution. 

We observe 12 Alfv\'en vortices ($\xi_{\parallel} < 0.35$) in the subset of 109 structures. All of them are characterized by propagation speeds different from zero (but smaller than $V_A$ in the limits of the errors) and are observed in plasma regions with $\beta_p$ of order of 1 and isotropic ions. Different characteristic spatial sizes are found, from $\sim 2 \rho_p$ to $\sim 8 \rho_p$, and typical diameters between $\sim 5 \rho_p$ and $\sim 17 \rho_p$.

Figure~\ref{fg:vor_com} shows an example of the most common coherent event found in our time interval. Panel (b) shows the magnetic fluctuations $\delta b_i$ in the minimum variance frame (direction of maximum variation in green, intermediate in red and minimum in black). One observes that $\delta b_{max} \gg \delta b_{int} \sim \delta b_{min}$, meaning that the minimum variance direction is not well defined. The direction of the maximum variance is quasi-perpendicular to ${\bf b_0}$ ($\theta_{max} \simeq 80^{\circ}$) and the intermediate is almost parallel to ${\bf b_0}$ ($\theta_{int} \simeq 16^{\circ}$). Panel (c) displays the same magnetic fluctuation $\delta b_i$, but in the $BV$--frame. This representation shows important fluctuations in the 3 components of ${\bf \delta b}$, with a significative compressive component ($\xi_{\parallel}=0.5$). We have also an impression that $\delta b_z$ (blue lines) is a bit more localized (within 0.8 s) than the transversal part (within 2 s). Moreover, for the observed structure, the current density, displayed in panel (d) of Figure~\ref{fg:vor_com}, is along ${\bf b_0}$. The velocity of propagation in the plasma frame is $\mathcal{V}_0=-(158 \pm 27)$~km/s with $\Theta_{nV_{\perp}}=16^{\circ}\pm 7^{\circ}$. The spatial scale is about $\sim 3 \rho_p$, while the diameter is larger than $\Delta \tau{'}$; in particular it is of the order of $1.5\Delta \tau{'}$, i.e. of the order of $\sim 10 \rho_p$. This vortex is observed for $\beta_p \simeq 0.5$ and an isotropic ion distribution.


At the moment there are no models to describe this kind of fluctuations, but some interpretations will be discussed in Section \ref{sec:fine}. We will call them compressive vortex-like structures.
  
In the subset of 109 structures, there are 40 compressive vortices, for which the compressibility parameter varies in the range $\xi_{\parallel} \in [0.35,1.1]$. Such structures are observed under different plasma conditions: $\beta_p \in [0.3,3]$ and both parallel and perpendicular temperature anisotropy. Moreover, they can propagate in the flow or can be convected by the wind. Their spatial scales vary between $\sim 1.5 \rho_p$ and $\sim 18 \rho_p$, while the diameters vary between $\sim 4.5 \rho_p$ and $\sim 32 \rho_p$.

\section{Statistical Study of Coherent Structures}
\label{sec:statistic}
In this section we will present, first, the results of the minimum variance analysis \cite[]{son98} applied to all the detected intermittent structures ($\sim 600$) on C1. Then, we analyze the detected structures using four {\em Cluster} satellites in order to estimate the normal of the structures  and their velocities in the plasma rest frame. This analysis was possible only for $109$ events from $600$. 

\begin{figure}
\begin{center}
\includegraphics [width=8.6cm]{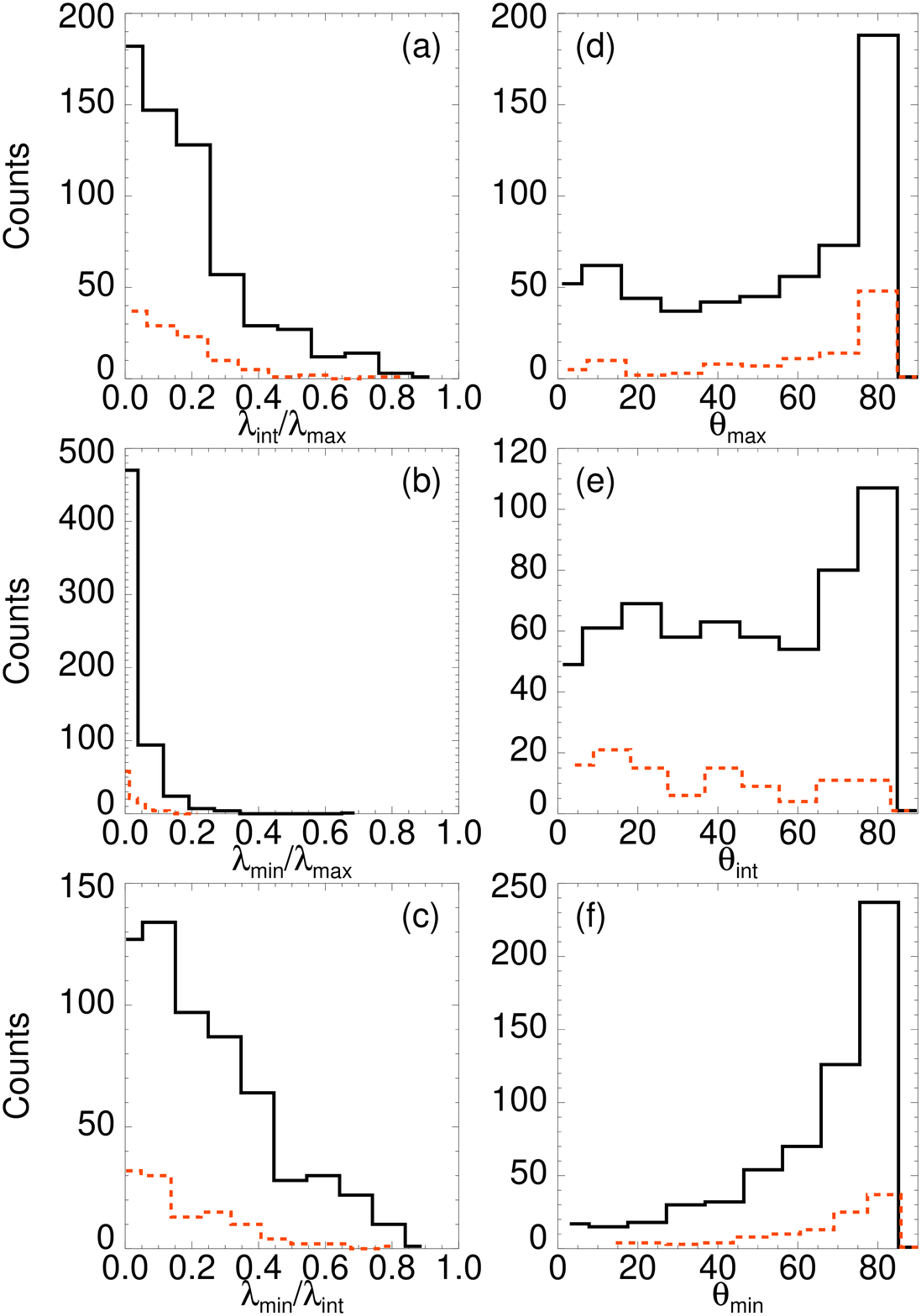}
\caption{Statistical analysis of the observed coherent structures in the variance frame: 600 selected coherent structures (black) and 109 coherent structures studied using the multi-satellite analysis (red). Left panels: Histograms for the intermediate (a) and minimum (b) eigenvalues, normalized to the maximum eigenvalue and for $\lambda_{min}/\lambda_{int}$ (c). Right panels: Histograms of the angles between the maximum (d), intermediate (e) and minimum (f) variance directions and  the local magnetic field.}
\label{fg:minvar}
\end{center}
\end{figure}

\subsection{Minimum variance analysis of 600 events}

The variance matrix \cite[]{son98} was calculated for each of the detected intermittent events during the time interval of $\Delta \tau{'}$ around each midpoint of the event (we recall that the temporal widths of the structures are $\Delta \tau{'} \in [0.75,7.5]$~s). Figure~\ref{fg:minvar} shows the results of this analysis: histograms in black display the results for all the events, while the dashed-red histograms represent the same results but for the structures which were possible to be studied with four satellites (see Section \ref{sub11}). 

The left column shows normalized values of the eigenvalues of the variance matrix: intermediate $\lambda_{int}$ and minimal $\lambda_{min}$ eigenvalues normalized by $\lambda_{max}$, in panels (a) and (b); panel (c) shows $\lambda_{min}/\lambda_{int}$. Most of the structures have $\lambda_{max} \gg \lambda_{int}, \lambda_{min}$ (i.e. 1D-fluctuations), however, some of them have $\lambda_{max} \geq \lambda_{int} \gg \lambda_{min}$ (2D-fluctuations). For most of the events, $\lambda_{min}/\lambda_{max} < 0.2$ and $\lambda_{min}/\lambda_{int} < 0.4$; meaning that, in general, the minimum variance direction is well defined.  However, for some selected events, there is a degeneracy $\lambda_{int} \sim \lambda_{min}$. 

The right column gives information on the orientation of the eigenvectors with respect to the local mean magnetic field, evaluated in the same interval in which we perform the minimum variance analysis, $\Delta \tau{'}$: $\Theta_{max}$ (panel d) is the angle between the maximal variance direction ${\bf\hat{e}}_{max}$ and ${\bf b_0}$. Respectively, we define, with similar definition, $\Theta_{int}$ (e) and $\Theta_{min}$ (f). A large number of selected structures have the direction of maximum variation almost in a direction perpendicular to ${\bf b}_0$ ($\theta_{max} > 65^{\circ}$), i.e. {\em Alfv\'enic} structures. However, a broad distribution of $\theta_{max}$ is obtained, including 25\% of structures with the direction of maximum variation almost along ${\bf b}_0$ ($\theta_{max} < 25^{\circ}$), i.e. {\em compressive} structures. The distribution of $\theta_{int}$ is almost uniform, with a peak at $80^{\circ}$. Finally, the distribution of $\theta_{min}$ shows that ${\bf\hat{e}}_{min}$ is almost perpendicular to ${\bf b}_0$. 

In the case of a planar structure, the direction of the minimum variance of the magnetic fluctuations is oriented parallel to the normal (or to the wavevector ${\bf k}$). However, when $\lambda_{min} \sim \lambda_{int}$, the use of ${\bf\hat{e}}_{min}$, as a predictor of the normal to the structure, can be erroneous \cite[]{son98,kne04}. In order to have robust information about the direction of the normal of the coherent structures, we need a multi-satellite analysis.

\subsection{Multi-satellite analysis of 109 events}
\label{sub1}
\begin{figure}
\begin{center}
\includegraphics [width=8.2cm]{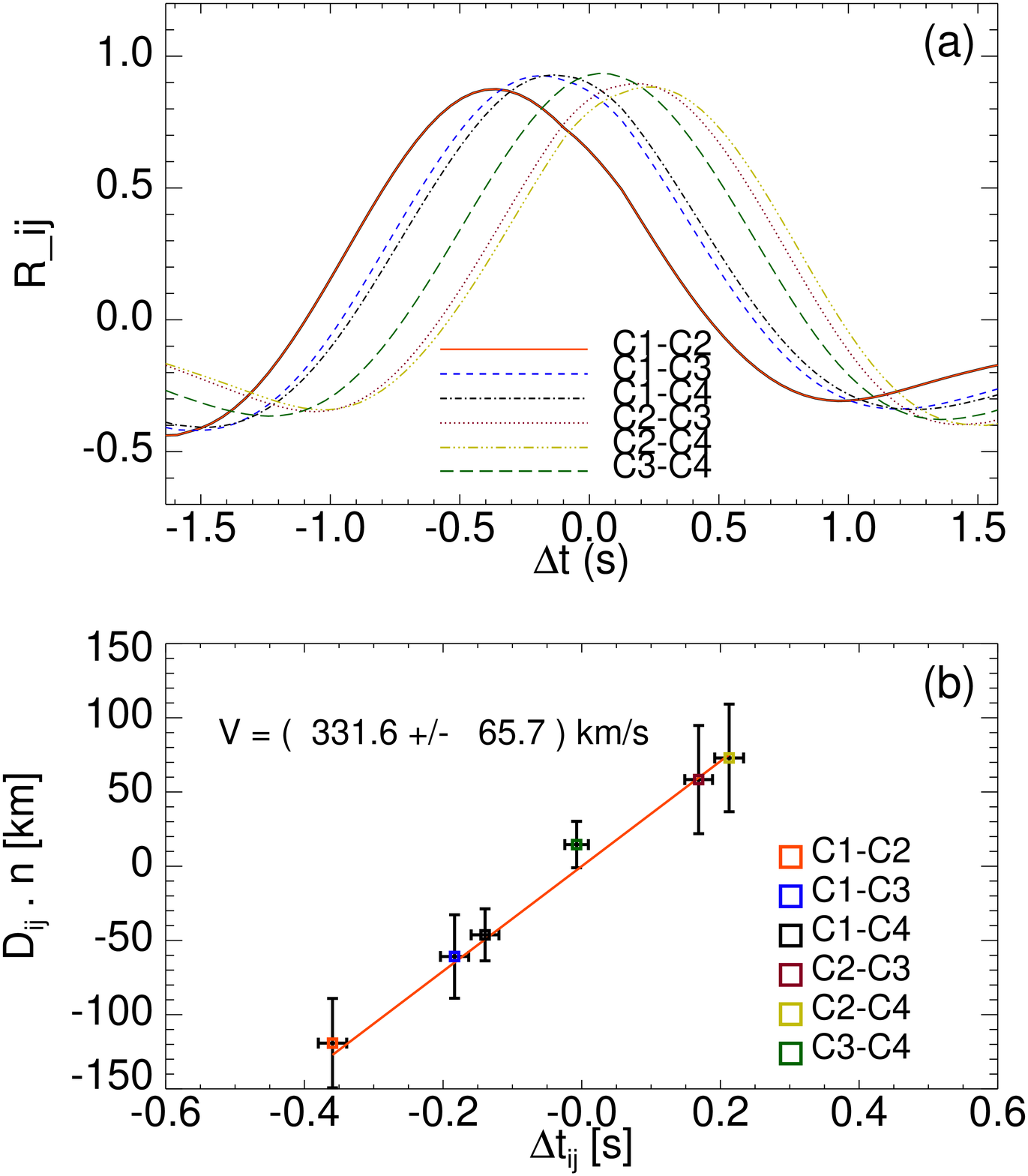}
\caption{Panel (a): Cross-correlation functions, $R_{ij}(\Delta t)$, for the six satellite pairs for the current sheet, shown in Figure~\ref{fg:cur}, as a function of the time lag. Panel (b): Satellite separations along the normal direction as a function of time delays $\Delta t_{ij}$ between the satellites. The liner fit is indicated by the red-solid line.}
\label{fg:cross}
\end{center}
\end{figure}

A one-satellite study provides the information only at a single point in space and no relation between temporal and spatial scales can be obtained except using the Taylor hypothesis. Thanks to the four spacecraft, it is possible to determine a normal to a locally planar structure, the speed along this normal and therefore the spatial scale of the structure without using the Taylor hypothesis. This information allows also the determination of the velocity of the structure in the plasma rest frame.

\subsubsection{Normal, velocity in the satellite frame and spatial scale of the structures }
\label{sub11}

The approach for determining a normal, ${\bf n}$, to the structure and the velocity along this normal, $\mathcal{V}$, is called the {\em timing method}. It is based on the time and space separations \cite[]{sch98}:
\begin{equation}
\label{eq:tim}
{\bf D}_{ij} \cdot \frac{{\bf n}}{\mathcal{V}} = \Delta t_{ij}, \ \ \ i,j=1,2,3,4; \  i \neq j
\end{equation}
where ${\bf D}_{ij} = {\bf D}_{j} - {\bf D}_{i}$ is a separation vector between the satellites $C_i$ and $C_j$, and $\Delta t_{ij}=t_j-t_i$ is a temporal delay between measurements on these two satellites. 
The satellite separations are known with an accuracy of $dD = 0.1$ km. The time separations $\Delta t_{ij}$ can be determined using the cross-correlation function between signals on 2 satellites, as was done, for example, in \cite{ale06}
\begin{equation}
\label{eq:cross_rel}
\mathcal{R}_{ij}(\Delta t)=\frac{\langle \delta {\bf B}_i(t) \cdot \delta {\bf B}_j(t+\Delta t) \rangle}{\sqrt{\langle \delta B^2_i \rangle \langle \delta B^2_j \rangle.}}  \ ;
\end{equation}
here, the angled brackets indicate the time average and $\Delta t$ is a time lag. The time lag, that corresponds to a maximum of $\mathcal{R}_{ij}$, gives the time delay between the satellites, $\Delta t_{ij}$. An error for $\Delta t_{ij}$ could be evaluated taking into account the shape of $\mathcal{R}_{ij}$ and the resolution of magnetic field data. In particular, using Taylor's expansion of $\mathcal{R}_{ij}$ around $\Delta t_{ij}$, this error, $d \Delta t_{ij}$, can be written as
\begin{equation}
d \Delta t_{ij} = \sqrt{\frac{2}{\mathcal{R}^{''}_{ij}(\Delta t_{ij})} \frac{\Delta B}{B}} \ ,
\end{equation}   
where $\mathcal{R}^{''}_{ij}$ is the second derivative of $\mathcal{R}_{ij}$, evaluated in $\Delta t_{ij}$, and $\Delta B/B=7.813 \cdot 10^{-3}$ is the relative error for the FGM instrument \cite[]{bal01}. An example for the determination of the time delays for the current sheet, shown in Figure~\ref{fg:cur}, is given in panel (a) of Figure~\ref{fg:cross}, that shows the six functions $\mathcal{R}_{ij}(\Delta t)$ for the six satellite pairs. The maxima of the functions are well defined and the time lags, corresponding to these maxima, give the time delays, $\Delta t_{ij}$. 

In order to be sure that all the satellites observe the same structure, the time delays have to satisfy the compatibility relation 
\begin{equation}
\label{eq:compat}
\Delta t_{ij}= \Delta t_{ik} + \Delta t_{kj}, \ \ \ \ \Delta t_{ij}=-\Delta t_{ji}
\end{equation}
If the compatibility relation is not satisfied, the different satellites most probably do not observe the same event, even if the maximum of $\mathcal{R}_{ij}$ is high \citep{ale06}.

Another important assumption of the timing method is the planarity of the structures, moving with constant velocity. These assumptions are somehow related to the compatibility condition, eq.~(\ref{eq:compat}). 
As an example, we report in Figure~\ref{fg:cross}(b) the dependence of separation vectors along the normal direction, ${\bf D}_{ij} \cdot {\bf n}$, on $\Delta t_{ij}$, for the same current sheet event. The horizontal error for time delays is given by $d \Delta t_{ij}$, while the error on the ordinate axis depends on the errors of the satellite separations and on the normal direction. The red-solid line represents the linear fit, whose slope gives an estimation of the constant speed of the structure across four satellites, which in this case is $\mathcal{V} = 331.6 \pm 65.7$ km/s.

Summarizing, the timing method keeps validity if the following conditions are satisfied: (i) the four satellites observe the same event; (ii) the compatibility relation (Eq. \ref{eq:compat}) for the time delays is satisfied for all triplets of satellites (that is verified for the locally planar structures propagating with a constant speed); (iii) the maxima of the cross-correlation function $\mathcal{R}_{ij}$ (Eq. \ref{eq:cross_rel}) and the corresponding time delays are well defined. 

In order to fulfill the conditions listed above, for each structure, we verify that:
(1) the time error on the compatibility relation is small: $|\Delta t_{ij} - \Delta t_{ik} - \Delta t_{kj} | < 3\delta t $, $\delta t=0.045$~s being the time resolution of the FGM instrument; (2) there are no zero values in the time separation vector $\Delta t_{ij}$; (3) the relative error on $\mathcal{V}$ is less than 20\%; (4) the difference between $\mathcal{V}$, determined by the timing method, and from the linear fitting in the plane $({\bf D}_{ij} \cdot {\bf n},\Delta t_{ij})$, as shown in Figure~\ref{fg:cross}(b), $\mathcal{V}_{fit}$, is less than twice the minimum of the errors on $\mathcal{V}$ and $\mathcal{V}_{fit}$, i.e. $|\mathcal{V}_{fit}-\mathcal{V}| < 2\cdot \min(d \mathcal{V}_{fit},d \mathcal{V})$\footnote{$d \mathcal{V}_{fit}$ is the 1-sigma fitting error in slope and intercept for $\mathcal{V}_{fit}$, while $d \mathcal{V}$ is obtained from the propagation of the errors for Eq. \ref{eq:tim}:
\begin{equation}
d \mathcal{V} = \mathcal{V} \left( \frac{\langle d \Delta t_{ij} \rangle}{\max{|(\Delta t_{ij})|}} + \frac{dD}{\max{(D_{ij}})} + dn \right),
\end{equation}
where $dn$ is the error on the normal determination (see details in Section \ref{sub12}).}. The four conditions are simultaneously verified by 109 structures from 600, and only for these events we are confident that we are able to determine properly ${\bf n}$ and $\mathcal{V}$ via the multi-satellite analysis.

The red histograms in Figure~\ref{fg:minvar} correspond to the 109 structures, for which ${\bf n}$ and $\mathcal{V}$ can be estimated. These red histograms look like the corresponding black ones (for the total number of structures). This means that the 109 structures belong to a representative subfamily of the 600 selected structures. In the following part of the paper, only the results for these 109 structures are shown.

\begin{figure}
\begin{center}
\includegraphics [width=7.5cm]{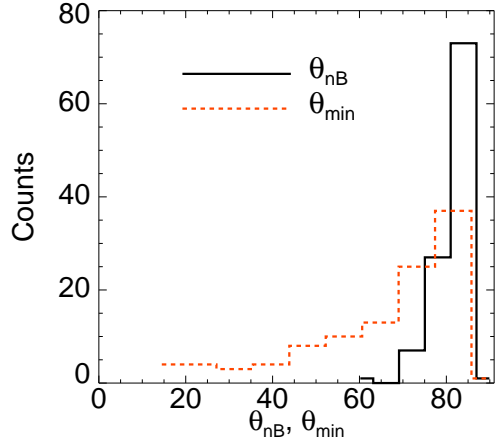}
\caption{Distribution of the angle between the normal of the structures and the local mean magnetic field, $\theta_{nB}$, in black solid line. The distribution of the angle $\theta_{min}$ between ${\bf e_{min}}$ and ${\bf b_0}$ (red dashed line) is given for comparison.}
\label{fg:stat}
\end{center}
\end{figure}

Figure~\ref{fg:stat} shows the angle between ${\bf n}$ and ${\bf b}_0$, $\theta_{nB}$ (black solid histogram), compared with $\theta_{min}$ (red dashed histogram). Even if the distribution of $\theta_{min}$ has its maximum around $\sim 80^{\circ}$, it covers also small angles. The results of the timing show that $\theta_{nB}$ is always close to $\sim 90^{\circ}$, without cases with small angles, i.e. all magnetic coherent structures have $k_{\perp} \gg k_{\parallel}$, a wave-vector anisotropy. The discrepancy between the directions of ${\bf e_{min}}$ and ${\bf n}$ comes from the fact that  ${\bf e_{min}}$ is not well defined for magnetic structures with $\lambda_{int}\simeq \lambda_{min}$. This comparison between $\theta_{nB}$ and $\theta_{min}$ shows that the four-satellite analysis is much more robust in the determination of the normal for the structures, especially for events with $\delta b_{max}\gg \delta b_{int}\sim \delta b_{min}$.  

\begin{figure}
\begin{center}
\includegraphics [width=8.6cm]{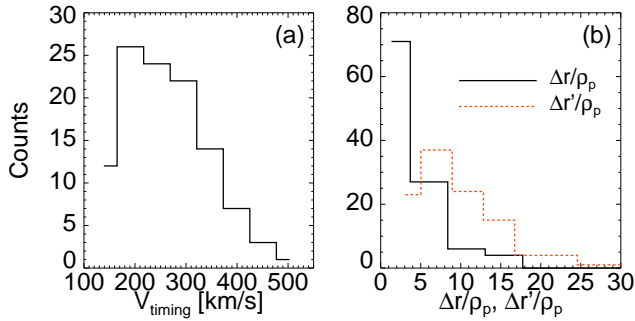}
\caption{Panel (a): Distribution of the velocity along the normal, $\mathcal{V}$, in the satellite frame. 
Panel (b): Distributions of the spatial scale, $\Delta r$ (black solid line), and of the total extension, $\Delta r{'}$  (red dashed line), normalized by $\rho_p$.}
\label{fg:timing}
\end{center}
\end{figure}

The velocity along the normal of the structures is $\mathcal{V} \in [150,500]$~km/s, see Figure~\ref{fg:timing}(a). Using this information, we estimate, along the normal, the spatial scale, $\Delta r$, and the total extension, $\Delta r{'}$, of the structures:
\begin{equation}
\label{eq:taglia}
\Delta r = \mathcal{V} \Delta \tau \ \ \ \ \mbox{and}\ \ \ \  \Delta r{'} = \mathcal{V} \Delta \tau{'}
\end{equation} 
$\Delta \tau$ and $\Delta \tau{'}$ being the time scale and the width of each structure, respectively, as defined above (see Section \ref{sec:met}). Figure~\ref{fg:timing}(b) shows the distributions of $\Delta r$ (black solid line) and $\Delta r{'}$ (red dashed line) normalized by $\rho_p$, with the proton Larmor radius estimated locally inside each structure. The typical scales of the analyzed structures are around $(2-8)\rho_p$, while the total extensions are around $(3-13)\rho_p$. By taking into account the normalization with the ion inertial length, $\lambda_p$ (not shown here), the behavior does not change, but the distributions are more peaked and in particular $\Delta r/\lambda_p$ is around 2 and 5, while $\Delta r{'}/\lambda_p$ is around 5 and 12.

\subsubsection{Velocity of the structures in the plasma frame}
\label{sub12}

\begin{figure}
\begin{center}
\includegraphics [width=9cm]{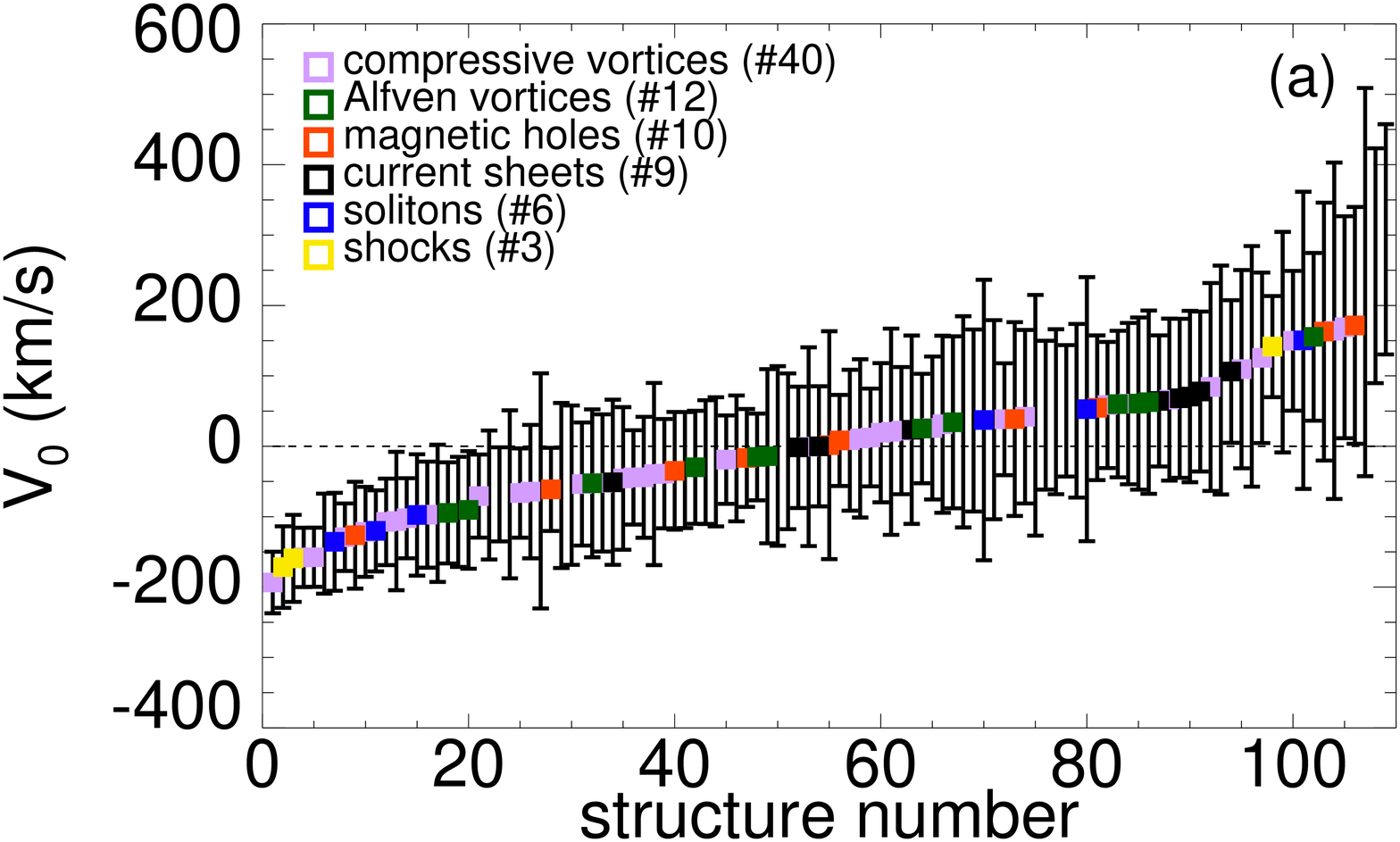}
\includegraphics [width=6.5cm]{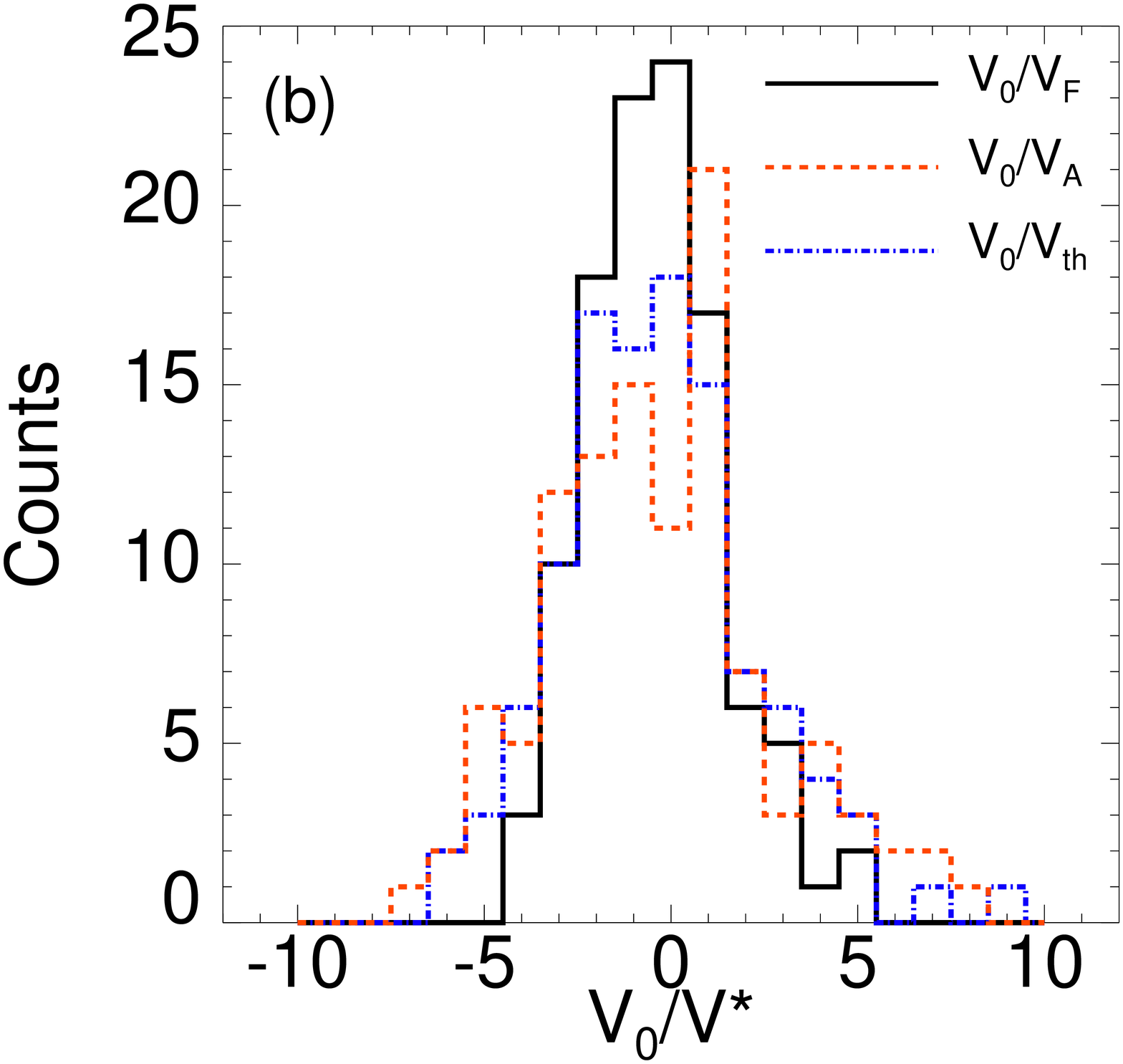}
\caption{Panel (a): Velocity of 109 structures in the plasma frame $\mathcal{V}_0$, with the corresponding error bars and arranged in increasing order of $\mathcal{V}_0$. The different families of the structures are indicated by different colors. The number of the structures, for each  family, is indicated in the legend.
Panel (b): Distribution of $\mathcal{V}_0$ normalized by fast magnetosonic speed, $V_F$, (black solid line), Alfv\'en speed, $V_A$, (green dot-dashed line) and proton thermal speed, $V_{th}$, (blue dashed line).} 
\label{fg:speed}
\end{center}
\end{figure}

Once $\mathcal{V}$ has been estimated by multi-point measurements, it is possible to determine for each coherent event the velocity component along the normal direction, $\mathcal{V}_0$, in the plasma frame:
\begin{equation}
\label{eq:vel}
\mathcal{V}_0 = \mathcal{V} - {\bf v}_{sw} \cdot {\bf n}
\end{equation} 
where ${\bf v}_{sw}$ is the local mean speed of the solar wind. 
The absolute error on $\mathcal{V}_0$ is
\begin{equation}
\label{eq:vel_err}
d\mathcal{V}_0 = d\mathcal{V} + dv_{sw} \cos{\theta_{nV}} + v_{sw} \sin{\theta_{nV}} d\theta_{nV} \ ,
\end{equation}  
where $d\mathcal{V}$ is the error on $\mathcal{V}$, as defined above; $dv_{sw}$ is the error for the CIS/HIA solar wind velocity measurements (about $5 \%$ of the bulk velocity) \cite[]{pas98}; the error $d\theta_{nV}$ includes the angular error for the solar wind velocity vector, that is $\sim 5^{\circ}$ \cite[]{rem97}, and the error on the normal determination (see Section $4.6$ in \cite{vog08}). The angular error of ${\bf n}$ is different for each structure. It has a distribution cone of uncertainty that peaks around $5^{\circ}$, with minimum and maximum values of about $1.5^{\circ}$ and $18^{\circ}$, respectively. 

Figure~\ref{fg:speed}(a) shows the results of the calculations of $\mathcal{V}_0$ for the 109 structures with the corresponding error bars. The structures are ordered by increasing $\mathcal{V}_0$. Although the majority of the structures (about $75 \%$) can be considered, within the limit of the errors, convected by the wind, the remaining part of the structures possesses significant velocities different from zero. In general, the nature of the different structures does not display any correlation with the value of $\mathcal{V}_0$. We have found that magnetic holes (red) and solitons (blue), but also the two kinds of vortices (compressive in lilac and Alfv\'enic in green), can propagate with $|\mathcal{V}_0| > 0$ or be convected by the wind $(\mathcal{V}_0 \simeq 0)$. Although almost all the current sheets (black) are convected, two examples of these structures have the velocity $\mathcal{V}_0$ different from zero. On the other hand, the three examples of shocks (yellow) have clear velocities much greater than zero.

Figure~\ref{fg:speed}(b) displays the distribution of $\mathcal{V}_0$, normalized by the speed for the fast modes $V_F$ (black histogram), by the Alfv\'en speed $V_A$ (red dashed line), and by the proton thermal speed $V_{th}$ (blue dot-dashed line). 
The characteristic velocities are calculated in the upstream region for each structure, which is known from the sign of $\mathcal{V}_0$. The narrowest distribution, which seems to be the most suitable, is found for the histogram of $\mathcal{V}_0/V_F$, showing that most of the structures propagates within $[-3,2]V_F$ limites. But, as discussed above, these limits are due to the error bars on $\mathcal{V}_0$, i.e. the structures from the central part of the histogram are, most probably, just convected by the wind.

\begin{figure}
\begin{center}
\includegraphics [width=9cm]{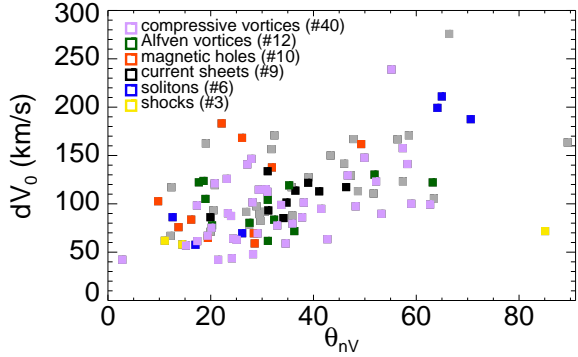}
\caption{The error $d\mathcal{V}_0$ (see eq. (\ref{eq:vel_err})) as a function of the angle between the normal to the structure and the local solar wind velocity, $\theta_{nV}$, for the 109 structures, separated in different classes (different colors). The number of structures, for each different nature, are also indicated in the legend.}
\label{fg:error}
\end{center}
\end{figure}

As one can see from Figure~\ref{fg:speed}(a) the errors $d\mathcal{V}_0$ can be very large. A rough estimate of $d\mathcal{V}_0$ (Eq. \ref{eq:vel_err}) shows that, even though $d\mathcal{V}$ is small, the other two terms can be large. These last terms are sensitive to the value of $\theta_{nV}$ (angle between the normal of the structure and the local solar wind speed). The two limit cases, for $\theta_{nV}$, are: (i) $\theta_{nV} \simeq 0^{\circ}$, which means $d\mathcal{V}_0 = d\mathcal{V}+ dv_{sw} \sim V_A$; (ii) $\theta_{nV} \simeq 90^{\circ}$, which gives $d\mathcal{V}_0 \sim d\mathcal{V}+ v_{sw} d\theta_{nV} \sim 1.3 V_A$ or $\sim 4 V_A$ (for the limit cases of $d\theta_{nV} = 6.5^{\circ}$ and $23^{\circ}$, respectively). All the other cases are a combination of the different contributions. The dependence of $d\mathcal{V}_0$ on $\theta_{nV}$ for all 109 structures is shown in Figure~\ref{fg:error}: higher values of $d\mathcal{V}_0$ are observed for higher values of $\theta_{nV}$. The different classes of structures are indicated by different colors (see legend). The structures whose nature is not clear are presented in grey.

\section{Conclusions and Discussions}
\label{sec:fine}

\begin{figure}
\begin{center}
\includegraphics [width=7.3cm]{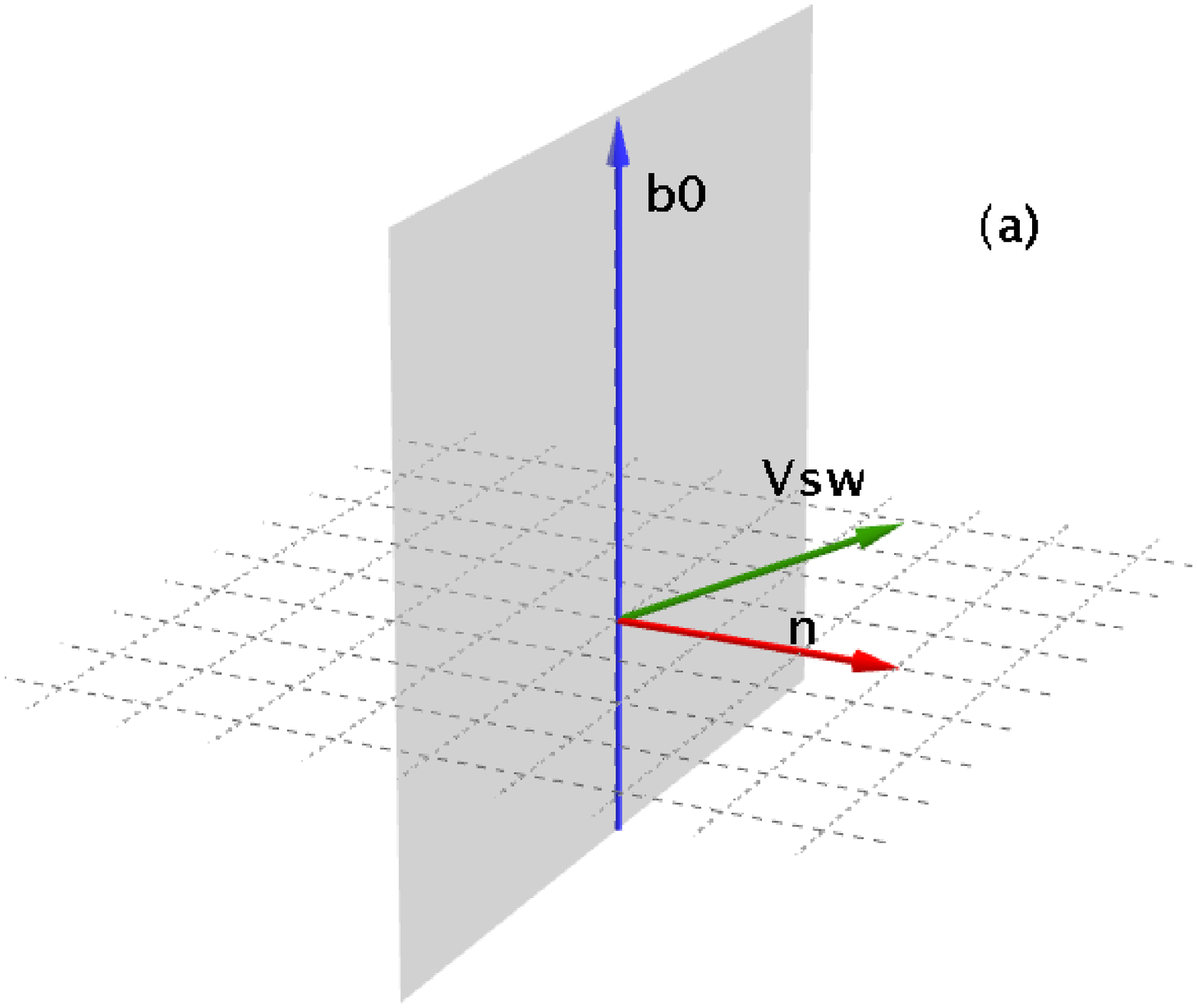}
\includegraphics [width=7.5cm]{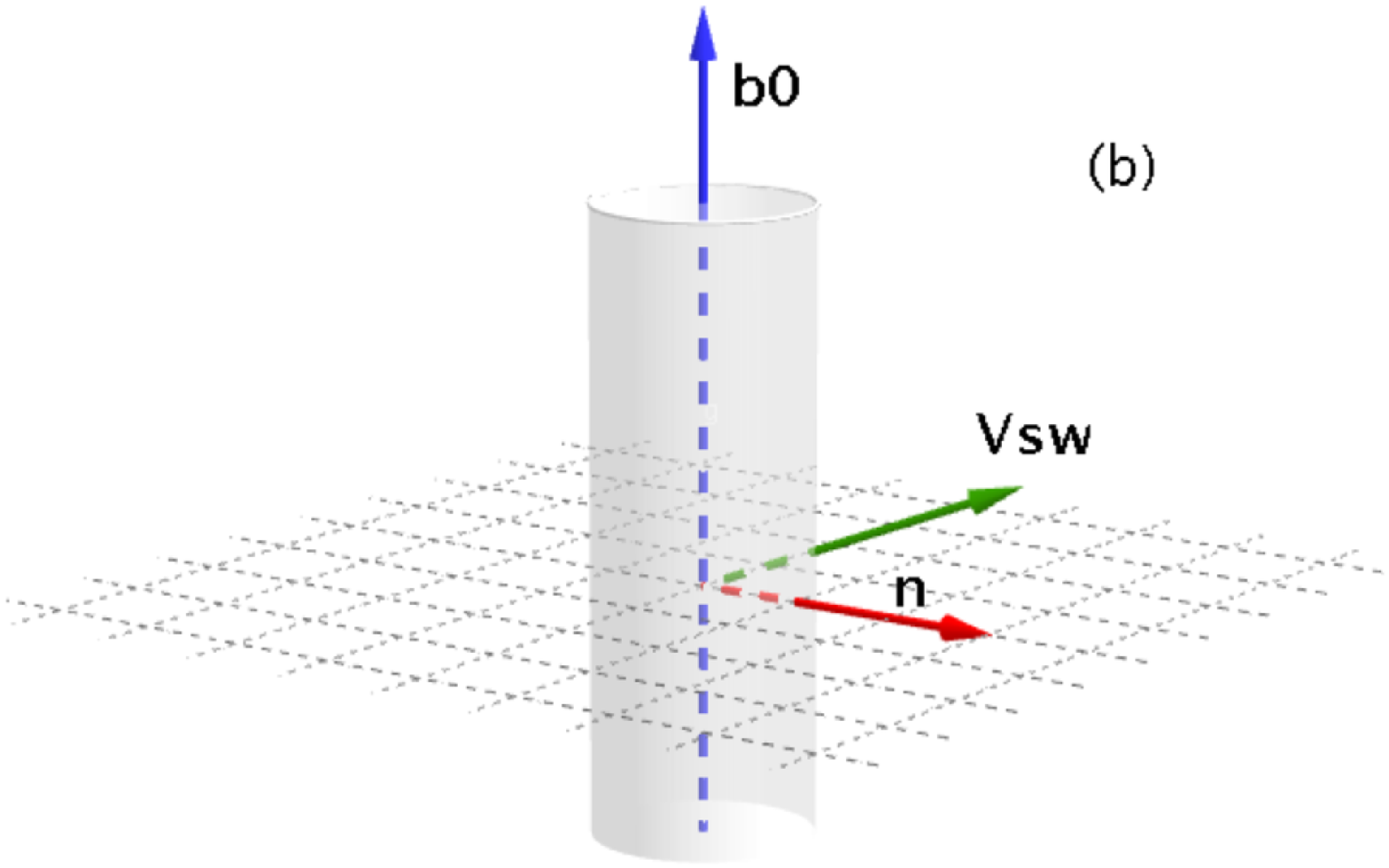}
\caption{Bi-dimensional (a) and three-dimensional (b) geometry models for the structures. The normal of the structures is indicated by the red arrow, the velocity of the flow is represented by the green arrow, while the local magnetic field is in blue.} 
\label{fg:geo}
\end{center}
\end{figure}

In this paper, motivated by an increase of magnetic compressibility around ion characteristic scales, we have studied compressive coherent structures, in a slow solar wind stream, in the frequency range $[0.1,2]$~Hz. The analyzed stream is characterized by a plasma beta which varies over a decade between $0.3$ and 5. 

Different families of structures have been detected. Among strongly compressive events with $\delta b_{\|}\gg \delta b_{\perp}$ we find: (1) magnetic solitons, (2) magnetic holes and (3) shock waves. Examples of Alfv\'enic structures, i.e. with $\delta b_{\perp} > \delta b_{\|}$, are (1) current sheets and (2) vortex-like structures, which can be Alfv\'enic with $\delta b_{\perp} \gg \delta b_{\|}$, but also with  $\delta b_{\perp} \gtrsim \delta b_{\|}$. In this interval of slow solar wind, we have found that the most frequent class of coherent structures are the vortex-like structures and in particular the structures that we have named here compressive vortices.

Thanks to a multi-satellite analysis, we have determined normals ${\bf n}$ to the structures, and the velocity along ${\bf n}$.
Independently of the nature of the structures, ${\bf n}$ is always perpendicular to ${\bf b_0}$. This means that the strongest gradients are in the plane perpendicular to ${\bf b_0}$, i.e. $k_{\perp} \gg k_{\parallel}$. Characteristic spatial scales of the structures along ${\bf n}$ vary between 2 and 20 $\rho_i$ and $\lambda_i$. 

The topology of the observed coherent structures can be (i) a quasi-planar isolated front (compatible with soliton, shock, current sheet) with the mean magnetic field in the plane of the front (see Figure~\ref{fg:geo}(a)); or (ii) a cylinder, or a cigar, with the axis along the local mean magnetic field and the normal parallel to the radius of the cylinder (see Figure~\ref{fg:geo}(b)). This last topology is compatible with the magnetic holes and vortices. 
To confirm the geometry of the structures, a comparison of the signals on the four satellites with different models for different structures should be done (that will be a subject of our future work).

Our multi-satellite analysis shows that the velocity of the structures along ${\bf n}$ in the plasma rest frame is zero (within the error-bars) for $75\%$ of the structures which we could study with four satellites. The remaining 25\% of the structures manifests significant velocities different from zero and may take values of several $V_F$ (fast magnetosonic speed). However, we point out that the errors on the estimated velocities can be huge, meaning that in the limit of these errors the velocities could be lower, but not zero.

The statistical study of all structures does not show any dependences between properties of the structures (size, amplitude, speed) and the plasma parameters. However, considering them by types, we realize that: 
\begin{enumerate}
\item Compressive vortices (40 examples of 109; see Figure~\ref{fg:vor_com}) are the most frequently observed structures, characterized by $\langle \delta b/B_0\rangle \sim 0.1$ and $\xi_{\parallel} > 0.35$ (see eq. (\ref{eq:com_par})). They can be found in the plasma region with both $\beta_p < 1$ and $\beta_p > 1$ ($\langle \beta_p \rangle \sim 1.5$) and for both $T_{\parallel} > T_{\perp}$ and $T_{\parallel} < T_{\perp}$ (proton temperature anisotropy $\langle A \rangle \sim 1.1$). The compressive vortices can propagate with $|\mathcal{V}_0| \in [1,4]V_A$ or be convected by the flow. Their size varies between $\sim 1.5 \rho_p$ and $\sim 18 \rho_p$, while the diameters between $\sim 4.5 \rho_p$ and $\sim 32 \rho_p$.  

\item Alfv\'en vortex-like structures (12 examples of 109; see Figures~\ref{fg:vor_alf} and \ref{fg:vor}) have $\langle \delta b/B_0 \rangle \sim 0.1$ and $\xi_{\parallel} < 0.35$. They are observed for $\langle \beta_p \rangle \sim 1.2$ and isotropic ions. The characteristic propagation speeds $|\mathcal{V}_0| \in [0.5,2] V_A$. Different sizes are also found: from $\sim 2 \rho_p$ to $\sim 8 \rho_p$ and typical diameters between $\sim 5 \rho_p$ and $\sim 17 \rho_p$.

\item Magnetic holes (10 examples of 109; see Figures~\ref{fg:hole} and \ref{fg:hole_bis}) usually have amplitudes $ \langle \delta b/B_0 \rangle \sim 0.06$ and $\langle \xi_{\parallel} \rangle \sim 1.7$, with current density strictly perpendicular to the local magnetic field. They appear in high beta plasma ($\langle \beta_p \rangle \sim 2$), and $T_{\perp} > T_{\parallel}$ $(\langle T_{\perp}/T_{\|} \rangle \sim 1.5)$. 
They are usually convected by the flow in the limits of errors. Typical size $\Delta r \in [2,10] \rho_p$.

\item Current sheets (9 examples of 109; see Figure~\ref{fg:cur}) have $\langle \delta b/B_0 \rangle \sim 0.25$ and are convected by the flow. The plasma parameters are characterized by values of $\beta_p \lesssim 1$ and $T_{\parallel} > T_{\perp}$. The sizes for the current sheets vary between  $\sim 4 \rho_p$ and $\sim 11 \rho_p$.

\item Solitons (6 examples of 109; see Figure~\ref{fg:sol}) have small amplitudes ($\langle \delta b/B_0 \rangle \sim 0.05$) and they are strongly compressive ($\langle \xi_{\parallel} \rangle \sim 1.7$). They are observed for a moderate ion beta ($\langle \beta_p \rangle \sim 1.2$) and almost isotropic ion distributions $(\langle T_{\perp}/T_{\|} \rangle \sim 1.2)$. These structures propagate with the typical velocity of the fast modes, $V_F$ and have characteristic sizes of $\sim 5-6 \rho_p$ and $\lambda_p$.

\item Shocks (3 examples of 109; see Figure~\ref{fg:sho}) have an amplitude of $\langle \delta b/B_0 \rangle \sim 0.05$ and propagate fast in the flow. The first example has $\mathcal{V}_0=(142 \pm 64)$~km/s, Mach numbers $M_F = 2.7$ and $M_A = 4.5$, size of $\sim 7 \rho_p$ and it is observed for $\beta_p \sim 1$ and $T_{\parallel} > T_{\perp}$ ($A \simeq 0.6$), while the second one (see Figure~\ref{fg:sho}) has $\mathcal{V}_0=-(172 \pm 58)$~km/s, Mach numbers $M_F = 2.8$ and $M_A = 4.9$, size of $\sim 4 \rho_p$ and it is found for $\beta_p > 1$ and $A >1$. The third example has almost the same characteristic of the shock in Figure~\ref{fg:sho}; in particular $\mathcal{V}_0=-(160 \pm 51)$~km/s, $M_F = 2.6$, $M_A = 4.3$ and typical size of $\sim 4 \rho_p$. Moreover, also this example of shock is found for $\beta_p > 1$ and $A >1$.
\end{enumerate}

In incompressible MHD theory, one expects to find current sheets and elongated structures, related to the intermittency of the magnetic field \cite[]{car90}. Recently, \cite{ser14b} have shown the existence, in the solar wind, of these equilibria, predicted by the MHD relaxation theory, that spontaneously emerge during the turbulent cascade. 

Most of the studies of plasma discontinuities are based on the use of the partial variance of increments (PVI) technique \cite[]{gre08} or Haar wavelet \cite[]{vel99}. These methods are oriented to catch planar/slab discontinuities in different regimes of the turbulence cascade. Around ion scales (as in the case of the present work), these studies, applied to both solar wind `in situ'  measurements and numerical simulations, reveal mostly the presence of current sheets \cite[]{gre12b,gre14}. Therefore, coherent current sheets were considered as the principal cause of intermittency in space plasma at ion scales.
Here, we have shown for the first time that in the case of the slow solar wind, very different types of coherent structures contribute to the intermittency at proton scales and current sheets are not the most common ones. In our study, we observe only $\sim 10\%$ of coherent structures in form of current sheets and an other $\sim 10\%$ in the form of Alfv\'enic vortices. Otherwise, a considerable part of the structures are compressive vortices ($\sim 35\%$), magnetic holes, solitons and shocks ($\sim 20\%$ in total). The remaining 25\% of the 109 structures are not well identified because of the interaction of adjacent structures with the selected intermittent events.

It is important to point out that the dominance of coherent structures in the form of vortices is not due to the choice of thresholding fluctuations of the parallel component of the magnetic field. In fact, by performing the same analysis in this interval of slow solar wind, but with a threshold on the total magnetic energy, preliminary results show that the nature of the most frequent structures does not change. In 140 observed structures, for which only $40\%$ are well identified, the most common structures remain the vortices ($\sim 26\%$) and not the current sheets ($\sim 11\%$). The main difference between the two analyses is that, in the case of total magnetic fluctuations, the level of the energy threshold is higher with respect to the compressive one. Therefore, in this case, the majority of compressive structures disappears, due to their low energetic nature, and only few examples of magnetic holes and solitons are recognized ($\sim 2\%$). Moreover, the dominant class of vortices in this case are Alfv\'enic ($\sim 80\%$ of the total structures in the form of vortices). 

Otherwise, different results are obtained if a stream of fast solar wind is considered (for example the same interval used in \cite{lea98} and \cite{lio15}). In this stream of fast solar wind observed by the {\it Wind} spacecraft, even if we consider thresholding compressive fluctuations, no compressive structures are found. For 254 structures detected around ion characteristic scales, we have been able to well identify $30\%$ of them, for which the nature of magnetic fluctuations appears to be characterized by the presence of coherent structures in the form of vortices ($\sim 17\%$, including $\sim 87\%$ of Alfv\'enic vortices) and current sheets ($\sim 13\%$). Examples of wave packets are also found ($\sim 8\%$). In conclusion, it seems that the presence of compressive structures are especially related to the characteristic velocity of the solar wind stream. For the future, we will study several intervals of solar wind data with different properties to have a general description of magnetic fluctuations around ion scales.

In the present paper, the most frequent structures (the so called compressive vortices) are characterized by the component of maximum variance almost perpendicular to ${\bf b_0}$ and the intermediate component almost parallel to ${\bf b_0}$. Moreover, the compressive component (as shown in panel (c) of Figure~\ref{fg:vor_com}) is a bit more localized within the structure, while the Alfv\'enic part is more delocalized, extending itself outside $\Delta \tau{'}$. This kind of structures could be described by the preliminary results of an hydrodynamic model of coherent vortices in which warm plasma ($\beta_p \sim \beta_e \sim 1$) with perpendicular spatial scales comparable to the ion characteristic scales. These preliminary results indicate that the Larichev-Reznik type of coherent dipolar structures \cite[]{lar76} may also appear in a high $\beta$ plasma. They are characterized by a very small parallel electric field and by a localized compressional component of the magnetic field in the interior of the vortex core, while the torsional components can have larger spatial extent outside the core edge. This work is in progress and the results will be published elsewhere \cite[]{jov15}. 

On the other hand, another possible explanation for these compressive vortices could be to consider them as a nonlinear evolution of the kinetic Alfv\'en waves (KAW). Recent observational \cite[]{sah10,pod13,rob13,rob15a} and numerical studies \cite[]{vas15} have shown that the ion characteristic scales in turbulent solar wind plasma are characterized by the presence of fluctuations described as KAW. These waves belong to the Alfv\'en branch, with wavelength comparable to the proton inertial length and wave vectors nearly perpendicular to the mean magnetic field. KAW are also characterized by compressive (parallel) magnetic field fluctuations and by a parallel electric field component \cite[]{lee94}. To test this idea, we can evaluate the expected ratio of magnetic compressibility for KAW, as defined in eq.~(55) of \cite{bol13}, inside the detected compressive vortices. The preliminary results show that for compressive vortices with $\xi_{\parallel} \in [0.47,0.7]$, the ratio of magnetic compressibility is in agreement with the prediction for KAW. On the other hand, when $\xi_{\parallel} < 0.47$ and $\xi_{\parallel} > 0.7$, the prediction is not in agreement with observations. Therefore, no definitive conclusion can be made at this point. A more detailed study to describe these structures should be done and this will be a subject of future works.


The other important contribution in our interval is given by one-dimensional (i.e. linearly polarized) compressive structures, such as magnetic holes and solitons. The detected magnetic holes are characterized by $T_{\perp} > T_{\parallel}$, high values of $\beta_p$ and velocity in the plasma frame, $\mathcal{V}_0$, almost zero in the limits of the errors. These characteristics are in agreement with the properties of the mirror mode structures. During the last decades, the structures in the form of solitary magnetic depression and humps, or a combination of both, observed in the interplanetary medium, have been mainly described as nonlinear mirror mode \cite[]{sou08,gen09}. 

In our case, however, we have found that the solitons are not convected by the flow. Therefore, we cannot interpret them as nonlinear mirror modes. Recently, \cite{nar15} proposed a new dissipation mechanism, related to the proton Landau damping of the quasi-perpendicular kinetic slow mode. This mode, linked to the oblique MHD slow mode, has shorter wavelengths going down to the proton inertial length and its phase velocity is the proton thermal speed. Moreover, the kinetic slow mode exhibits a compressive character that is similar to the MHD slow mode (with magnetic field fluctuations nearly aligned with the mean magnetic field). The compressive solitons, described in our paper, propagate perpendicular to the magnetic field with finite velocities in the plasma frame, comparable to the velocity for the fast modes and/or to the proton thermal speed. Unfortunately, due to the errors on the evaluation of the velocity, it is not possible to confirm that the velocity of propagation is exactly the proton thermal speed. However, it could be realistic that the magnetic solitons observed here can be described as a nonlinear evolution of the kinetic slow mode. 

The quasi-perpendicular kinetic slow mode can also lead to the efficient heating of the protons in the parallel direction by the Landau resonance mechanism and maybe in the perpendicular direction by the pitch angle scattering. Due to the low resolution of the particle measurements on {\it Cluster} (4 seconds), there are just one or two points of measurements within an event. Therefore, it is impossible to conclude anything about the heating process at the moment.
A more detailed study of our compressive structures is needed and it will be investigated in future works using kinetic simulations.

The understanding of the physical mechanisms that generate coherent structures and how these events contribute to dissipation in collisionless plasma could provide key insights into the general problem of the solar wind heating. Recently, solar wind measurements and numerical simulations have shown that a connection between kinetic processes and intermittent turbulence exists. Particle heating and acceleration and temperature anisotropy appear localized in and near coherent structures \cite[]{wu13,per13,per14}. This means that the connection between intermittent turbulence, coherent structures and kinetic effects on particle distribution functions cannot be ignored \cite[]{per13b}. However, unfortunately, the existing particle `in situ' measurements have several limitations in temporal, energy and angular resolutions. In particular, due to low time resolution, there are not enough measurements within the structures to study the heating processes at kinetic scales and sometimes this low resolution can generate unphysical effects due to the procedure of data sampling and averaging \cite[]{per14b}. Moreover, accelerated particles, which appear as a beam in the distribution functions, can be resolved only if the resolution in velocity space is sufficiently high. This means that  higher energy and angular resolutions of particle distributions are also crucial. The recent space mission MMS has improved the temporal resolution for the particle measurements, but angular/energy resolution still remains insufficient to resolve solar wind ions. An important contribution in studying dissipation mechanisms in the solar wind, with the best measurements in terms of temporal, energy and angular resolutions for 3D particle distributions, might be provided by the future mission, currently under study, THOR. Such measurements are required to study the connection between coherent structures and kinetic effects on the particle distribution functions.


\section*{acknowledgments}
All Cluster data are obtained from the ESA Cluster Active Archive.We thank the FGM, CIS, WHISPER and PEACE instrument teams and the ESA Cluster Science Archive. D. P. would like to acknowledge 
S. Lion for helpful conversations. 




\begin{thebibliography}{}

\bibitem[Alexandrova et al.(2004)]{ale04} 
Alexandrova, O., Mangeney, A., Maksimovic, Lacombe, C., M., Cornilleau-Wehrlin, N., Lucek, E. A., D\'ecr\'eau, P. M. E., 
Bosqued, J.-M., Travnicek, P. \& Fazakerley, A. N. 2004,
J. Geophys. Res., 109, A05207 

\bibitem[Alexandrova et al.(2006)]{ale06} 
Alexandrova, O., Mangeney, A., Maksimovic, M., Cornilleau-Wehrlin, N., Bosqued, J.-M. \& Andr\'e, M. 2006,
J. Geophys. Res., 111, A12208 

\bibitem[Alexandrova et al.(2007)]{ale07}
Alexandrova, O., Carbone, V., Veltri, P., \& Sorriso-Valvo, L. 2007,
Planet. Spa. Sci., 55, 2224

\bibitem[Alexandrova(2008)]{ale08_n}
Alexandrova, O. 2008, 
Nonlin. Processes Geophys., 15, 95

\bibitem[Alexandrova et al.(2008)]{ale08}
Alexandrova, O., Carbone, V., Veltri, P., \& Sorriso-Valvo, L. 2008,
Astrophys. J., 674, 1153
 
\bibitem[Alexandrova \& Saur(2008)]{ale08b}
Alexandrova, O., \& Saur, J. 2008,
 Geophys. Res. Lett., 35, 15102 

\bibitem[Alexandrova et al.(2009)]{ale09} 
Alexandrova, O., Saur, J., Lacombe, C, Mangeney, A., Mitchell, J., Schwartz, S., J., \& Robert, P. 2009, 
Phys. Rev. Lett., 103, 165003

\bibitem[Alexandrova et al.(2012)]{ale12}
Alexandrova, O., Lacombe, C., Mangeney, A., Grappin, R., \& Maksimovic, M. 2012
Astrophys. J., 760, 121


\bibitem[Bale et al.(2003)]{bal03} 
Bale, S. D., Mozer, F. S., \& Horbury, T. S. 2003, 
Phys. Rev. Lett., 91, 26

\bibitem[Bale et al.(2005)]{bal05} 
Bale, S. D., Kellogg, P. J., Mozer, F. S., Horbury, T. S., \& Reme, H. 2005, 
Phys. Rev. Lett., 94, 215002

\bibitem[Balogh et al.(2001)]{bal01} 
Balogh, A., Carr, C. M., Acu\~na, M. H., Dunlop, M. W., Beek, T. J., Brown, P., Fornacon, K.-H., 
Georgescu, E., Glassameier, K.-H., Harris, J., Musmann, G., Oddy, T., \& Schwingenschuh, K. 2001, 
Ann. Geophys., 19, 1207

\bibitem[Baumg\"artel(1999)]{bau99}
Baumg\"artel, K. 1999,
J. Geophys. Res., 104, 28295

\bibitem[Biskamp(1993)]{bis93}
Biskamp, D. 1993, Nonlinear Magnetohydrodynamics (Cambridge: Cambridge Univ. Press)
 
\bibitem[Boldyrev et al.(2013)]{bol13} 
Boldyrev, S., Horaites K., Xia, Q., \& Perez, J. C. 2013, 
Astrophys. J., 777, 41

\bibitem[Bourouaine et al.(2012)]{bou12} 
Bourouaine, S., Alexandrova, O., Marsch, E., \& Maksimovic, M. 2012, 
Astrophys. J., 749, 102

\bibitem[Bruno et al.(2001)]{bru01}
Bruno, R., Carbone, V., Veltri, P., Pietropaolo, E., \& Bavassano, B. 2001,
Planet. Spa. Sci., 49, 1201

\bibitem[Bruno et al.(2003)]{bru03}
Bruno, R., Carbone, Sorriso-Valvo, L., \& Bavassano, B. 2003,
J. Geophys. Res., 108, 1130

\bibitem[Bruno \& Carbone(2005)]{bru05} 
Bruno, R., \& Carbone, V. 2005, 
Living Rev. Solar Phys., 2, 4

\bibitem[Bruno et al.(2007)]{bru07}
Bruno, R., D'Amicis, R., Bavassano, B., Carbone, V., \& Sorriso-Valvo, L. 2007,
Planet. Spa. Sci., 55, 2233

\bibitem[Bruno et al.(2014)]{bru14}
Bruno, R., Trenchi, L., \& Telloni, D. 2014,
Astrophys. J. Lett., 793, L15

\bibitem[Burlaga(1991)]{bur91}
Burlaga, L. F. 1991, 
J. Geophys. Res., 96, 5847

\bibitem[Burlaga(1993)]{bur93}
Burlaga, L. F. 1993, 
J. Geophys. Res., 98, 17467

\bibitem[Carbone et al.(1990)]{car90}
Carbone, V., Veltri, P., \& Mangeney, A. 1990,
Phys of Fluids A2, 8, 1487

\bibitem[Carbone et al.(1995)]{car95}
Carbone, V., Veltri, P., \& Bruno, R. 1995,
Phys. Rev. Lett., 75, 3110

\bibitem[Carbone et al.(1996)]{car96}
Carbone, V., Bruno, R., \& Veltri, P. 1996,
Geophys. Res. Lett., 23, 121

\bibitem[Chasapis et al.(2015)]{cha15}
Chasapis, A., Retin\`o, A., Sahraoui, F., et al. 2015,
Astrophys. J. Lett., 804, L1 

\bibitem[D\'ecr\'eau et al.(2001)]{dec01}
D\'ecr\'eau, P. M., Fergeau, P., \& Krasnoselskikh, V. 2001,
Ann Geophys., 19, 1241

\bibitem[Dunlop et al.(1988)]{dun88}
Dunlop, M. W., Southwood, D. J., Glassmeier, K.-H.\& Neubauer, F. M. 1988,
Adv. Space Res., 8, 273

\bibitem[Dunlop et al.(2002)]{dun02}
Dunlop, M. W., Balogh, A., Glassmeier, K.-H.\& Robert, P. 2002,
J. Geophys. Res., 107, 1384

\bibitem[Erd\"os \& Balogh(1996)]{erd96}
Erd\"os, G., \& Balogh, A.. 1996,
J. Geophys. Res., 101, 1 

\bibitem[Farge(1992)]{far92}
Farge, M. 1992,
Annual Rev. Fluid Mech., 24, 395

\bibitem[Frisch(1995)]{fri95}
Frisch, U. 1995, 
Turbulence: The legacy of A. N. Kolmogorov (Cambridge: Cambridge Univ. Press)

\bibitem[G\'enot et al.(2009)]{gen09} 
G\'enot , V., Budnik, E., Hellinger, P., Passot, T., Belmont, G., Tr\'avn\'icek, P. M., Sulem, P.-L., Lucek, E.  \& Dandouras, I. 2009, 
Ann. Geophys., 27, 601 

\bibitem[Gosling et al.(2009)]{gos09} 
Gosling, J. T., McComas, D. J., Roberts, D. A., \& Skoug, R. M. 2009, 
Astrophys. J. Lett., 695, L213 

\bibitem[Greco et al.(2008)]{gre08}
Greco, A., Chuychai, P., Matthaeus, W. H., Servidio, S., \& Dmitruk, P. 2008,
Geophys. Res. Lett., 35, L19111

\bibitem[Greco et al.(2012a)]{gre12}
Greco, A., Matthaeus, W. H., D'Amicis, R., Servidio, S., \& Dmitruk, P. 2012a,
Astrophys. J., 749, 105

\bibitem[Greco et al.(2012b)]{gre12b}
Greco, A., Valentini, F., Matthaeus, W. H., Servidio, S., \& Dmitruk, P. 2012b,
Phys. Rev. E, 86, 066405 

\bibitem[Greco \& Perri(2014)]{gre14}
Greco, A., \& Perri, S. 2014,
Astrophys. J., 784, 163

\bibitem[Gustafsson et al.(1997)]{gus97}
Gustafsson, G., Bostrom, R., Holback, B., et al. 1997, 
Space Sci. Rev. 79, 137

\bibitem[Hamilton et al.(2008)]{ham08}
Hamilton, K., Smith, C. W., Vasquez, B. J., \& Leamon, R. L. 2008,
J. Geophys. Res., 113, A01106

\bibitem[Haynes et al.(2015)]{hay15}
Haynes, C. T., Burgess, D., Camporeale, E., \& Sundberg, T. 2015,
Phys. Plasma, 22, 012309

\bibitem[He et al.(2011)]{he11}
He, J., Marsch, E., Tu, C., Yao, S. \& Tian, H. 2011,
Astrophys. J., 731, 85 

\bibitem[He et al.(2012)]{he12}
He, J., Tu, C., Marsch, E., \& Yao, S. 2012,
Astrophys. J. Lett., 745, L8 

\bibitem[Horbury et al.(1997)]{hor97}
Horbury, T. S., Balogh, A., Forsyth, R. J., \& Smith, E. J. 1997,
Adv. Space Res., 19, 847

\bibitem[Jovanovic et al.(in preparation)]{jov15}
Jovanovic, D., Perrone, D., Alexandrova, O., \& Maksimovic, M.,
in preparation 

\bibitem[Kellogg \& Horbury(2005)]{kel05}
Kellogg, P. J., \& Horbury, T. S. 2005,
Ann. Geophys., 23, 3765

\bibitem[Kiyani et al.(2013)]{kiy13}
Kiyani, K. H., Chapman, S. C., Sahraoui, F., Hnat, B., Fauvarque, O., \& Khotyaintsev, Yu. V. 2013,
Astrophys. J., 763, 10

\bibitem[Knetter et al.(2004)]{kne04}
Knetter, T., Neubauer, F. M., Horbury, T., \& Balogh, A. 2004,
J. Geophys. Res., 109, A06102

\bibitem[Kolmogorov(1941)]{kol41} 
Kolmogorov, A. N. 1941, 
Dokl. Akad. Nauk SSSR, 30, 9

\bibitem[Krasnoselskikh et al.(2013)]{kra13}
Krasnoselskikh, V., Balikhin, M., Walker, S. N., Schwartz, S., Sundkvist, D., Lobzin, V., Gedalin, M., Bale, S. D., Mozer, F., Soucek, J., Hobara, Y. \& Comisel, H. 2013,
Space Sci. Rev. 178, 535

\bibitem[Lacombe et al.(2014)]{lac14} 
Lacombe, C., Alexandrova, O., Matteini, L., Santol\'ik, O., Cornilleau-Wehrlin, N., Mangeney, A., 
de Conchy, Y.. \& Maksimovic, M. 2014,
Astrophys. J., 796, 5
 
\bibitem[Larichev \& Reznik(1976)]{lar76}
Larichev, V., \& Reznik, G. 1976,
Rep. USSR Acad. Sci., 231, 1077

\bibitem[Leamon et al.(1998)]{lea98}
Leamon, R. J., Smith, C. W., Ness, N. F., Matthaeus, W. H., \& Wong, H. K. 1998, 
J. Geophys. Res., 103, 4775

\bibitem[Leamon et al.(2000)]{lea00} 
Leamon, R. J. , Matthaeus, W. H., Smith, C. W., Zank, G. P., Mullan, D. J., \& Oughton, S. 2000, 
Astrophys. J., 537, 1054
 
\bibitem[Lee et al.(1994)]{lee94}
Lee, L. C., Johnson, J. R., \& Ma, Z. W. 1994, 
J. Geophys. Res., 99, A9, 17405

\bibitem[Lion et al.(in press)]{lio15} 
Lion, S., Alexandrova, O., \& Zaslavsky, A. 2016, 
Astrophys. J.,  arXiv:1602.07213v1 (in press)

\bibitem[Marsch \& Tu(1994)]{mar94}
Marsch, E., \& Tu, C.-Y. 1994,
Ann. Geophys., 12, 1127

\bibitem[Marsch(2006)]{mar06} 
Marsch, E. 2006, 
Living Rev. Solar Phys., 3, 1

\bibitem[Matteini et al.(2014)]{mat14}
Matteini, L., Horbury, T. S., Neugebauer, M., \& Goldstein, B. E. 2014, 
Geophys. Res. Lett., 41, 259

\bibitem[Matthaeus \& Goldstein(1982)]{mat82}
Matthaeus, W. H., \& Goldstein, M. L. 1982, 
J. Geophys. Res., 87, 6011

\bibitem[Narita \& Marsch(2015)]{nar15}
Narita, Y., \& Marsch, E. 2015, 
Astrophys. J., 805, 24

\bibitem[Newell et al.(2001)]{new01}
Newell, A. C., Nazarenko, S. V., \& Biven, L. 2001,
Physica D 152, 520 

\bibitem[Osman et al.(2011)]{osm11}
Osman, K. T., Matthaeus, W. H., Greco, A., \& Servidio, S. 2011, 
Astrophys. J. Lett. 727, L11

\bibitem[Osman et al.(2012)]{osm12} 
Osman, K. T., Matthaeus, W. H., Hnat, B., \& Chapman, S. C. 2012, 
Phys. Rev. Lett. 108, 261103

\bibitem[Paschmann et al.(1998)]{pas98}
Paschmann, G. J., Fazakerley, A. N., \& Schwzartz, S. J. 1998,
Moments of plasma velocity distributions,
in {\em Analysis Methods for Multi-Spacecraft Data},
ISSI Sci. Rep., ESA Publ. Div., Noordjwick, Netherlands

\bibitem[Pedersen(1995)]{ped95} 
Pedersen, A. 1995, 
Ann. Geophys. 13, 118

\bibitem[Pedersen et al.(2001)]{ped01} 
Pedersen, A., D\'ecr\'eau, P., Escoubet, C.-P., Gustafsson, G., Laakso, H., Lindqvist, 
P.-A., Lybekk, B., Masson, A., Mozer, F., \& Vaivads, A. 2001, 
Ann. Geophys. 19, 1483

\bibitem[Perri et al.(2012)]{per12}
Perri, S., Goldstein, M. L., Dorelli, J. C., \& Sahraoui, F. 2012,
Phys. Rev. Lett. 109, 191101

\bibitem[Perrone et al.(2013a)]{per13} 
Perrone, D., Valentini F., Servidio, S., Dalena, S., \& Veltri, P. 2013a,
Astrophys. J., 762, 99

\bibitem[Perrone et al.(2013b)]{per13b}
Perrone, D., Dendy, R. O., Furno, I., Sanchez, R., Zimbardo, G., Bovet, A., Fasoli, A., 
Gustafson, K., Perri, S., Ricci, P., \& Valentini, F. 2013b, 
Space Sci. Rev. 178, 233

\bibitem[Perrone et al.(2014a)]{per14} 
Perrone, D., Valentini, F., Servidio, S., Dalena, S., \& Veltri, P. 2014a,
Eur. Phys. J. D68, 209

\bibitem[Perrone et al.(2014b)]{per14b}
Perrone, D., Bourouaine, S., Valentini, F., Marsch, E., \& Veltri, P. 2014b,
J. Geophys. Res. 119, 2400

\bibitem[Petviashvili \& Pokhotelov(1992)]{pet92}
Petviashvili, V. I., \& Pokhotelov, O. 1992, 
Solitary Waves in Plasmas and in the Atmosphere (Gordon \& Breach Science Pub)

\bibitem[Pin{\c c}on \& Lefeuvre(1991)]{pin91}
Pin{\c c}on, J.-L., \& Lefeuvre, F. 1991,
J. Geophys. Res., 96, 1789

\bibitem[Podesta \& Gary(2011)]{pod11}
Podesta, J. J., \& Gary, S. P. 2011,
Astrophys. J., 734, 15
 
\bibitem[Podesta(2013)]{pod13}
Podesta, J. J. 2013,
Sol Phys., 286, 529

\bibitem[Rees et al.(2006)]{ree06}
Rees, A., Balogh, A., \& Horbury, T. S. 2006, 
J. Geophys. Res., 111, A10106

\bibitem[R\`eme et al.(1997)]{rem97}
R\`eme et al. 1997,
The Cluster Ion Spectrometry (CIS) experiment,
in {\em The Cluster and Phoenix Missions},
ESA Publ. Div., Noordjwick, Netherlands

\bibitem[R\`eme et al.(2001)]{rem01}
R\`eme, H., Aoustin, C., Bosqued, J. M. et al. 2001, 
Ann. Geophys., 19, 1303

\bibitem[Retin\`o et al.(2007)]{ret07}
Retin\`o, A., Sundkvist, D., Vaivads, A. et al. 2007,
Nature Phys., 3, 236

\bibitem[Roberts et al.(2013)]{rob13}
Roberts, O. W., Li, X., \& Li, B. 2013,
Astrophys. J., 769, 58
 
\bibitem[Roberts et al.(2015)]{rob15a}
Roberts, O. W., Li, X., \& Jeska, L. 2015,
Astrophys. J., 802, 2

\bibitem[Roberts et al.(submitted, 2016)]{rob15} 
Roberts, O. W., Li, X., Alexandrova, O., \& Li, B. 2016
J. Geophys. Res., arXiv:1602.07410v1 (in press)

\bibitem[Sahraoui et al.(2010)]{sah10} 
Sahraoui, F., Goldstein, M. L., Belmont, G., Canu, P., \& Rezeau, L. 2010, 
Phys. Rev. Lett., 105, 131101

\bibitem[Sahraoui et al.(2013)]{sah13} 
Sahraoui, F., Huang, S. Y., Belmont, G., Goldstein, M. L., Retin\`o, A., Robert, P., \& De Patoul, J. 2013, 
Astrophys. J., 777, 15

\bibitem[Salem et al.(2009)]{sal09}
Salem, C., Mangeney, A., Bale, S. D., \& Veltri, P. 2009,
Astrophys. J., 702, 537

\bibitem[Salem et al.(2012)]{sal12}
Salem, C. S., Howes, G. G., Sundkvist, D., Bale, S. D., Chaston, C. C., Chen, C. H. K., \& Mozer, F. S. 2012,
Astrophys. J. Lett., 745, L9

\bibitem[Schwartz(1998)]{sch98}
Schwartz, S. J. 1998,
Shock and discontinuity normals, Mach numbers,
and related parameters,
in {\em Analysis Methods for Multi-Spacecraft Data},
ISSI Sci. Rep., ESA Publ. Div., Noordjwick, Netherlands

\bibitem[Servidio et al.(2012)]{ser12} 
Servidio, S., Valentini, F., Califano, F., \& Veltri, P. 2012, 
Phys. Rev. Lett., 108, 045001

\bibitem[Servidio et al.(2014a)]{ser14} 
Servidio, S., Osman, K. T., Valentini, F., Perrone, D., Califano, F., 
Chapman, S. C., Matthaeus, W. H., \& Veltri, P. 2014a, 
Astrophys. J. Lett., 781, L27

\bibitem[Servidio et al.(2014b)]{ser14b} 
Servidio, S., Gurgiolo, C., Carbone, V., \& Goldstein, M. L. 2014b, 
Astrophys. J. Lett., 789, L44

\bibitem[She et al.(1990)]{she90} 
She, Z.-S., Jackson, E., \& Orszag, S. A. 1990, 
Nature, 344, 226

\bibitem[Smith et al.(2006)]{smi06}
Smith, C. W., Hamilton, K., \& Vasquez, B. J. 2006, 
Astrophys. J. Lett., 645, L85


\bibitem[Sonnerup \& Scheible(1998)]{son98}
Sonnerup, B., \& Scheible, M. 1998,
Minimum and maximum variance analysis,
in {\em Analysis Methods for Multi-Spacecraft Data},
ISSI Sci. Rep., ESA Publ. Div., Netherlands


\bibitem[Sorriso-Valvo et al.(2001)]{sor01}
Sorriso-Valvo, L., Carbone, Giuliani, P., V., Veltri, P., Bruno, R., Antoni, V., \& Martines, E. 2001,
Planet. Space Sci., 49, 1193

\bibitem[Sorriso-Valvo et al.(2005)]{sor05}
Sorriso-Valvo, L., Carbone, V., \& Bruno, R. 2005,
Space Sci. Rev., 121, 49

\bibitem[Soucek et al.(2008)]{sou08}
Soucek, J., Lucek, E., \& , Dandouras, I. 2008,
J. Geophys. Res., 113, A04203

\bibitem[Stasiewicz et al.(2003)]{sta03}
Stasiewicz, K., Shukla, P. K., Gustafsson, G., Buchert, S., Lavraud, B., Thidé, B., \& Klos, Z. 2003,
Phys. Rev. Lett., 90, 085002

\bibitem[Stevens \& Kasper(2007)]{ste07}
Stevens, K. L., \& Kasper, J. C. 2007
J. Geophys. Res., 112, A05109

\bibitem[Sundkvist et al.(2005)]{sun05}
Sundkvist, S., Krasnoselskikh, V., Shukla, P. K., Vaivads, A., Andr\'e, M., Buchert, S. \& R\`eme, H. 2005
Nature, 436, 825

\bibitem[Sundkvist et al.(2007)]{sun07}
Sundkvist, S., Retin\`o, A., Vaivads, A. \& Bale, S. D. 2007
Phys. Rev. Lett., 99, 025004

\bibitem[Szita et al.(2001)]{szi01}
Szita, S., Fazakerley, A. N., Carter, P. J., James, A. M., Travnicek, P., Watson, G., Andr\'e, M., Eriksson, A. \& Torkar, K. 2001
Ann. Geophys., 19, 1721

\bibitem[Tessein et al.(2013)]{tes13}
Tessein, J. A., Matthaeus, W. H., Wan, M., Osman, K. T., Ruffolo, D., \& Giacalone, J. 2013,
Astrophys. J. Lett., 776, L8

\bibitem[Torrence \& Combo(1998)]{tor98}
Torrence, C., \& Combo, G. P. 1998,
Bull. Am. Meteorol. Soc., 79, 61

\bibitem[Tu \& Marsch(1995)]{tu95}
Tu, C. Y., \& Marsch, E. 1995, 
Space Sci. Rev., 73, 1

\bibitem[Turner et al.(1977)]{tur77}
Turner, J. M., Burlaga, L. F., Ness, N. F., \& Lemaire, J. F. 1977,
J. Geophys. Res., 82, 1921
 
\bibitem[V\'asconez et al.(2015)]{vas15}
V\'asconez, C. L., Pucci, F., Valentini, F., Servidio, S., Matthaeus, W. H., \& Malara, F. 2015,
Astrophys. J., 815, 7


\bibitem[Veltri \& Mangeney(1999)]{vel99}
Veltri, P., \& Mangeney, A. 1999, 
in AIP Conf. Proc. 471, Solar Wind IX ed. S. Habbal (USA), 543

\bibitem[Veltri(1999)]{vel99b}
Veltri, P. 1999,
Plasma Phys. Control. Fusion, 41, A787

\bibitem[Verkhoglyadova et al.(2003)]{ver03}
Verkhoglyadova, O. P., Dasgupta, B., \& Tsurutani, B. T. 2003, 
Nonlinear Proc. in Geophys., 10, 335

\bibitem[Vogt et al.(2008)]{vog08}
Vogt, J., Paschmann, G., Chanteur, G. 2008,
Reciprocal vectors,
in {\em Multi-Spacecraft Analysis Methods Revisited},
ISSI Sci. Rep., ESA Publ. Div., Noordjwick, Netherlands

\bibitem[Wan et al.(2012a)]{wan12a}
Wan, M., Osman, K. T., Matthaeus, W. H., \& Oughton, S. 2012a,
Astrophys. J., 744, 171

\bibitem[Wan et al.(2012b)]{wan12b}
Wan, M., Matthaeus, W. H., Karimabadi, H., Roytershteyn, V., Shay, M. A., Wu, P. , 
Daughton, W., Loring, B., \& Chapman, S. C. 2012b
Phys. Rev. Lett., 109, 195001

\bibitem[Winterhalter et al.(1994)]{win94}
Winterhalter, D., Neugebauer, M., Goldstein, B. E., Smith, E. J., Bame, S. J., \& Balogh, A. 1994
J. Geophys. Res., 99, 23371

\bibitem[Wu et al.(2013)]{wu13}
Wu, P., Perri, S., Osman, K. T., Wan, M., Matthaeus, W. H., Shay, M. A., Goldstein, M. L.,
Karimabadi, H., \& Chapman, S. C. 2013,
Astrophys. J. Lett. 763, L30 

\bibitem[Zhdankin et al.(2012)]{zhd12}
Zhdankin, V., Boldyrev, S., Mason, J., \& Perez, J. C. 2012,
Phys. Rev. Lett., 108, 175004



\end{thebibliography}
\end{document}